\documentclass[longauth]{aa}  

\usepackage{comment}
\usepackage{graphicx}
\usepackage{multirow}
\usepackage{booktabs}
\usepackage{textgreek}
\usepackage{xargs}
\usepackage[dvipsnames]{xcolor}
\usepackage{xspace}
\usepackage{txfonts}
\usepackage{hyperref}
\hypersetup{
    colorlinks=true,
    linkcolor=blue,
    citecolor=blue,
    filecolor=magenta,      
    urlcolor=cyan,
    }
\usepackage{subcaption}
\usepackage{adjustbox}

\newcommand{\source}{{\sc MARTA-4327}\xspace}
\newcommand{\sourceshort}{{\sc M4327}\xspace}

\let\oldtextsigma\textsigma
\renewcommand{\textsigma}{\oldtextsigma\xspace}
\let\oldtextalpha\textalpha
\renewcommand{\textalpha}{\oldtextalpha\xspace}
\let\oldAA\AA
\renewcommand{\AA}{\text{\oldAA}\xspace}
\let\oldtextdegree\textdegree
\renewcommand{\textdegree}{\oldtextdegree\xspace}

\newcommand{\MSun}{\ensuremath{{\rm M}_\odot}\xspace}

\newcommand{\AV}{\ensuremath{A_\mathrm{V}}\xspace}

\newcommandx{\Mout}[2][1=,2=]{\ensuremath{M_{\mathrm{out}{#2}}^{#1}}\xspace}
\newcommandx{\Mdotout}[2][1=,2=]{\ensuremath{\dot{M}_{\mathrm{out}{#2}}^{#1}}\xspace}

\newcommandx{\fluxdcgs}[1][1=-20]{\ensuremath{\mathrm{10^{#1}~erg~s^{-1}~cm^{-2}~\AA^{-1}}}\xspace}
\newcommandx{\fluxcgs}[1][1=-20]{\ensuremath{\mathrm{10^{#1}~erg~s^{-1}~cm^{-2}}}\xspace}
\newcommandx{\powercgs}[1][1=44]{$\times 10^{#1}$~erg~s$^{-1}$\xspace}
\newcommand{\Av}{\ensuremath{A_V}\xspace}

\newcommand{\Te}{\ensuremath{T_\text{e}}\xspace}
\newcommand{\Tiii}{\Te[\ion{O}{iii}]\xspace}
\newcommand{\Tii}{\Te[\ion{O}{ii}]\xspace}

\newcommand{\Ne}{n$_{\text{e}}$\xspace}

\newcommand{\EWr}{EW$_{\text{0}}$\xspace}

\newcommand{\jwst}{\textit{JWST}\xspace}

\newcommand{\Halpha}{\text{H\textalpha}\xspace}
\newcommand{\Hbeta}{\text{H\textbeta}\xspace}
\newcommand{\Hgamma}{\text{H\textgamma}\xspace}
\newcommand{\Hdelta}{\text{H\textdelta}\xspace}

\newcommandx{\permittedEL}[6][1=O,2=III,3=,4=,5=,6=]{\text{{#1}\,{\sc {#2}}{#3}{#4}{#5}{#6}}\xspace}
\newcommandx{\semiforbiddenEL}[6][1=O,2=III,3=,4=,5=,6=]{\text{{#1}\,{\sc{#2}}]{#3}{#4}{#5}{#6}}\xspace}
\newcommandx{\forbiddenEL}[6][1=O,2=III,3=,4=,5=,6=]{\text{[{#1}\,{\sc{#2}}]{#3}{#4}{#5}{#6}}\xspace}

\newcommandx{\NVL}[1][1=1243]{\permittedEL[N][v][\textlambda][#1]}
\newcommandx{\NVall}{\permittedEL[N][v][\textlambda][\textlambda][1239,][1243]}

\newcommandx{\NIVL}[1][1=1486]{\semiforbiddenEL[N][iv][\textlambda][#1]}

\newcommand{\CIV}{\permittedEL[C][iv]}
\newcommandx{\CIVL}[1][1=1550]{\permittedEL[C][iv][\textlambda][#1]}

\newcommand{\HeII}{\permittedEL[He][ii]}
\newcommandx{\HeIIL}[1][1=1640]{\permittedEL[He][ii][\textlambda][#1]}

\newcommandx{\OIIIL}[1][1=1666]{\semiforbiddenEL[O][iii][\textlambda][#1]}

\newcommandx{\OIIIoptL}[1][1=5007]{\forbiddenEL[O][iii][\textlambda][#1]}

\newcommandx{\NIIIL}[1][1=1750]{\semiforbiddenEL[N][iii][\textlambda][#1]}

\newcommandx{\CIII}{\semiforbiddenEL[C][iii]}
\newcommandx{\CIIIL}[1][1=1909]{\semiforbiddenEL[C][iii][\textlambda][#1]}

\newcommand{\CIIIp}{\permittedEL[C][iii]}

\newcommand{\NIIp}{\permittedEL[N][ii]}
\newcommand{\NIIIp}{\permittedEL[N][iii]}
\newcommand{\NIVp}{\permittedEL[N][iv]}
\newcommand{\NVp}{\permittedEL[N][v]}
\newcommand{\SiIIIp}{\permittedEL[Si][iii]}

\newcommandx{\NeIVL}[1][1=2424]{\forbiddenEL[Ne][iv][\textlambda][#1]}

\newcommandx{\MgIIL}[1][1=2803]{\permittedEL[Mg][ii][\textlambda][#1]}

\newcommandx{\NeVL}[1][1=3426]{\forbiddenEL[Ne][v][\textlambda][#1]}

\newcommand{\OII}{\forbiddenEL[O][ii]}
\newcommandx{\OIIL}[1][1=3727]{\forbiddenEL[O][ii][\textlambda][#1]}

\newcommand{\OI}{\forbiddenEL[O][i]}
\newcommandx{\OIL}[1][1=6300]{\forbiddenEL[O][i][\textlambda][#1]}

\newcommand{\OIIaur}{\forbiddenEL[O][ii][\textlambda][\textlambda][7320,][7330]}

\newcommand{\SII}{\forbiddenEL[S][ii]}
\newcommandx{\SIIL}[1][1=6725]{\forbiddenEL[S][ii][\textlambda][#1]}

\newcommandx{\SIIIL}[1][1=9068]{\forbiddenEL[S][iii][\textlambda][#1]}

\newcommandx{\NeIIIL}[1][1=3869]{\forbiddenEL[Ne][iii][\textlambda][#1]}

\newcommandx{\ArIIIL}[1][1=7135]{\forbiddenEL[Ar][iii][\textlambda][#1]}

\newcommand{\NII}{\forbiddenEL[N][ii]}
\newcommandx{\NIIL}[1][1=6585]{\forbiddenEL[N][ii][\textlambda][#1]}

\newcommandx{\FeIIIL}[1][1=4658]{\forbiddenEL[Fe][iii][\textlambda][#1]}
\newcommandx{\FeIII}{\forbiddenEL[Fe][iii]}

\graphicspath{{./}{figs/}}

\begin{document} 

   \title{MARTA: The connection between chemical enrichment, feedback, and dust in a Wolf-Rayet galaxy at z$\sim$2}

   \titlerunning{A Wolf-Rayet galaxy at z$\sim$2}
   
   \author{Mirko Curti\inst{\ref{eso}}
          \thanks{E-mail: mirko.curti@eso.org}          
\and
Elisa Cataldi\inst{\ref{UNIFI},\ref{arcetri}}\and
Francesco Belfiore\inst{\ref{eso},\ref{arcetri}}\and
Bianca Moreschini\inst{\ref{UNIFI},\ref{arcetri}} \and
Magda Arnaboldi\inst{\ref{eso}} \and
Martyna Chru\'sli\'nska\inst{\ref{eso}} \and
Filippo Mannucci\inst{\ref{arcetri}} \and
Alessandro Marconi\inst{\ref{UNIFI},\ref{arcetri}} \and
Quirino D'Amato\inst{\ref{arcetri}} \and
Stefano Carniani\inst{\ref{SNS}} \and
William M. Baker\inst{\ref{dark}} \and
Annalisa De Cia\inst{\ref{eso}} \and
Nimisha Kumari\inst{\ref{aura}} \and
Amirnezam Amiri\inst{\ref{arkansas}} \and
Giovanni Cresci\inst{\ref{arcetri}} \and
Chiaki Kobayashi\inst{\ref{herts}} \and
Fergus Cullen\inst{\ref{ROE}} \and
Anna Feltre\inst{\ref{arcetri}} \and
Roberto Maiolino\inst{\ref{cavensish}, \ref{kicc}} \and
Irene Shivaei\inst{\ref{CAB}} 
}
   \institute{
\label{eso} European Southern Observatory, Karl-Schwarzschild Straße 2, D-85748 Garching bei München, Germany \and
\label{UNIFI} Università di Firenze, Dipartimento di Fisica e Astronomia, via G. Sansone 1, 50019 Sesto Fiorentino, Florence, Italy \and
\label{arcetri} INAF -- Arcetri Astrophysical Observatory, Largo E. Fermi 5, I-50125, Florence, Italy \and
\label{SNS} Scuola Normale Superiore, Piazza dei Cavalieri 7, I-56126 Pisa, Italy \and
\label{dark} DARK, Niels Bohr Institute, University of Copenhagen, Jagtvej 155A, DK-2200 Copenhagen, Denmark
\label{aura} AURA for European Space Agency, Space Telescope Science Institute, 3700 San Martin Drive. Baltimore, MD, 21210 \and
\label{arkansas} Department of Physics, University of Arkansas, 226 Physics Building, 825 West Dickson Street, Fayetteville, AR 72701, USA \and
\label{herts} Centre for Astrophysics Research, Department of Physics, Astronomy and Mathematics, University of Hertfordshire, Hatfield AL10 9AB, UK \and
\label{ROE} Institute for Astronomy, University of Edinburgh, Royal Observatory, Edinburgh, EH9 3HJ, UK \and
\label{cavensish} Cavendish Laboratory, University of Cambridge, 19 JJ Thomson Avenue, Cambridge, CB3 0HE, UK \and
\label{kicc} Kavli Institute for Cosmology, University of Cambridge, Madingley Road, Cambridge, CB3 0HA, UK \and
\label{CAB} Centro de Astrobiología (CAB), CSIC-INTA, Carretera de Ajalvir km 4, Torrejón de Ardoz, E-28850, Madrid, Spain
}
   \authorrunning{M. Curti et al.}
   \date{}

  \abstract
  
   \date{}
   
\abstract
{We present the analysis of the stellar and interstellar medium (ISM) properties of \source, a star-forming galaxy at z=2.223 observed by means of deep JWST/NIRSpec spectroscopy in both medium- and high-resolution gratings as part of the "Measuring Abundances at high Redshift with the Te Approach" (MARTA) program. 
We report one of the highest-redshift detections of the Wolf–Rayet (WR) blue and red bumps in a non-lensed system. 
The broad \HeIIL[4686] feature is consistent with a young ($\sim$5--6~Myr) burst dominated by WN stars, although both  stellar population synthesis models and empirical templates struggle to reproduce nitrogen stellar features at $\approx4640\AA$.  
Based on the relative strength of the available optical stellar features, we disfavor the presence of very massive stars (VMS) in this system. 
Elemental abundance ratios such as Ne/O, N/O, and Ar/O align with observations of local star-forming galaxies (including WR galaxies), suggesting that any impact of the WR population on the chemical enrichment of the ISM is strongly localized. 
However, the gas-phase Fe/O ratio appears enhanced compared to local galaxies of similar metallicity, which we interpret as evidence for reduced Fe depletion onto dust grains, possibly linked to localized destruction in WR-driven wind environments. In addition, we detect a broad ($\sim445$ km s$^{-1}$) and blueshifted ($\sim70$ km s$^{-1}$) H$\alpha$ component, suggesting the presence of an ionized outflow with a mass loading factor $\eta \sim 0.2$. Finally, we report the robust detection of O\,{\sc i}$\lambda8446$ emission (among the firsts at high redshift), which we interpret as originating from Ly$\beta$ fluorescence and/or collisional excitation in dense clumps. Overall, \source represents a unique system for studying the role of massive stars in shaping the ISM properties of galaxies at Cosmic Noon. 
}


   \maketitle
%
%________________________________________________________________

\section{Introduction}
Characterizing the physics of galaxies at redshift z$\sim 2-3$, during the peak of cosmic star-formation, feedback, and black hole activity, has represented one of the most relevant goals for the astronomical community over almost the past three decades.
With the advent of 8-10m-class telescopes, paramount efforts have been devoted to perform near-infrared (NIR) spectroscopy from the ground, aimed at observing the optical spectral features required to probe the physical conditions of gas and stars in these systems \citep[e.g.,][]{Pettini_LBGs_2001, steidel_lyman_2003, erb_h_2006, maiolino_amaze_2008, mannucci_lsd_2009,shapley_physical_2011,  steidel_strong_2014, sanders_mosdef_2015, shapley_mosdef_2015, strom_measuring_2018, kashino_fmos-cosmos_2019}. 
Because of the large cosmological distance of these sources, however, in most cases the investigation was limited to the brightest emission lines detectable in the spectra, originating from either recombination of hydrogen atoms or collisional excitation of some of the most abundant metal ions (e.g., oxygen, nitrogen, sulfur) present in the gas-phase of the interstellar medium (ISM) of galaxies.
Therefore, despite the tremendous advancement in our understanding of galaxy properties at these epochs provided by such extensive programs, several spectral features remained just too faint to be probed by ground-based instrumentation, hiding within the noise of even the deepest observations.

Among such features, auroral lines of metal ions, sensitive to the electron temperature (\Te) of the gas and hence key probes of the physics of chemical enrichment in galaxies, have been pursued for a long time.
Although a handful of individual low signal-to-noise (S/N) measurements have been reported in the past using ground-based telescopes \citep[e.g.,][]{christensen_gravitationally_2012, sanders_mosdef_2016-1, patricio_testing_2018}, these were not representative of the "typical" galaxy population at z$\sim$2. 
Information on the metallicity of sources at Cosmic Noon has therefore been inferred by means of indirect tracers constituted by ratios of bright emission lines calibrated on local samples of galaxies \citep[e.g.,][]{pettini_oiiinii_2004, maiolino_amaze_2008, curti_new_2017}, or on samples of analogs of high-$z$ sources \citep[e.g.,]{bian_ldquodirectrdquo_2018}, despite the potential biases induced by the different physical conditions among which nebular emission occurs in early galaxies compared to the local Universe.

The advent of the \textit{James Webb} Space Telescope (JWST) rapidly changed the landscape of the field, as several auroral line detections have been reported within the first years of operations \citep[e.g.,][]{curti_smacs_2023, nakajima_mzr_ceers_2023, sanders_calibrations_2023, laseter_auroral_jades_2023}, allowing for a more "direct" probe of the metallicity of the gas in high-$z$ sources.
Notably, the vast majority of such detections concerned the high ionization \OIIIoptL[4363] auroral line, which was preferentially detected in spectra of z$\gtrsim$5 galaxies as possibly favored by the combination of high ionization parameter and low metallicity, coupled with the higher sensitivity of JWST/NIRSpec beyond 2$\mu$m.

Among JWST projects approved in Cycle 1, the "Measuring Abundances at high Redshift with the Te Approach" (MARTA) Programme (PID 1879) was primarily aimed at delivering high S/N rest-frame optical spectra of individual z$\sim$2 sources by means of very deep integrations in G140M/F100LP and G235M/F170LP grating/filter configurations, allowing a detailed characterization of the physical conditions of the ISM in $z\sim2$ galaxies leveraging the detection of both bright and faint emission lines \citep{Cataldi_MARTA_2025}.
Thanks to this and other dedicated observational efforts such as the CECILIA \citep{Strom_Cecilia_2023}, AURORA \citep{Shapley_AURORA_BPT_2025}, and EXCELS \citep{Carnall_EXCELS_quiescent_2024} surveys, the number of auroral line measurements (including low-ionization ionic species) at $z\sim2-5$ is rapidly increasing, allowing not only to directly test the behavior of strong-line metallicity diagnostics over this largely unexplored redshift range \cite[e.g.,][]{Chakraborty_strong_lines_2024, Scholte_EXCELS_2025, Cataldi_MARTA_2025, Sanders_aurora_calibrations_2025}, but also to get insights into the star-formation history of galaxies from detailed studies of multiple elemental abundance patterns \citep[e.g.,][]{Rogers_Cecilia_z3_2024, Bhattacharya_argon_2025, Arellano-Cordova_CNO_EXCELS_2024, Stanton_excels_2025}, shedding light on the role of massive stars in driving early chemical enrichment.

In this context, galaxies hosting Wolf–Rayet (WR) stars offer a rare opportunity to study the impact of massive stellar populations on the ISM. At low redshift, WR galaxies exhibit localized chemical enrichment \citep[e.g.,][]{lopez_sanchez_WR_abundances_2010}, shock-excited gas \citep[e.g.,][]{brinchmann_bpt_highz_2008}, and enhanced Fe emission lines \citep{Izotov_J0811_2018, Kojima2021}, suggesting active metal injection and dust processing. 
However, little is known about their counterparts at z$\sim2$, when star formation and feedback processes were most intense, as WR features tend to be elusive (also in light of the short lifetimes of these stars) and have only been detected at high-$z$ in strongly lensed systems such as the "Sunburst Arc" \citep{Rivera_Thorsen_WR_sunburst_arc_2024}.   
The impact of WR populations on high-$z$ galaxy properties has also been invoked by means of indirect evidence, for example, on specific line ratios and chemical abundance patterns such as enhanced N/O \citep[e.g.,][]{curti_GSz9_2025, Gunawardhana_WR_GNz11_2025}. 

In this paper, we present the analysis of a z$=2.224$ galaxy from the MARTA program, namely \source (\sourceshort hereafter), where we report the highest-redshift detections of the WR "blue" and "red" bump spectral features in a non-lensed galaxy. We present a detailed study of the chemical enrichment and dust properties of the system, leveraging also a direct determination of the gas-phase iron (Fe) abundance.
We outline observations and data processing in Sect.~\ref{sec:data}, and the spectral fitting procedure in Sect.~\ref{sec:fitting}.
Sect.~\ref{sec:physical_prop} details the methods employed to derive the physical quantities of interests, while we interpret our findings and discuss their physical implications in Sect.~\ref{sec:discussion}. Finally, we summarize our results in Sect.~\ref{sec:summary}.
Throughout this work, we assume a \cite{planck_2020}
cosmology, with $H_0$ = 67.4~km~s$^{-1}$~Mpc$^{-1}$, $\Omega_M$ = 0.315, and $\Omega_{\Lambda}$ = 0.685. Solar abundances are taken from \cite{Asplund_solar_2021}.

\section{Observations and data processing}
\label{sec:data}

\begin{figure}
    \centering
    \includegraphics[width=0.9\columnwidth]{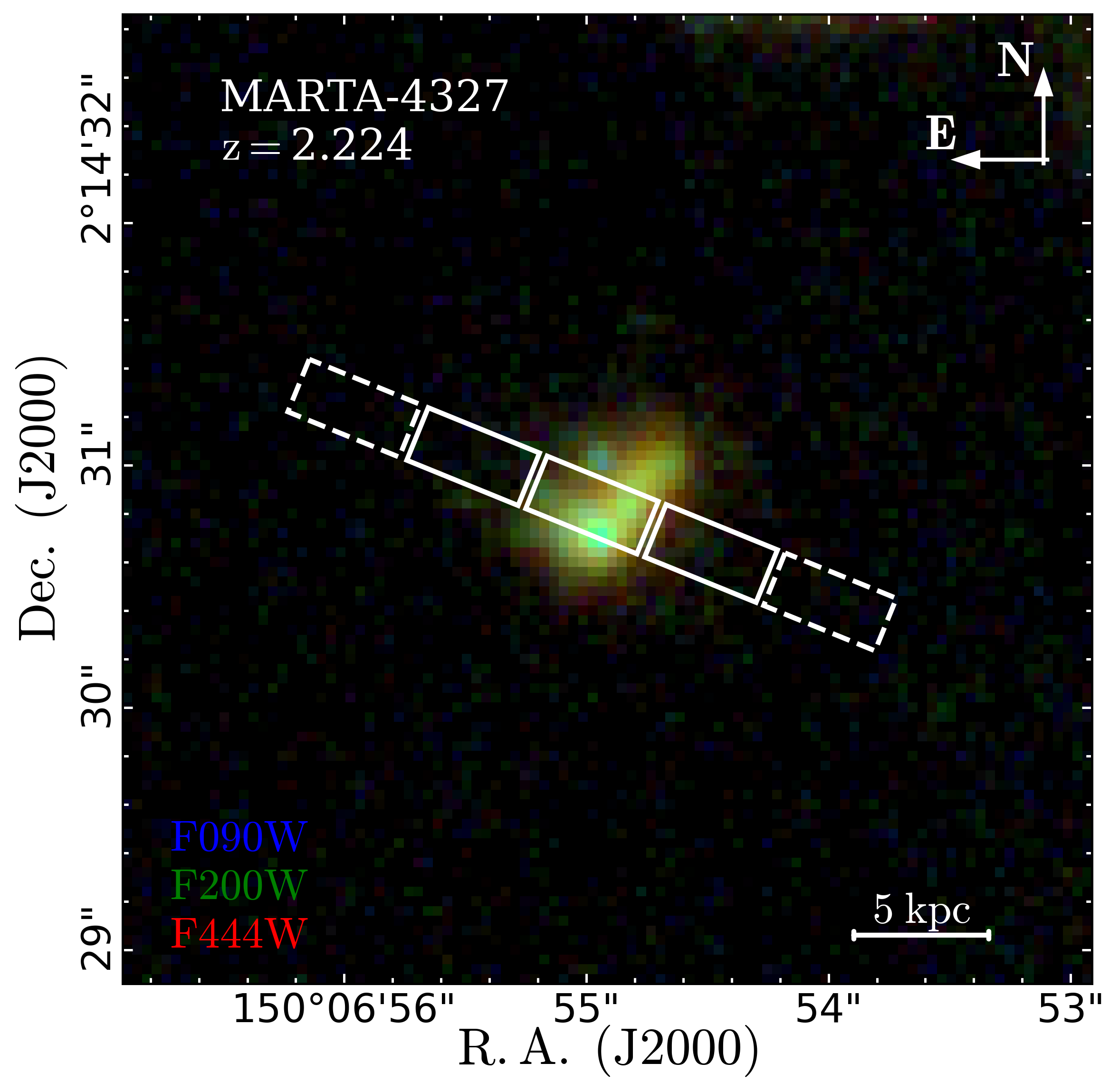}
    \caption{False color RGB image of \sourceshort. The three-shutters long NIRSpec slitlet and the effective five-shutters locations spanned by the three-nodding pattern are overlaid. }
    \label{fig:cutout}
\end{figure}

The requirement for deepest integration in MARTA (32 hours in G140M/F100LP) was driven by the goal of reaching a line sensitivity of $\sim 10^{-19}$erg~s$^{-1}$~cm$^{-2}$, aimed at detecting \OIIIoptL[4363] in the high-priority targets of our sample. An additional seven hours of integration were performed in G235M/F170LP to cover the rest-frame optical wavelength range up to $\sim 1\mu$m, which includes not only strong nebular lines such as \Halpha, \NII, and \SII, but also the \OIIaur doublet for all targets at $z\lesssim3$. 
Finally, three hours of integration with the G235H/F170LP configuration were spent, aimed at performing a better kinematical decomposition and search for faint and broad components under the \Halpha line.    
MARTA targeted in total 127 galaxies with one NIRSpec pointing in the COSMOS field. Details of sample selection and MSA design prioritization are given in \cite{Cataldi_MARTA_2025}. 

For the analysis presented in this paper, we have taken advantage of the most recent realization of the data reduction pipeline provided by the NIRSpec GTO team \citep[described in][]{Scholtz_JADES_DR4_2025}, which extends the extracted and calibrated spectral traces on the detector significantly beyond the nominal red-end boundary for each grating/filter configuration.
This enabled us to recover more emission lines compared to the standard data reduction pipeline, as, for example, the G235M/F170LP spectrum now extends up to $\sim 4\mu{m}$.
To assess consistency in the flux calibration, we compared the spectra of the two medium resolution gratings in their overlapping region, where the median flux densities are found to be in agreement within 1 percent.  
Furthermore, we checked for any residual, relative flux calibration issue as a function of wavelength (possibly due to non-adequate pathlosses correction, which are modeled by default assuming point-like source geometry) by comparing the synthetic photometry extracted from the spectrum with the in-slit photometry derived from available NIRCam images in F115W, F150W, F200W, F270W (i.e., the broadband filters overlapping with our NIRSpec filter choice), finding good agreement and the need for only minor correction. The complete procedure is described in Appendix~\ref{sec:appendix_A}.
Finally, we note that, in order to maximize the number of observable sources within one pointing, in MARTA we allowed moderate overlapping of spectral traces onto the detector, as the filling factor of emission lines on the detector is very low (hence, it is the probability of emission lines overlap from two different sources).
However, given the depth of our observations, in many cases significant continuum emission is detected, potentially contaminating the upper and/or lower shutters of a given slitlet and preventing the adoption of the standard "nodded" background subtraction.
In the specific case of \sourceshort, however, no strong contamination from other sources is present, and the source does not extend significantly beyond the central shutter of the MSA slitlet in any of the NIRCam filters; therefore, a standard nodding approach for background subtraction has been adopted.

\begin{figure*}
    \centering
    \includegraphics[width=0.9\textwidth]{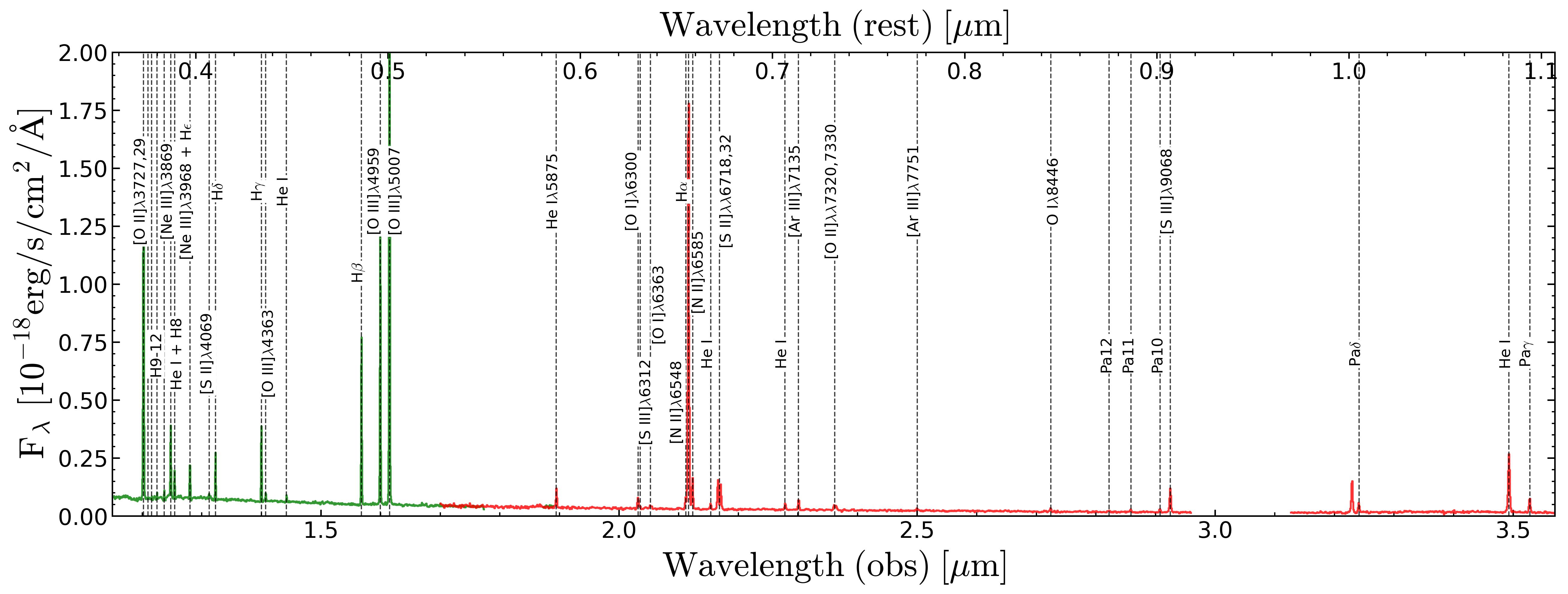}\\
    \includegraphics[width=0.9\textwidth]{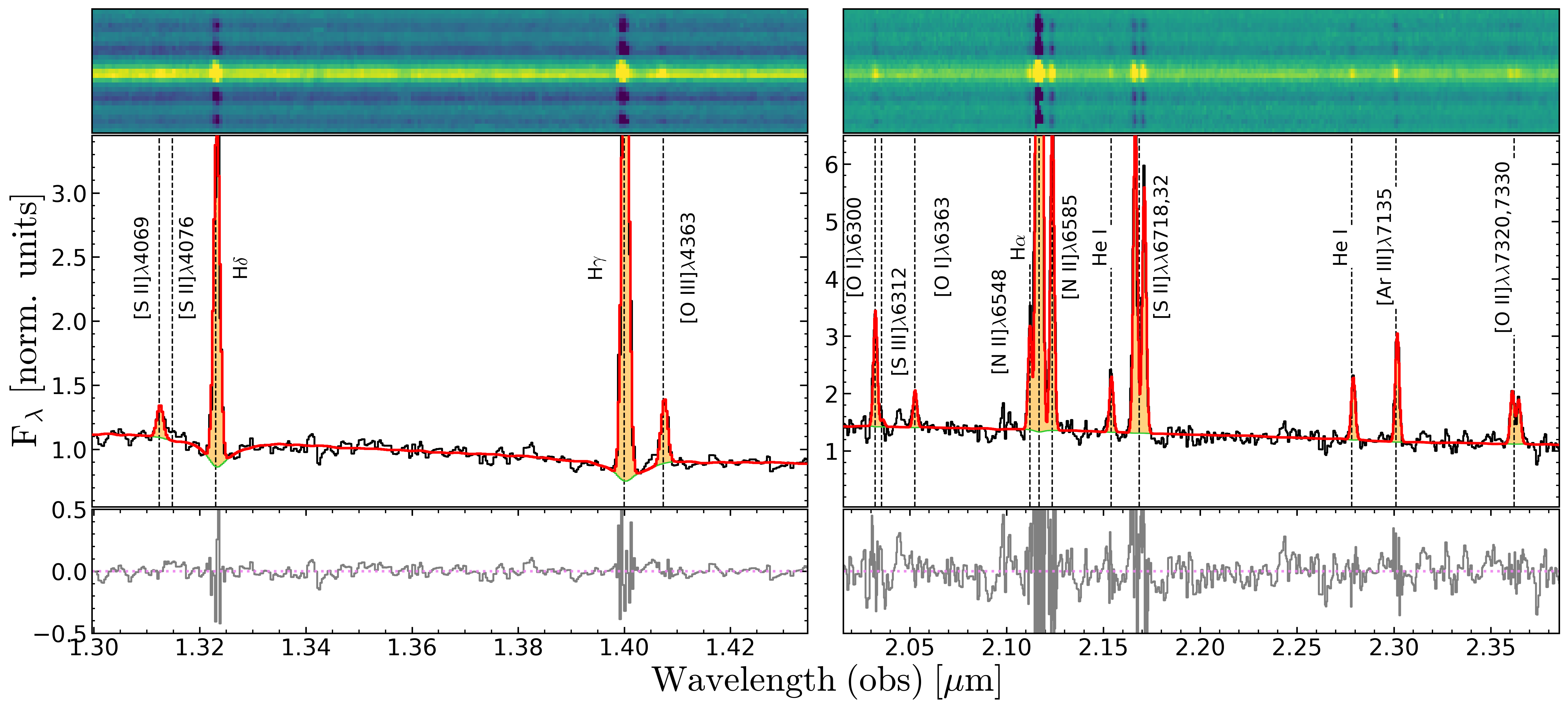}\\
    \includegraphics[width=0.9\textwidth]{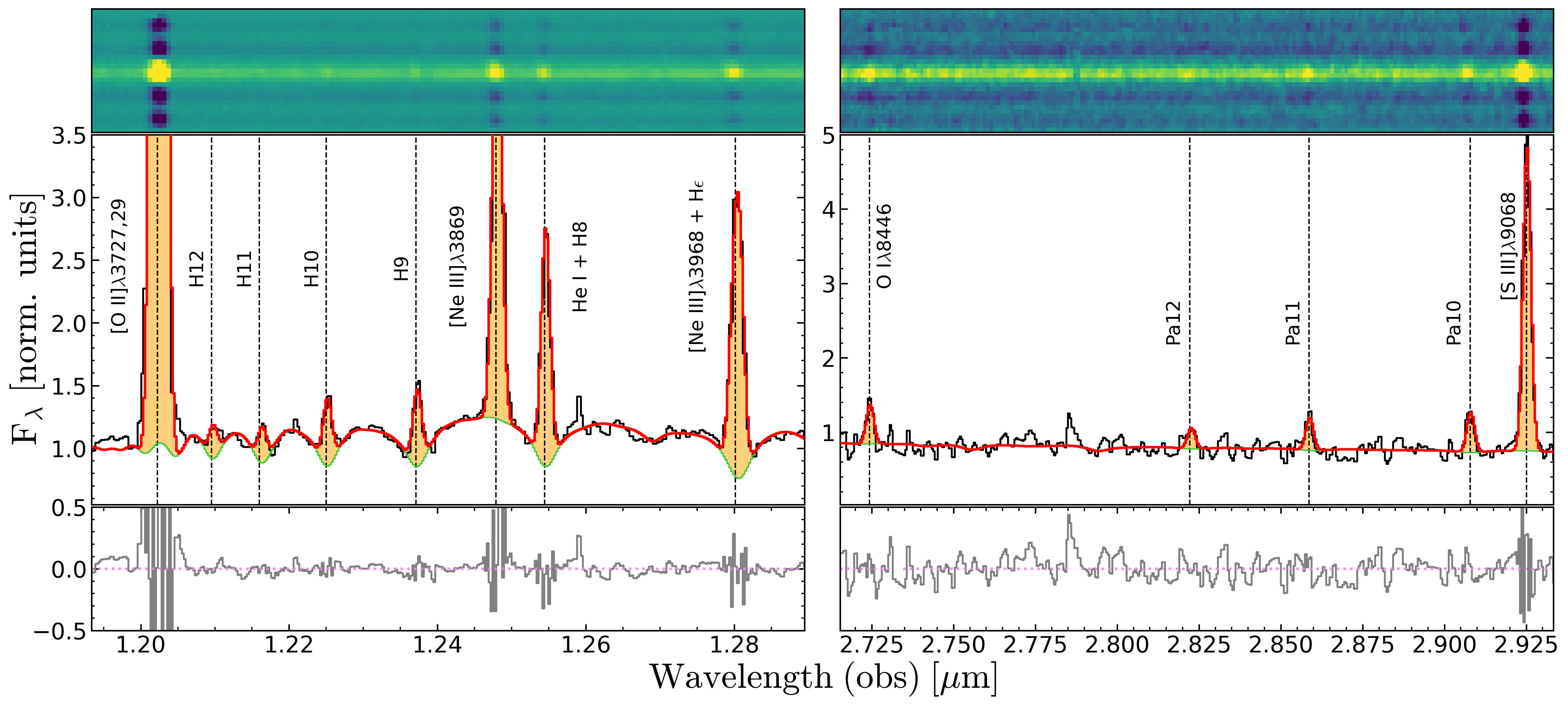}
    \caption{\jwst/NIRSpec spectrum of \source. \textit{Top panel} : Combined G140M/F100LP (green) and G235M/F170LP (red) spectrum. The main emission lines detected are reported. 
    \textit{Middle panels}: Zoom-in onto the region of the auroral lines, i.e., \OIIIoptL[4363] and \SIIL[4069] in G140M (left), \SIIIL[6312] and \OIIaur in G235M (right). The \textsc{ppxf} best-fit to the spectrum is overlaid in red (continuum in green, emission lines highlighted in yellow), while the 2D spectra and the fit residuals are shown in the upper and lower inset panels, respectively.
    \textit{Bottom panels}: Zoom-in onto the region of high-order Balmer lines (left), and Paschen and O I~$\lambda8446$ lines (right).}
    \label{fig:plot_spectra}
\end{figure*}

The top panel of Figure~\ref{fig:plot_spectra} reports the reduced, calibrated, and extracted 1D spectrum of \sourceshort, with the spectra of G140M/F100LP (green) and G235M/F170LP (red) gratings plotted beside each other. 
The emission lines detected in the spectrum are also marked.

\section{Spectral analysis}
\label{sec:fitting}

\subsection{Full spectral fitting}

We summarize here the spectral fitting steps implemented for \sourceshort, while more details on the full fitting procedure for MARTA galaxies are given in \cite{Cataldi_MARTA_2025}.

The full spectral fitting of the extracted 1D spectrum has been performed using the \textsc{python} implementation of the \textsc{ppxf} software \citep{cappellari_parametric_2004, cappellari_improving_2017, cappellari_ppxf_2022}, which enables simultaneous modeling of both emission lines and the stellar continuum. 
The fitting was carried out separately over the full wavelength ranges of the G140M and G235M gratings.

To model the stellar continuum, we employed the Binary Population and Spectral Synthesis (BPASS) templates v2.2.1 \citep{Stanway_Eldrige_BPASS_2018} with a \citet{chabrier_galactic_2003} initial mass function and an upper-mass cut-off of 300~\MSun. 
We convolved the continuum templates with a wavelength-dependent Gaussian kernel based on the instrumental line spread function (LSF) curves for each grating, as retrieved from the JWST documentation\footnote{\url{https://jwst-docs.stsci.edu/jwst-near-infrared-spectrograph/nirspec-instrumentation/nirspec-dispersers-and-filters\#gsc.tab=0}}.
To account for the effect of residual flux calibration issues and template mismatch, we applied a 10-degree multiplicative polynomial to the continuum model. 
Emission lines were included as gas templates, modeled by individual Gaussian profiles, and fitted simultaneously with the stellar continuum. This approach is critical, especially in spectral regions where stellar absorption features overlap with line emission. 
We verified that the inferred fluxes of the faintest lines of interest for this work (e.g., the \OIIIoptL[4363] auroral line) remained stable within $2\%$ among different choices of degrees of multiplicative polynomial, ensuring that the continuum adjustments did not introduce significant biases. 
The results of our fitting procedure are shown in the middle and bottom panels of Fig.~\ref{fig:plot_spectra}: the observed spectrum is shown in black, while the full \textsc{ppxf} best-fit spectrum is overlaid in red ( with the best-fit stellar component in green and the area underlying detected emission lines is shaded in gold).
Beside classical bright emission lines such as \OIIL[3727,29], \OIIIoptL[5007], \Hbeta, \Halpha, \NIIL[6548,84], and \SIIL[6717,32], we report the detection of auroral lines of both oxygen (\OIIIoptL[4363], \OIIaur) and sulfur (\SIIL[4069], \SIIIL[6312]) ions, together with other semi-faint forbidden lines such as, for example,  \ArIIIL[7135] and \SIIIL[9058], the OI$\lambda8446$ permitted line, hydrogen recombination lines down to H12, five lines from the Paschen series (Pa$\gamma$, Pa$\delta$, Pa10, Pa11, Pa12, the former two recovered thanks to the new data reduction\footnote{Pa8 and Pa9 lines, while formally covered by the extended G235M/F170LP grating, fall within the MSA detector gap}, see Sect.~\ref{sec:data}), and several helium recombination lines (including He~I$\lambda 1.032\mu{m}$).
Specifically, the middle panels of Fig.~\ref{fig:plot_spectra} focus on the spectral region surrounding auroral lines, while the bottom panels zoom-in onto the regions of \OIIL[3727] and high-order Balmer lines (left), He~I$\lambda 1.032\mu{m}$ and Paschen lines (right), respectively. 

\subsection{Detection of blue and red optical bumps}

\begin{figure*}
    \centering
    \includegraphics[width=0.95\textwidth]{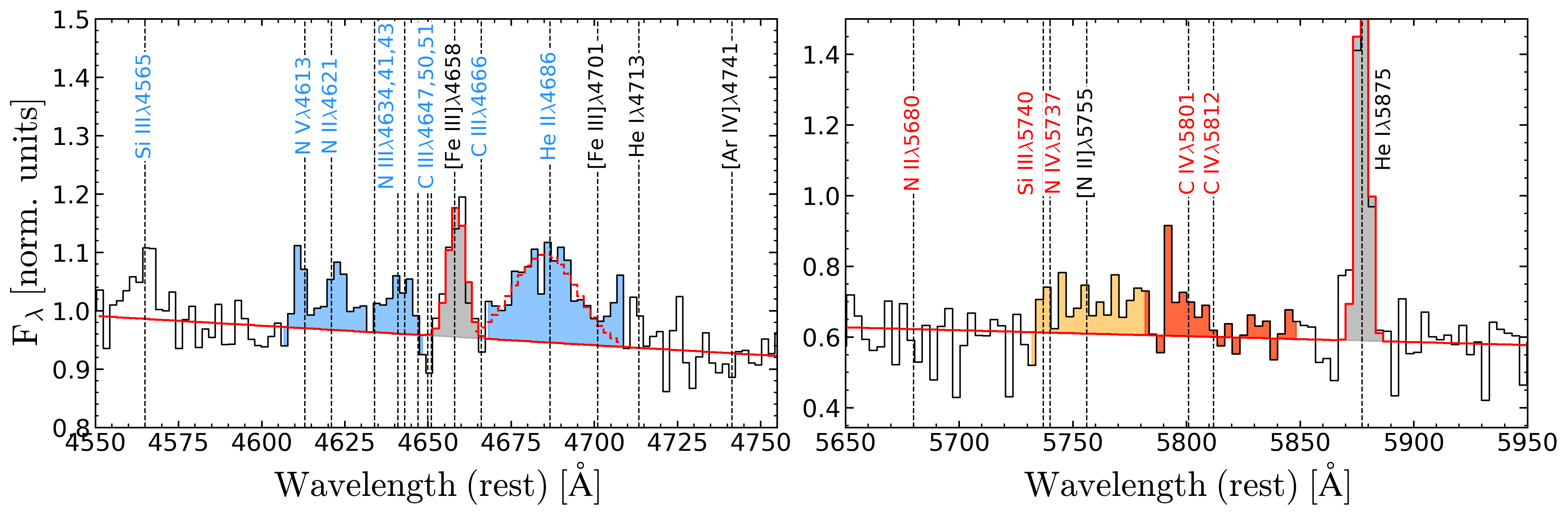} \\
    \caption{Blue and red bump features in \sourceshort. The plots show a zoom-in on the spectral region between 4550 - 4750 $\AA$ (left panel), and 5650 - 5950 $\AA$ (right panel), highlighting both stellar and nebular features. Features associated with the so-called blue and red bumps are marked in blue and red, respectively, with nebular line emission in black. The shaded regions highlight the spectral ranges adopted to derive the EW of the associated stellar features (as reported in Table~\ref{tab:properties}), with the "extended red bump" region defined by the combined orange and red area.  
    A single Gaussian fit to the broad \HeIIL[4686] emission is shown by the dashed line, while the fit to the nebular \FeIIIL[4658] and HeI$\lambda$5875 emission lines are marked by solid lines (and shaded in gray).}
    \label{fig:WR_features}
\end{figure*}

Figure~\ref{fig:WR_features} shows a zoom-in of the G140M and G235M spectra around the $\lambda \in$ [4575,4750] and [5650,5950] wavelength regions, showing evidence for the presence of the so-called blue and red bump features associated with WR stars.
More specifically, in the left-hand panel a broad feature, likely from stellar \HeIIL[4686] emission, is observed, together with flux excess between $\approx4613\AA$ and $\approx4650\AA$, tracing a complex of stellar features including metal lines from \NVp, \NIIp, \NIIIp, and \CIIIp. 
The narrow line at $4658\AA$ is instead nebular in origin and traces Fe~\textsc{iii} in emission.
A fit to the continuum and the narrow \FeIIIL[4658] emission line is overlaid (solid red line), while a single-Gaussian fit to the broad \HeIIL[4686] emission is shown as the dashed red line, giving a full width at half maximum (FWHM)~$=1460\pm170$~km~s$^{-1}$.
By integrating over the shaded blue regions (while excluding at the same time nebular emission), we derive a rest-frame equivalent width (EW) for the [\NIIIp+\CIIIp] complex ($4605-4650\AA$) of \EWr(N-C)~$=2.5\pm0.5 \AA$, and for the \HeII emission of \EWr(\HeII)~$=4.2 \pm 1 \AA$. 

The red bump instead is primarily a blend of \CIV transitions at $\lambda$5801,5812, and appears much less prominent in the spectrum of \sourceshort, as shown in the right-hand panel of Figure~\ref{fig:WR_features}.
However, an excess emission over the continuum level between $\approx 5740$-$5850 \AA$ is detected. Integrating the spectrum over the entire shaded region ($5730-5850\AA$, orange plus red in Fig.~\ref{fig:WR_features}) produces EW[red bump]~$=12\pm4 \AA$, whereas integrating only over the red region around the expected \CIV emission ($5780-5850\AA$) delivers EW~$=6 \pm 3 \AA$.
Based on extensive discussion about their origin in the literature \citep[see, e.g.,][and references therein for a review on the topic]{Crowther_WR_review_2007}, we suggest that these features could originate from a population of WR stars of the nitrogen type (WN), in that the (\NIIIp + \CIIIp) complex has a strength comparable to the \HeII bump, while the red bump is only marginally detected.
A more quantitative analysis and discussion are presented in Sect.~\ref{ssec:WR_discussion}.

\subsection{Detection of O\textsc{i}$\lambda8446$}
\label{ssec:OI_detection}

In addition to classical nebular lines, we report the presence of the permitted emission line O\textsc{i}$\lambda$8446, detected at $\sim 8\sigma$ significance (see bottom-right panel of Fig.~\ref{fig:plot_spectra}).
This detection represents one of the first of its kind in individual star-forming galaxies at high redshift, having been reported to date only in stacked spectra \citep{Strom_Cecilia_2023} or in highly magnified systems such as the `Godzilla' stellar complex in the Sunburst Arc \citep{Choe_SArc_godzilla_2025}, while it is generally more commonly observed in local active galactic nuclei  \citep[AGNs; e.g.,][]{Rudy_OI_AGN_1992, Rodriguez-Ardilla_AGN_2004, Matsuoka_OI_AGN_2007}). 
As such an emission line is rarely observed, its origin in the context of high-z galaxies such as \sourceshort warrants closer examination: we discuss possible physical mechanisms powering this line in Sect.~\ref{ssec:discussion_OI8446}.

\subsection{Signatures of broad components in the high-resolution grating}
\label{sec:g235h_fit}

We also performed emission-line fitting on the G235H/F170LP spectrum. The higher spectral resolution of such a grating allows us to search for the presence of fainter, broad components under \Halpha (associated with different kinematic components, e.g., tracing outflowing gas). 
In this case, given the shallower exposure time (three hours on-source) and higher spectral resolution, no detailed stellar continuum features are observed, and we model the continuum with a simple power-law.
The emission lines are still modeled by individual Gaussian components (convolved with the G235H LSF curve), with the exception of \Halpha, for which an additional broad component is included. 

\begin{figure}
    \centering
    \includegraphics[width=0.99\columnwidth]{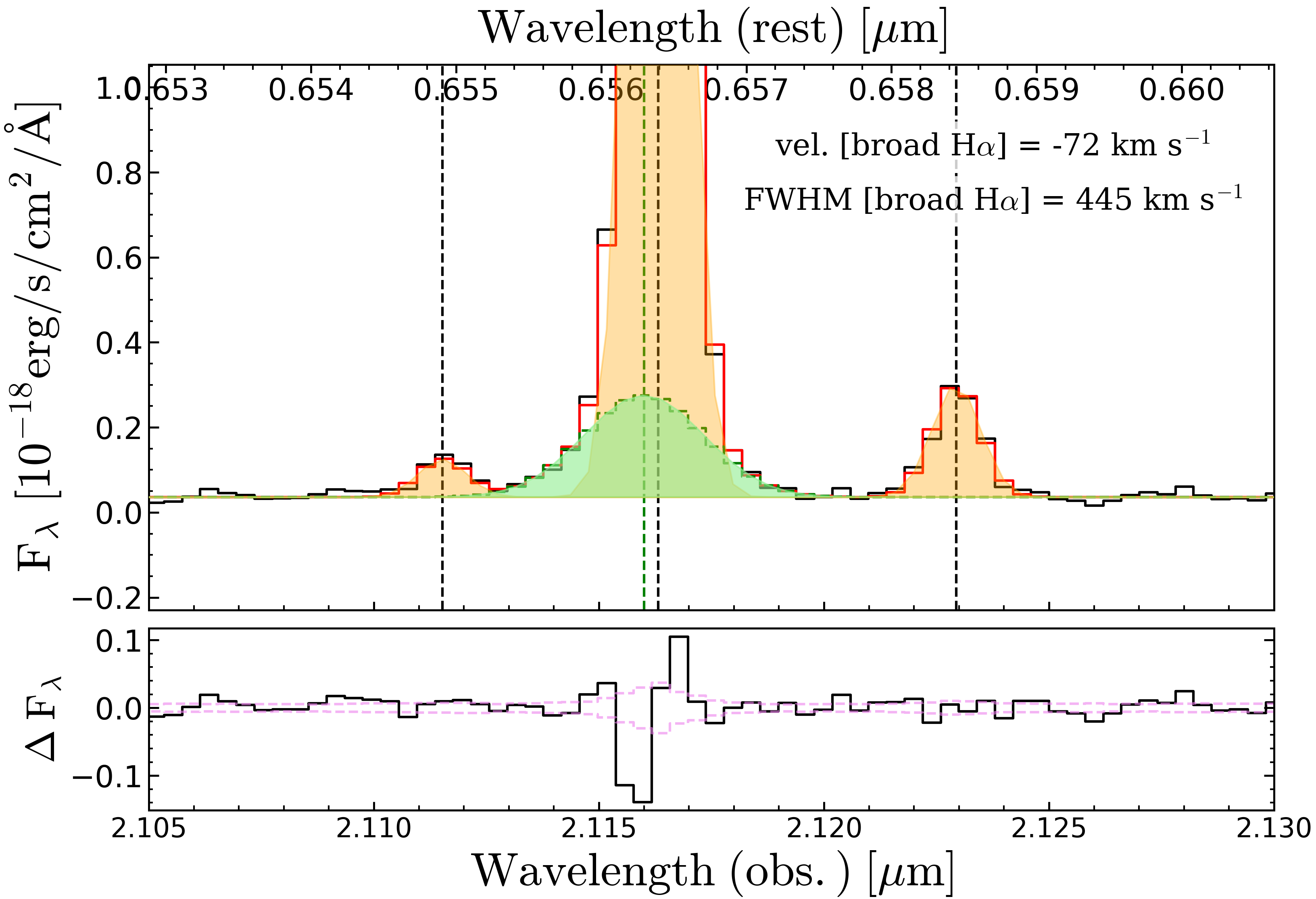}
    \caption{Zoom-in on the \Halpha and \NIIL emission lines in the G235H/F170LP grating. The best-fit model is overlaid in red, with the \Halpha broad component highlighted in green. The inclusion of such a component significantly improves the fit, with a $\Delta$BIC~$\gg10$ compared to the one-component fit. 
    The \Halpha broad component (FWHM~$= 445$~~km~s$^{-1}$) is blueshifted from the systemic redshift by $\sim70$~km~s$^{-1}$, possibly tracing outflowing gas in the galaxy.}
    \label{fig:g235h_fit}
\end{figure}

The results of the G235H/F170LP fitting are shown in Figure~\ref{fig:g235h_fit}. We detect, with high significance \citep[based on the BIC criterion, with $\Delta$BIC~$\gg$10, ][]{Schwarz_BIC_1978}, a broad component under \Halpha, blueshifted by $71$~km~s$^{-1}$ with respect to the systemic redshift and with FWHM$=445$~km~s$^{-1}$.
We discuss a possible interpretation of such a component as tracing outflowing gas in Sect.~\ref{ssec:discussion_outflow}, whereas an alternative, more speculative scenario involving the presence of an active black hole is discussed in Appendix~\ref{sec:appendix_D}.  

\section{Derivation of physical properties}
\label{sec:physical_prop}

\subsection{Nebular attenuation}
\label{ssec:neb_attenuation}

\begin{figure}
    \centering
    \includegraphics[width=0.99\columnwidth]{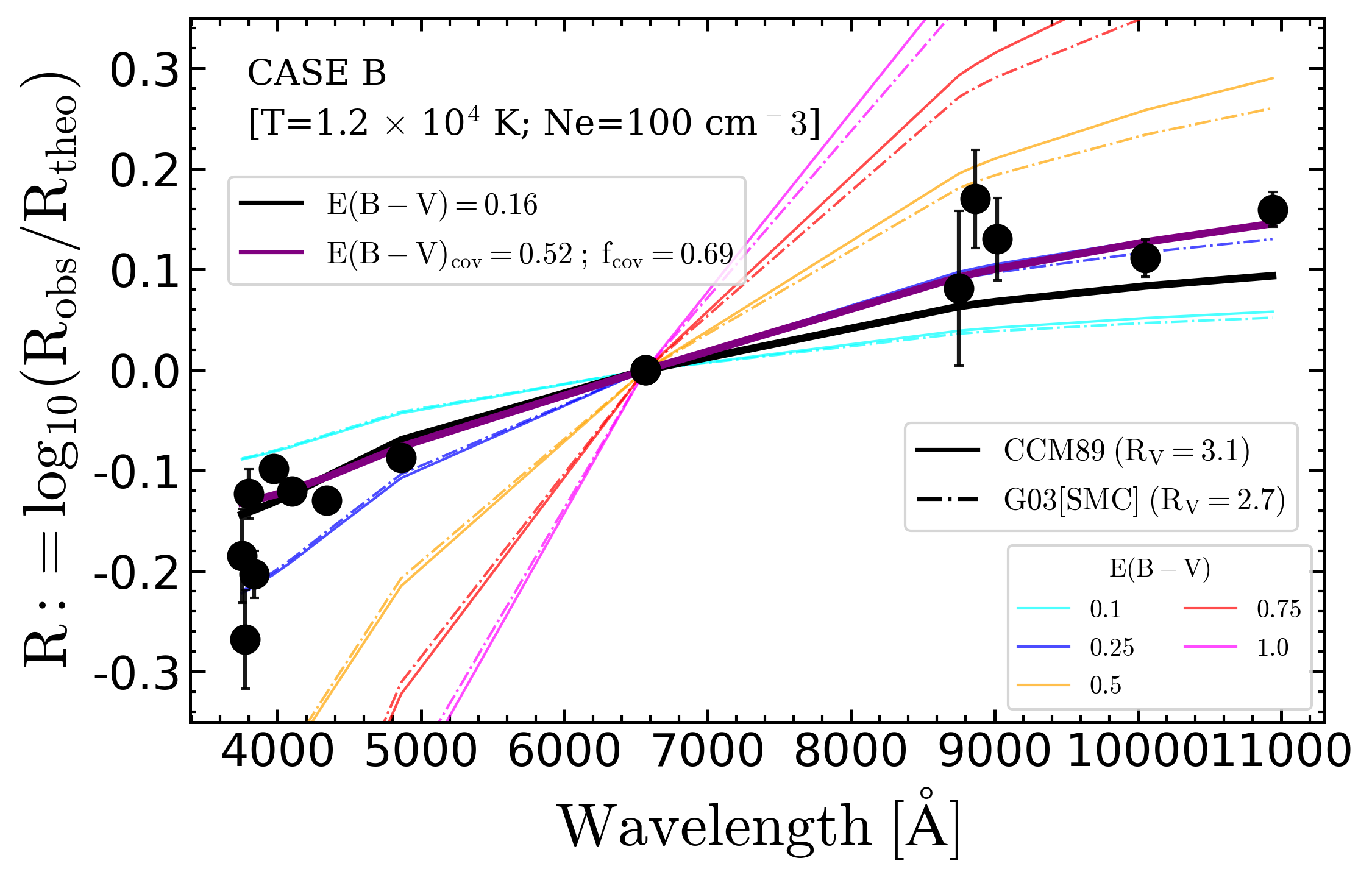}
    \caption{$R$ versus wavelength for \sourceshort.
    The logarithm of the ratio between observed and theoretical (Case B) line ratios of hydrogen lines to \Halpha, $R$, is plotted as a function of wavelength. Colored lines represent the $R$ vs $\lambda$ relationship predicted for different values of $E(B-V)$, given the \citet{cardelli_relationship_1989} (solid lines, $R_V=3.1$) and \citet{gordon_LMC_attenuation_2003} Small Magellanic Cloud (dashed lines, $R_V=2.74$) attenuation curves, respectively. The black curve reports the relationship obtained by performing a least-square fit to all available hydrogen lines ratios (for a \citealt{cardelli_relationship_1989} law), whereas the purple curve represents the best-fit $E(B-V)_{cov}$ and $f_{cov}$ values obtained by adopting the dust-covering fraction model by \citet{Reddy_AURORA_dust_fcov_2025}.
    }
    \label{fig:R_vs_lambda}
\end{figure}

We begin by studying the dust attenuation properties of \sourceshort,  
leveraging the detection of multiple hydrogen recombination lines, including the Balmer lines, down to H12 and Paschen lines Pa12, Pa11, Pa10, Pa$\delta$, and Pa$\gamma$.
In Fig.~\ref{fig:R_vs_lambda} we report the quantity $R$, defined as the logarithm of the ratio between the observed and theoretical ratio of a given hydrogen line to \Halpha, $R=\text{log}_{10}(\text{R}_{\text{obs}}/\text{R}_{\text{theo}})$ (where $\text{R}_{\text{theo}}$ is computed for a fiducial Case B recombination, \Te=$1.2\times10^4$~K, \Ne=$100$~cm$^{-3}$), as a function of wavelength.
The colored lines represent the $R$ vs $\lambda$ relationship predicted for different values of $E(B-V)$, given the \cite{cardelli_relationship_1989} (solid lines) and \cite{gordon_LMC_attenuation_2003} (dashed lines) extinction curves for the Milky Way (MW, $R_V = 3.1$) and Small Magellanic Cloud (SMC, $R_V = 2.74$), respectively.
The solid black curve reports the $R$ vs $\lambda$ relationship for the best-fit value $E(B-V)$ obtained by performing a least-square fit to all available hydrogen line ratios (assuming the law \citealt{cardelli_relationship_1989}), and corresponds to $E(B-V)=0.16\pm0.01$ (\AV$=0.50\pm0.03$).
The black curve matches the Balmer lines of the highest S/N, whereas the high-order Balmer lines H10, H11, H12, as well as Paschen lines, deviate from the best-fit predictions. 
For what concerns high-order Balmer lines (H10, H11, H12), we note that their measured flux is strongly sensitive to the shape of the stellar continuum in that region and, in particular, to the depth of the underlying stellar absorption features, which introduces model-dependent systematics. Furthermore, possible deviations from the Case B recombination for high-order Balmer lines have been discussed, for example, in \cite{Mesa-Delgado_HH202_2009}.

Interestingly, discrepancies between the Balmer and Paschen lines might reveal clues to the properties and geometric distribution of dust. 
Recently, \cite{Reddy_AURORA_dust_fcov_2025} analyzed a large sample of $z\sim2-3$ galaxies from the AURORA survey, finding evidence for an offset between $E(B-V)$ values inferred from Balmer and Paschen lines, when these two subsets of lines are fitted separately.
A possible turnaround invokes the assumption of a dust-covering fraction lower than one ($f_{cov}<1$), such that part of the light escapes from the nebula unattenuated.
This practically translates into a wavelength-dependent, light-weighted reddening, i.e., the $E(B-V)$ depends on which lines are adopted in its derivation.
In this scenario, the $E(B-V)$ computed from the Balmer
lines at bluer wavelengths is lower than that derived from the Paschen lines, as emission lines at shorter wavelengths are weighted more heavily toward the un-reddened sightlines, and vice versa. 

We therefore performed a fit to our set of Balmer and Paschen lines adopting the \cite{Reddy_AURORA_dust_fcov_2025} formalism (and the \citet{cardelli_relationship_1989} law): this is marked by the purple solid line in Fig.~\ref{fig:R_vs_lambda} and delivers a best-fit $E(B-V)_{cov}=0.52\pm0.20$ and $f_{cov}=0.69\pm0.06$.
We observe that such a framework provides a much better simultaneous match to both Paschen and the brightest Balmer lines (H$\alpha$, H$\beta$, H$\gamma$, H$\delta$, H$\epsilon$) (while still struggling to reproduce all high-order Balmer lines).
Therefore, throughout the rest of the analysis we model the nebular attenuation as produced by a dust geometry with variable $f_{cov}$ (i.e., the effective $E(B-V)$ does depend on wavelength), as we note that the impact of a variable $f_{cov}$ on the inferred physical properties of galaxies (especially those sensitive to ratios of emission lines widely spaced in wavelength, such as the temperature \Tii of the O$^{+}$ ion, see Sect.~\ref{ssec:te_ne_av}) can be significant. 
We defer a more thorough analysis of the comparison between different assumptions on the dust properties, shape of the attenuation curve, and their impact on the derivation of ISM properties in high-z galaxies to forthcoming work.

\subsection{Simultaneous modeling of temperature, density, and \AV}
\label{ssec:te_ne_av}

We simultaneously infer \Te, \Ne, and \Av by forward-modeling multiple emission line ratios and comparing them with the observed values within a Bayesian framework. 
Because theoretical Case B values for different hydrogen line ratios depend (mildly) on temperature and density, this would affect the derivation of \AV and, as a consequence, the temperatures and density themselves as derived from reddening-corrected emission line ratios.  
Following what was discussed in Sect.~\ref{ssec:neb_attenuation}, we model the nebular attenuation following the formalism by \cite{Reddy_AURORA_dust_fcov_2025} and include the dust-covering fraction of dust $f_{cov}$ as an additional free parameter.

We defined the likelihood as $L$ = $\exp{(- \chi^{2}/2)}$, with $\chi^{2}$ equal to
\begin{equation}
    \label{eq:H_chi2}
    \chi^2 = \sum \ \frac{(\frac{I(\lambda_1)}{I(\lambda_2)}_{\text{mod}} - \frac{I(\lambda_1)}{I(\lambda_2)}_{\text{obs}})^2}{\sigma^2_{\lambda_{1,2}}} \ + \text{ln}(2 \pi \ \sigma^2_{\lambda_{1,2}} ) \,,
\end{equation}
and where $\sigma_{\lambda_{1,2}}$ is the uncertainty on the flux ratio between the emission lines at $\lambda_1$ and $\lambda_2$.
The predicted flux ratios are modeled as 
\begin{equation}
    \label{eq:line_ratios_model}
    \frac{I(\lambda_1)}{I(\lambda_2)}_{\text{mod}} = \frac{E_{\lambda_1}(n_{e}, T_e)}{E_{\lambda_2}(n_{e}, T_e)} \ \times  \ \frac{(1-f_{cov}) +  f_{cov} \times 10^{-0.4 \ A_V \  k(\lambda_1)}}{(1-f_{cov}) +  f_{cov} \times 10^{-0.4 \ A_V \  k(\lambda_2)}} ,
\end{equation}
where $E$ is the emissivity (which is a function of density and temperature), $k(\lambda)$ is the reddening law, \AV is the extinction in magnitudes in the V band and $f_{cov}$ the dust-covering fraction.
The $\chi^{2}$ includes co-adding multiple terms that comprise ratios between hydrogen lines from both Balmer and Paschen series, auroral-to-nebular line ratios of the same ion (hence, that do not depend on relative abundances), which are strongly sensitive to \Te (with a milder, secondary dependence on \Ne), and density-sensitive line ratios.

For hydrogen, all line ratios are computed with respect to \Halpha, and the ratios between relative emissivities are modeled under the assumption of Case B recombination. The following hydrogen emission lines are included in the model: \Halpha, \Hbeta, \Hgamma, \Hdelta, H7\footnote{Which is de-blended from the \NeIIIL[3968] emission by assuming a fixed \NeIIIL[3968]/\NeIIIL[3869] = 0.31 line ratio.}, H9, H10, H11, and H12 from the Balmer series (we exclude the H8+HeI$\lambda3889$ blended complex at this stage, but see Section~\ref{ssec:helium_abund}), as well as Pa$\gamma$ at $\lambda~1.0941~\mu$m, Pa$\delta$ at $\lambda~1.0052~\mu$m, Pa10 at $\lambda~9017~\AA$, Pa11 at $\lambda~8865~\AA$, and Pa12 at $\lambda~8752~\AA$.
For what concerns temperature-sensitive diagnostics, we included the \OIIIoptL[4363]/\OIIIoptL[5007] and \OIIL[7325]\footnote{Taken as the sum of the \OIIL[7320,7330] doublet.}/\OIIL[3727,29] auroral-to-nebular line ratios: in our model, we assume these ratios trace the temperature of two different emitting zones, such that
in Equation~\ref{eq:line_ratios_model} $\lambda_1$ and $\lambda_2$ are $\lambda7325$ and $\lambda7328$ for \Te = $T_2$, and $\lambda4363$ and $\lambda5007$ for \Te = $T_3$, respectively.
Finally, we include a term to model the ratio among the lines of the \SII[6718,6732] doublet, strongly sensitive to the gas density \Ne (and for which we adopt \Te=$T_2$ to compute the emissivity, as we assume that S$^{+}$ traces the same region of the O$^{+}$ ion responsible for the \OII transitions).

The free parameters in our model are therefore \Ne, $T_{2}$, $T_{3}$, \Av, and $f_{cov}$, and we run a Markov chain Monte Carlo (MCMC)
to explore the parameter space via the \textsc{emcee} implementation \citep{emcee_2013}, with 24 walkers, 1500 steps, and 20\% burn-in. 
At each MCMC step, the temperature assumed to model the emissivities of hydrogen lines is taken as the average between $T_{2}$ and $T_{3}$, and we adopt the 
\cite{cardelli_relationship_1989} attenuation law with $R_V = 3.1$ for dust reddening.
The following uniform priors are imposed on the parameters: $0.5 \leq \text{log}{_{10}}(n_e/cm^{-3}) \leq 4 \ ; 0 \leq A_V \leq 5 \ ; 0 \leq f_{cov} \leq 1 \ ; 0.5 \leq T_{2,3}/10^4~K \leq 2.5$.

We obtain \Tii = 12260$\substack{+480 \\ -520}$~K, \Tiii = 11430 $\substack{+150 \\ -160}$~K, n$_{\text{e}}$ = 80 $\substack{+50 \\ -40}$~cm$^{-3}$, \AV = 1.53 $\substack{+0.15 \\ -0.16}$, and $f_{cov}$ = 0.74 $\substack{+0.01 \\ -0.01}$.
The corner plots of the MCMC run and the comparison between the MAP model and observed line ratios are shown in Appendix~\ref{ssec:appendix_B_te_ne_av}.
Throughout the rest of the paper, we assume the values inferred as detailed in this section as our fiducial estimates on temperatures, density, and nebular attenuation. 

\subsection{Te-based chemical abundances}
\label{ssec:abundances}

The simultaneous detection of both nebular and auroral lines in the spectrum enables us to perform a detailed study of chemical abundance patterns in this galaxy, employing the "direct" electron temperature (T$_e$) method. In fact, once the temperature and density of the ionized gas are known, we can derive the relative ionic abundance of two elements comparing the intensity $I(\lambda)$ in the emission lines of each species while taking into account the different temperature and density dependences of the volumetric emissivity of the transitions.

For this part of the analysis, we start from the posterior distributions of physical parameters derived in Section~\ref{ssec:te_ne_av}, and use them to infer "dependent" quantities, such as ionic abundances and their uncertainties, by applying the same methodology to all samples in the chain.
When using emission lines not already folded in the original likelihood (such as \NIIL[6584], used to derive N/O), we generated 100 realizations of the flux per MCMC sample from a Gaussian distribution centered on the measured value and with $\sigma$ equal to the observational error, such that the uncertainty on the flux is propagated onto the final derived quantity. 
The final corner plots resulting from this procedure are shown in Appendix~\ref{ssec:appendix_B_abundances}.
Throughout this section, unless stated otherwise, we adopt \textsc{pyneb} routines for the derivation of temperature and chemical abundances, with atomic data taken as default for all atoms but for O$^{++}$ and He species, for which we adopt data from \cite{palay_improved_2012} and \cite{Porter_He_emiss_2013}, respectively.

\subsubsection{Oxygen, nitrogen, sulfur, neon, argon}

To derive the total oxygen abundance, for each element of the MCMC chain, we compute the abundance of O$^{++}$/H and O$^{+}$/H from the dust-corrected \OIIIoptL/\Hbeta (assuming t=\Tiii) and \OII/\Hbeta (assuming t=\Tii) ratios, respectively.
The total oxygen abundance is then O/H = O$^{++}$/H + O$^{+}$/H, delivering 12+log(O/H) = 8.15 $\substack{+0.02 \\ -0.02}$.
For other elemental species, we followed the same procedure described in the previous section to derive the ionic abundances of nitrogen (N), sulphur (S), neon (Ne), and argon (Ar) relative to those of oxygen (O).
Where needed, we applied ionization correction factors (ICF) as parametrized by \cite{amayo_ICFs_2020} on the basis of the oxygen ionization fraction, defined as $\omega$ = O$^{++}$/(O$^{++}$ + O$^{+}$) = $0.70 \substack{+0.03 \\ -0.04}$.

For nitrogen we inferred the abundance of N$^{+}$/O$^{+}$ by comparing the (dust-corrected) intensity \NIIL[6584]with that of \OII[3727], assuming t = \Tii and the density set by \Ne; after applying the ICF[N$^{+}$/O$^{+}$], we obtain a total log(N/O) = -1.14 $\substack{+0.03 \\ -0.03}$. 
We observe sulfur emission originating from two different ionization states, namely S$^{+}$ and S$^{++}$. 
We derived the abundance of  S$^{+}$/O$^{+}$ by comparing \SII to \OII[3727], assuming t = \Tii and \Ne, while we infer  S$^{++}$/O$^{++}$ from the \SIIIL[9068] / \OIIIoptL[5007] ratio, assuming t = \Tiii.
The total S/O abundance is then given by S/O = (S$^{++}$/O$^{++}$ + S$^{+}$/O$^{+}$) $\times$ ICF[S$^{++}$/O$^{++}$ + S$^{+}$/O$^{+}$], delivering log(S/O) = -1.74 $\substack{+0.02 \\ -0.02}$.
Finally, we compute the Ne$^{++}$/O$^{++}$ and Ar$^{++}$/O$^{++}$ abundance ratios comparing the intensity of \NeIIIL[3869] and \ArIIIL with that of \OIIIoptL[5007], respectively, assuming the \Tiii temperature for both ions and inferring log(Ne/O) = -0.51 $\substack{+0.01 \\ -0.01}$ and log(Ar/O) = -2.49 $\substack{+0.02 \\ -0.02}$.

\subsubsection{Helium abundance}
\label{ssec:helium_abund}

To measure He$^{+}$/H$^{+}$ abundance (y$^{+}$), we followed a similar approach to that of Section~\ref{ssec:te_ne_av} and Equation~\ref{eq:line_ratios_model}, modeling the intensity ratio in a given He I line compared to \Hbeta as
\begin{equation}
    \label{eq:He_abund}
    \frac{I(\lambda)}{I(H\beta)} = y^{+} \ \frac{E_{\lambda}(n_{e}, T)}{E_{H\beta}(n_{e}, T)} \ \times \ \text{dust model} \ \times \ f_{\lambda}(n_{e}, T, \tau) \frac{1 + \frac{C}{R}(\lambda)}{1 + \frac{C}{R}(H\beta)} \,,
\end{equation}
where $E$ is the emissivity, $\text{``dust model''}$ refers to the \cite{Reddy_AURORA_dust_fcov_2025} formalism already implemented in Sect.~\ref{ssec:te_ne_av}, $f_{\lambda}$ is a correction term to take into account photons reabsorbed or scattered out of the line of sight (function of temperature, density, and optical depth $\tau$), and the  collisional-to-recombination ratio (C/R) corrects for the amount of neutral hydrogen and helium atoms excited to
higher-energy states due to collisions with free electrons, being a function of the ratio of neutral-to-ionized hydrogen density $\xi$.
Compared to other studies
\citep[e.g.,][]{Hsyu_PLEK_2020, Aver_Helium_LeoP_2021}, here we do not include any corrections for the underlying absorption, as this is already accounted for in our continuum modeling and spectral fitting procedure. 

In our analysis, we consider the following He I transitions: He I $\lambda$4472, $\lambda$4687, $\lambda$4923, $\lambda$5017, $\lambda$6679, $\lambda$7067, $\lambda$7283, and $\lambda$10832 (the latter being particularly sensitive to variations in electron density, see, e.g., \citealt{berg_aurora_helium_2025}).
In addition, we include the He I$\lambda3889$+H8 complex as an individual emission line, taking into account radiative transfer effects for both transitions.
We set up a similar MCMC as in Section~\ref{ssec:te_ne_av}, and include hydrogen lines in our modeling too, leaving again \Av and $f_{cov}$ free to vary; the other free parameters in our model are y$^{+}$, $\tau$, and $\xi$.

Further details about the formalism adopted to include optical depth and C/R corrections, as well as on the adopted priors and results from the MCMC run, are provided in Appendix~\ref{sec:appendix_B}.
From marginalized posterior distributions, we infer an He$^{+}$ abundance y$^{+}$ = 0.085$\substack{+0.002\\ -0.002}$.
Because of our interpretation of the \HeIIL[4686] emission as primarily stellar in origin, we did not use its flux to correct for the doubly-ionized helium state, and we assume $y^{+}$ as indicative of the total helium abundance $y$ in the gas-phase, while noting that this could more likely represent a lower limit.
Converting this value to the He mass fraction $Y=4y(1 - 20\times \text{O/H})/(1 + 4y)$, we obtain $Y=0.253\pm0.008$, in good agreement with the measurements of local and high-z systems of metallicity comparable to \sourceshort \citep{Aver_Helium_LeoP_2021, berg_aurora_helium_2025}.

\subsubsection{Gas-phase and total iron-to-oxygen abundance}
\label{sssec:gas_Fe_O}

We derived the gas-phase (Fe/O)$_{\text{gas}}$ abundance from the \FeIIIL emission line.
First, we infer the Fe$^{++}$/O$^{+}$ relative abundance from the  \FeIIIL/\OII[3728] ratio, assuming T=\Tii and the density derived from the \SII doublet, which we then correct assuming the ICF(Fe) scheme provided by \cite{Rodriguez_Rubin_Fe_ICF_2005} based on the degree of ionization measured by O$^{++}$/O$^+$, obtaining log(Fe/O)$_{\text{gas}}$ = $-2.07\substack{+0.06 \\ -0.07}$. Furthermore, we derive the gas-phase Fe/N abundance by combining (Fe/O)$_{\text{gas}}$ with N/O, obtaining log(Fe/N)$_{\text{gas}} = -0.93\substack{+0.07 \\ -0.08}$.

The determination of Fe abundance in the ISM is subject to several uncertainties, mostly related to the effect of iron depletion on dust grains and to the correction of unseen ionization states. In particular, the ICF for Fe$^{++}$, based on photoionization models or on direct observations of emission from Fe$^+$, Fe$^{3+}$, and Fe$^{4+}$ ions, is affected by large scatter \citep{Rodriguez_Rubin_Fe_ICF_2005, izotov_chemical_2006}. 
Other sources of uncertainty include uncertainties in the atomic parameters of Fe, and the effects of dust depletion, which can dramatically alter the observed metal abundances, especially for the most refractory elements \citep[e.g.,][]{DeCia_dust_LyA_absorbers_2016, Roman-Duval_METAL_dust_depletion_2022}. For example, in the Milky Way ISM, 90 to 99\% Fe is missing from the gas-phase but incorporated into the dust \citep{jenkins_depletion_2009}.
For a more thorough discussion of the systematics associated with the choice of ICF and other sources of uncertainties in the measurement of nebular Fe abundance, we refer to the discussion in \citealt{Mendez-Delgado_Fe_O_2024}. Here, we conservatively account for such systematic uncertainties by incorporating an additional relative uncertainty of $20\%$ on the ICF(Fe$^{++}$), providing a final fiducial estimate of log(Fe/O)$_{\text{gas}} = -2.07\substack{+0.14 \\ -0.15}$ and log(Fe/N)$_{\text{gas}} = -0.93\substack{+0.15 \\ -0.15}$.

\begin{figure*}
    \centering
    \includegraphics[width=0.48\textwidth]{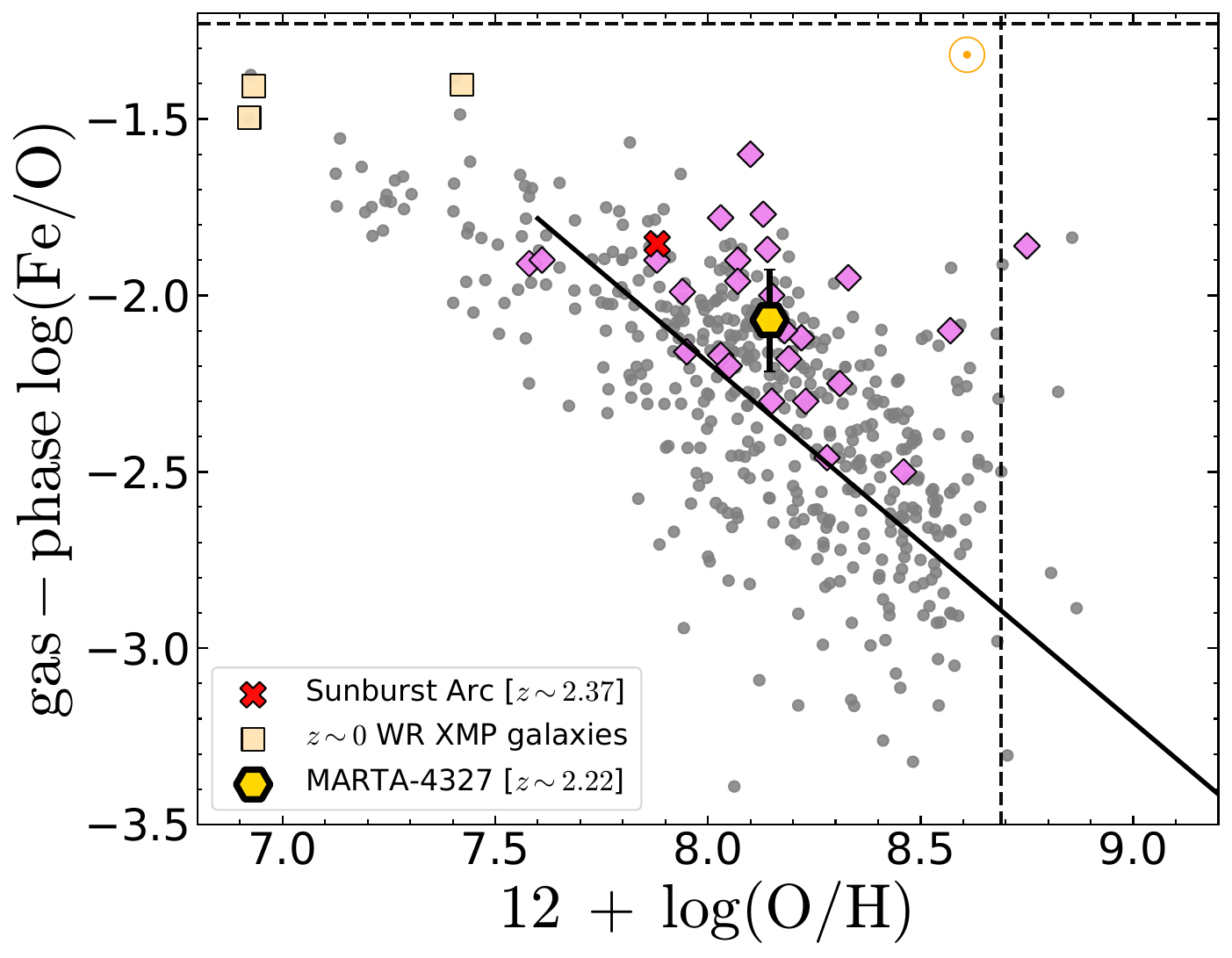}
    \includegraphics[width=0.48\textwidth]{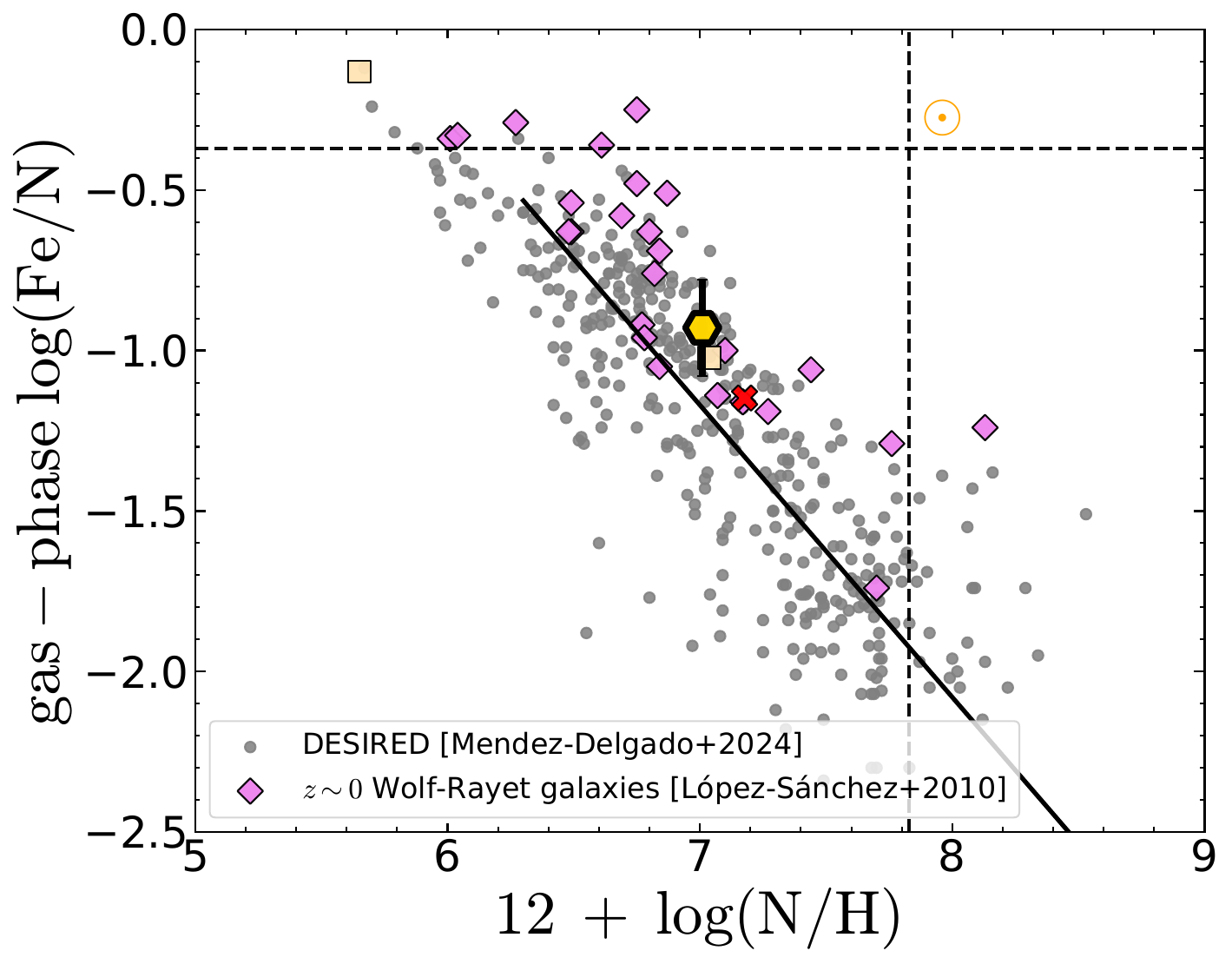}
    \caption{Relative gas-phase Fe, O, and N abundance patterns. The position of \sourceshort in the log(Fe/O) vs 12+log(O/H) (left panel) and log(Fe/N) vs 12+log(N/H) (right panel) diagrams is compared with the distribution of local star-forming nebulae from the DESIRED project \citealt{Mendez-Delgado_Fe_O_2024}, star-forming galaxies hosting WR features \citep{lopez_sanchez_WR_abundances_2010}, extremely metal-poor galaxies with signatures of WR activity (J1205+4551, \citealt{Izotov2017}; J0811+4730, \citealt{Izotov_J0811_2018}; J1631+4426, \citealt{Kojima2021}), and with the Sunburst Arc at $z\sim2.37$ \citep{Rivera_Thorsen_WR_sunburst_arc_2024}. Notably, almost all galaxies with WR signatures are characterized by an Fe-enhancement in the gas-phase compared to the average trend of local nebulae.
    }
    \label{fig:Fe_O}
\end{figure*}

In Fig.~\ref{fig:Fe_O} we compare the position of \sourceshort on the (Fe/O)$_{\text{gas}}$ versus O/H plane with a distribution of star-forming galaxies with Te-based abundance measurements.
In particular, gray points mark the sample compiled and analyzed by \cite{Mendez-Delgado_Fe_O_2024} in the framework of the DESIRED project\footnote{Here we consider the abundance values derived under the assumption of no temperature fluctuations ($t^2=0$) for consistency with our fiducial measurements in \sourceshort.}, in violet we report a sample of starburst galaxies showcasing clear signatures of Wolf-Rayet stellar features from \cite{lopez_sanchez_WR_abundances_2010}, whereas squares mark extremely-metal poor (XMP) galaxies with WR activity from \cite{Izotov2017, Izotov_J0811_2018, izotov_xmp_2021, Kojima2021} and, 
finally, the location of the LyC emitting complex in the lensed system Sunburst Arc at $z\sim2.37$ is marked by the red cross; in both panels, the black line marks the best-fit to the trend in the abundance patterns for the DESIRED sample \citep{Mendez-Delgado_Fe_O_2024}.

According to Fig.~\ref{fig:Fe_O}, \sourceshort appears relatively enhanced in (Fe/O)$_{\text{gas}}$ compared to the bulk of local star-forming nebulae, occupying the upper envelope of the distribution of the DESIRED sample; remarkably, we note that such a behavior is common to almost all galaxies with WR features in their spectra.  
At the same time, despite showing evidence for enhancement in the (Fe/N)$_{\text{gas}}$ vs N/H diagram too, the offset of
\sourceshort (and of WR galaxies in general) from the average trend of local galaxies is less pronounced (with \sourceshort being consistent with the best-fit curve of DESIRED galaxies within the uncertainties), reflecting the overall tighter relationship (as observed also in samples of MW stars) expected given the more similar production timescales of the two elements \citep{Mendez-Delgado_Fe_O_2024}. We return to the possible implications of these findings in Sect.~\ref{sssec:WR_and_dust}.

Taking advantage of the tighter relationship between Fe/N and N/H gas-phase abundances to establish a link between the relative dust depletion properties (as harder radiation fields, which scales inversely with Fe/H, are more likely to destroy dust grains), we can assume N/H as a proxy of the fraction of total Fe locked into dust grains.
Applying Equation~1 from \cite{Mendez-Delgado_Fe_O_2024}, we infer a Fe$_{\text{dust}}$/Fe$_{\text{total}}$=0.65, from which we derive a total Fe/O abundance of log(Fe/O)$_{\text{total}}$=-1.61$\pm$0.22\footnote{Where we have conservatively applied a further relative 20\% uncertainty on the dust depletion correction factor.} or, equivalently, [O/Fe]~$=0.38\pm0.22$. 
Under the assumption that the Fe/O abundance measured in the ISM is representative of the stellar metallicity of the young stellar populations responsible for powering line emission, this would correspond to an alpha-enhancement approximately of a factor $2.4\times$ the solar value.

\begin{table}
    \caption{Physical properties of \source.}
    \centering
    \setlength{\tabcolsep}{4pt}
    \begin{tabular}{l|c}
    \toprule
    Property & Value \\
    % \vspace{0.2cm}
    \midrule
 MARTA ID & 4327 \\
 RA & 150.1152501 \\
 Dec & 2.2419069 \\
 Redshift (z) & 2.22344 $\pm$ 0.00007 \\
log(M$_{\star}$/M$_{\odot}$)~(SED; total phot.) & 9.36$\pm$0.10 \\
SFR~[M$_{\odot}$yr$^{-1}$]~(H$\alpha$; in-slit) & 7.28$\pm$0.01 \\
log(sSFR/yr$^{-1}$)~(H$\alpha$; in-slit) & -8.50$\pm$ 0.10 \\
\midrule
\multicolumn{2}{c}{ISM properties} \\
\AV & 0.50$\pm$0.03 \\
E(B-V) & 0.16$\pm$0.01 \\
\AV$_{cov}$$^{\dagger}$ & 1.53$\substack{+0.15 \\ -0.16}$ \\
E(B-V)$_{cov}$$^{\dagger}$ & 0.49$\substack{+0.05 \\ -0.05}$ \\
$f_{cov}$$^{\dagger}$ & 0.74$\substack{+0.01 \\ -0.01}$ \\
\Tiii~[K] & 11430$\substack{+150 \\ -160}$ \\
\Tii~[K] & 12260$\substack{+480 \\ -520}$ \\
\Te~HeI~[K] & 9915$\pm$ 3070 \\
log(O$^{+}$/H) & -4.38$\substack{+0.07 \\ -0.06}$ \\
log(O$^{++}$/H) & -4.01$\substack{+0.02 \\ -0.02}$ \\
12+log(O/H) & 8.15$\substack{+0.02 \\ -0.02}$ \\
O$^{++}$/(O$^{++}$+O$^{+}$) & 0.70$\substack{+0.03 \\ -0.04}$ \\
He$^{+}$/H$^{+}$ (y$^{+}$) & 0.085$\substack{+0.002 \\ -0.002}$ \\
log(N/O) & -1.14$\substack{+0.03 \\ -0.03}$ \\
log(S/O) & -1.71$\substack{+0.02 \\ -0.01}$ \\
log(Ne/O) & -0.51$\substack{+0.01 \\ -0.01}$ \\
log(Ar/O) & -2.49$\substack{+0.02 \\ -0.02}$ \\
log(Fe/O)$_{gas}$ & -2.07$\substack{+0.06 \\ -0.07}$ ($\substack{+0.14 \\ -0.15}$ syst.$^{\ddagger}$) \\
log(Fe/N)$_{gas}$ & -0.93$\substack{+0.07 \\ -0.08}$ ($\substack{+0.15 \\ -0.15}$ syst.$^{\ddagger}$) \\
Fe dust-to-total ratio & 0.65$\pm$0.13 \\
log(Fe/O)$_{total}$ & -1.61$\substack{+0.22 \\ -0.22}$ \\
$[$O/Fe$]$ & 0.38$\pm$0.22 \\
\midrule
\multicolumn{2}{c}{Blue and red bump features} \\
FWHM~(\HeIIL[4686])~[km~s$^{-1}$] & 1460 $\pm$ 170 \\
EW$_0$~\HeIIL[4686]~[\AA] & 4$\pm$1 \\
EW$_0$~[4605-4650]~[\AA] & 2.4$\pm$0.5 \\
EW$_0$~[5780-5850]~[\AA] & 6$\pm$3 \\
EW$_0$~[5730-5850]~[\AA] & 12$\pm$4 \\
    \bottomrule
    \end{tabular}
  \tablefoot{
  \tablefoottext{$\dagger$}{Adopting the \citet{Reddy_AURORA_dust_fcov_2025} formalism for non-unity covering fraction of dust.}\\
  \tablefoottext{$\ddagger$}{Including additional 20 \% relative uncertainty on the ICF(Fe$^{++}$/O$^{+}$).}
  }
    \label{tab:properties}
\end{table}

\section{Discussion}
\label{sec:discussion}

\subsection{The Wolf-Rayet spectral features}
\label{ssec:WR_discussion}

\subsubsection{Analysis of the blue and red bump}
\label{sssec:WR_bumps}

\begin{figure*}
    \centering
    \includegraphics[width=0.98\textwidth]{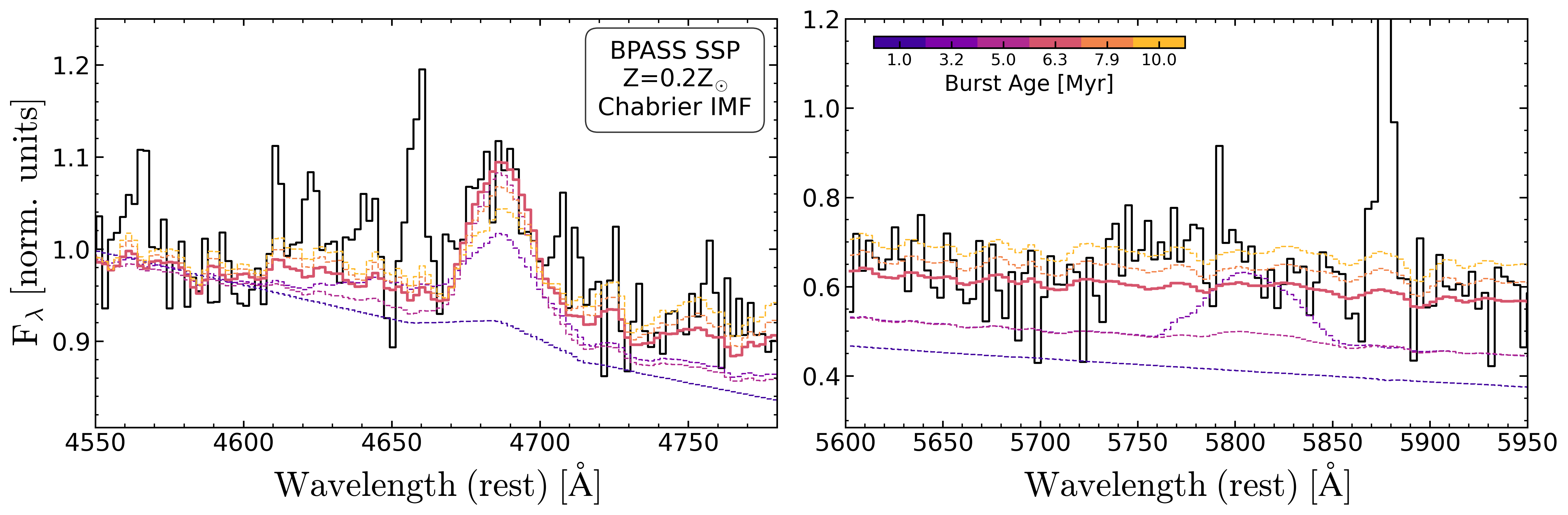}\\
    \includegraphics[width=0.98\textwidth]{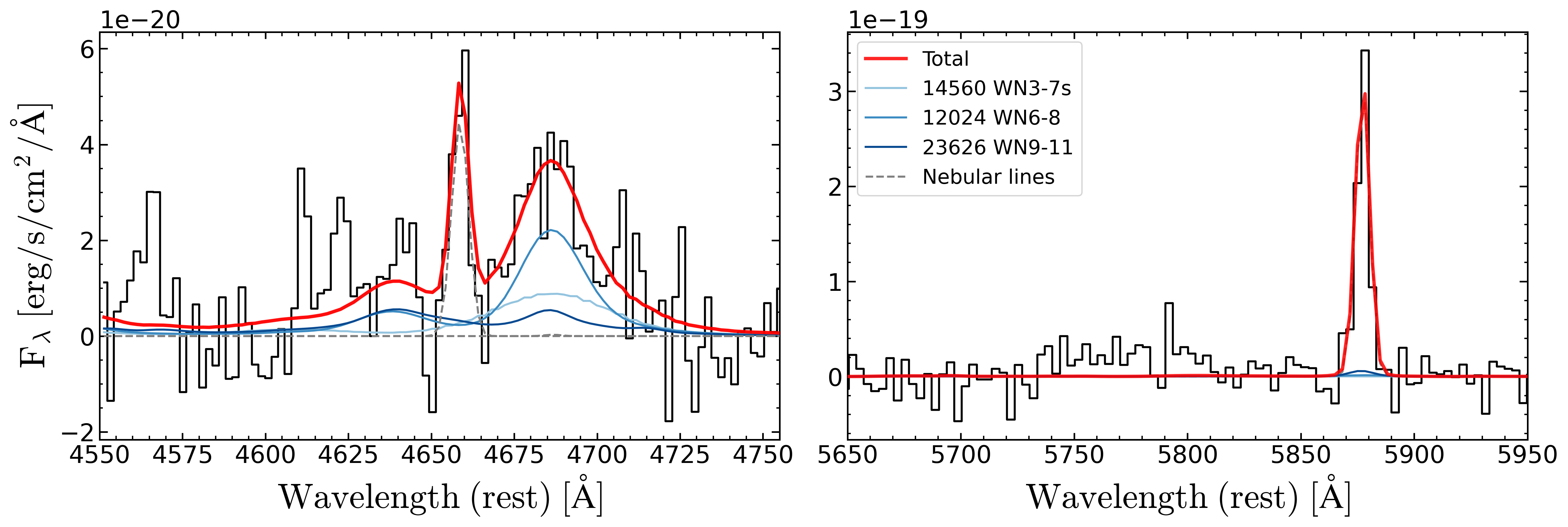}\\
    \includegraphics[width=0.98\textwidth]{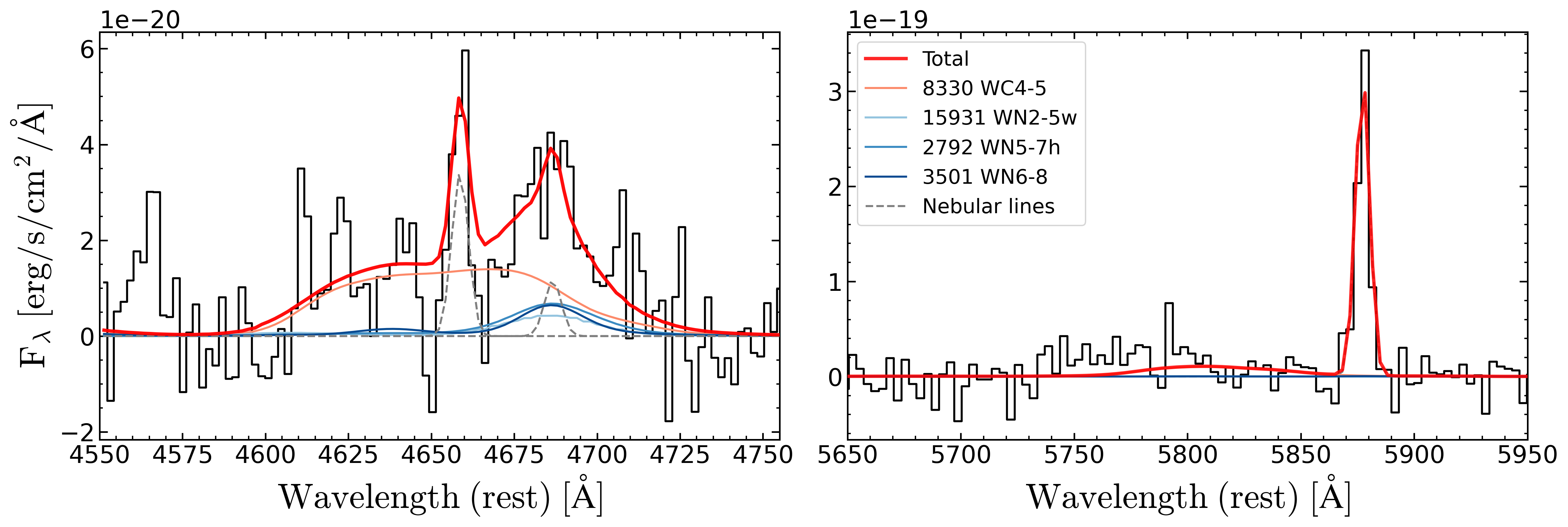}\\   
    \caption{Comparison of blue and red bump features in \sourceshort with SSP models and empirical templates.
    \textit{Upper panels}:  \sourceshort spectrum is compared with a set of SSP models from \textsc{bpass} v2.2.1 for different burst ages, assuming a stellar metallicity of $20\%$~Z$_{\odot}$, a \citet{chabrier_galactic_2003} IMF, an alpha-enhancement [$\alpha$/Fe]$=0.4$, and including binary evolution. The broad \HeII feature, if assumed purely stellar in origin, is well reproduced by a $t\sim6$~Myr-old burst (solid line), although these models under-predict the strength of the blue nitrogen features at $\approx4620$-$4640 \AA$.
    \textit{Middle and bottom panels}:  \sourceshort spectrum is fitted with a combination of empirical WR template spectra from \cite{Crowther_WR_LMC_2023}. The best-fit model in the middle panels include only stars of the WN type, whereas in the bottom panels we have included also WC stars; nebular emission lines are fitted separately with individual Gaussian components.
    }
    \label{fig:WR_bumps}
\end{figure*}

\begin{figure*}[!h]
    \centering
    \includegraphics[width=0.9\textwidth]{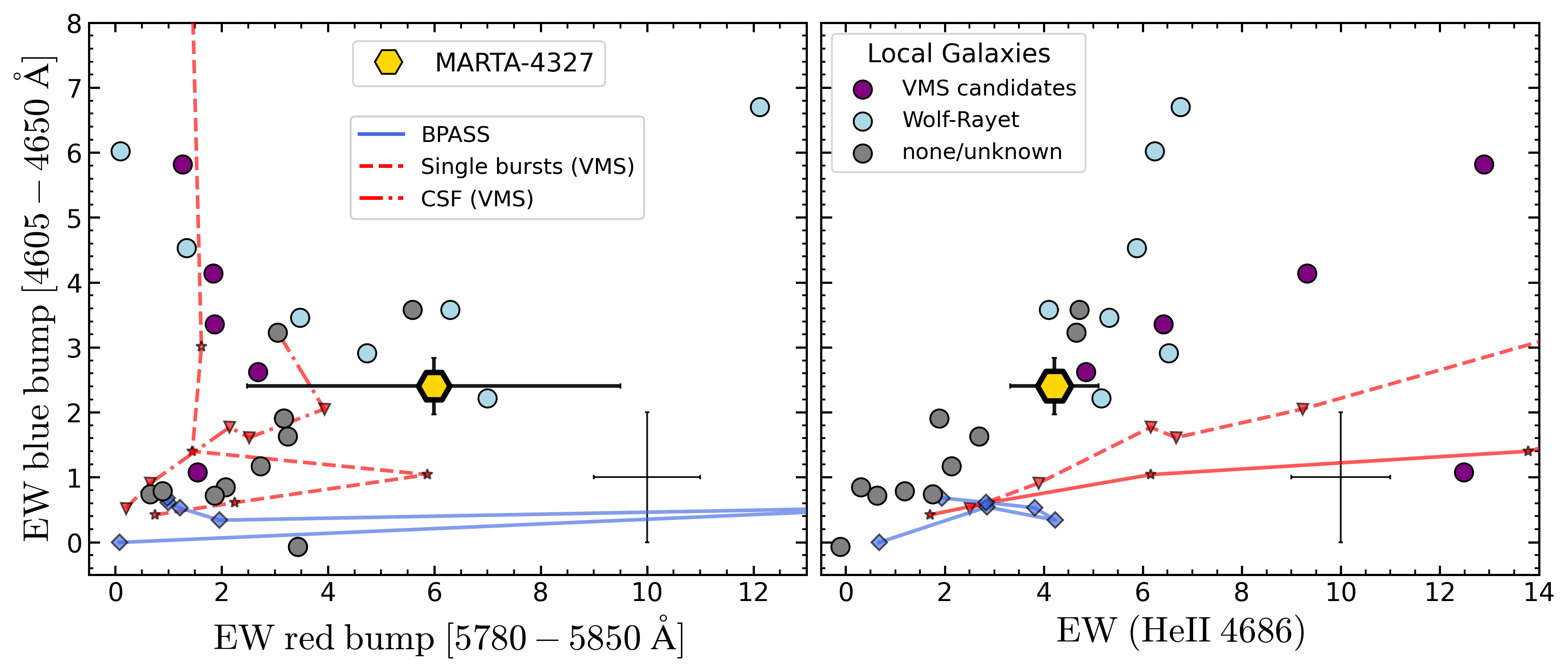}
    \caption{Comparison between the EWs of the blue and red bump features. 
\textit{Left panel}: EW of the blue bump (specifically, the $4605-4650\AA$ complex) versus the EW of the red bump complex (measured between 5780-5850~$\AA$). \textit{ Right panel}: EW[$4605-4650\AA$] versus EW(\HeIIL[4686]).
The circle points report similar measurements for a sample of local galaxies where the presence of VMS (purple) of WR stars (blue) has been identified (or suggested). Gray points mark systems with no signatures of WR or VMS, or galaxies with \HeIIL[4686] detection but no robust classification. Blue tracks are BPASS models for single bursts of star-formation of different ages (from 1 to 10~Myrs), whereas dashed and dot-dashed red tracks are models including VMS contribution for single bursts and constant star-formation, respectively. Data and VMS models compiled from \citealt{Martins_VMS_2023} (see also references therein).
    }
    \label{fig:VMS_vs_WR}
\end{figure*}

In the top panels of Figure~\ref{fig:WR_bumps} we visually compare the NIRSpec spectrum of \sourceshort in the region of the blue and red bumps with a set of synthetic single stellar population (SSP) models from \textsc{BPASS} v2.2.1 and v2.2.3 \citep[with binary evolution enabled, ][]{Stanway_Eldrige_BPASS_2018, Byrne_BPASS_2023}, the latter allowing the elemental abundance ratios to vary compared to the solar composition, in particular regarding the enhancement of $\alpha$-elements relative to Fe.
In particular, the model spectra are generated by assuming a \cite{chabrier_galactic_2003} IMF with high-mass cutoff~$=300$\MSun, stellar metallicity Z~$=0.20$~Z$_{\odot}$, and [$\alpha$/Fe]~$=0.4$\footnote{We note that different choices of [$\alpha$/Fe] do not significantly impact the shape of the WR bumps, which are primarily sensitive to the total stellar metallicity and age of the burst.} (matching the oxygen abundance measured from the \Te-method (Section~\ref{ssec:abundances}) and following values typically observed at $z\sim2$, e.g. ,\citealt{topping_mosdef-lris_2020_ii, Cullen_NIRVANDELS_alpha_enhance_2021}).
Synthetic spectra from individual bursts of different ages (at various steps between $1$ and $10$~Myr, as detailed in the panel) are shown, after they have been normalized (at flux density at $4550\AA$), convolved, and resampled to match the resolution of the NIRSpec data.
We find that the \HeII feature is best reproduced by a single burst of age $\sim 6$~Myr (solid lines), whose height and width nicely overlap with that of the observed bump; this assumes that the \HeII is purely stellar in origin, neglecting possible contributions from narrow, nebular \HeII components originating from photo-ionized gas.
Younger bursts (with fixed other parameters) underpredict the strength of the stellar \HeII bump while, on the other hand, predicting a more prominent red bump at $5800\AA$ (which peaks at $\sim3$~Myr), which is not observed. The same applies to considering higher and/or lower metallicities at fixed age of the burst.

The best-matching $6$~Myr-old \textsc{bpass} model contains a relative fraction of H-rich WR stars (WNh), nitrogen-rich (WN), and carbon-rich (WC) stars (over the total number of WR stars) of $62\%$, $29\%$, and $9\%$, respectively.
However, we observe that such a model is not capable of reproducing the intensity of the nitrogen features observed at $\approx4620$-$4640 \AA$, suggesting some level of nitrogen enhancement in the stellar atmospheres compared to the model spectrum.
This is common to several SPS models \citep[as noted already by ][]{Eldridge_BPASS_2017}, in that the strength of the nitrogen emission is often underestimated compared to that observed in WR-host galaxies \citep[e.g.,][]{Brinchmann_WR_2008} as possibly caused by a non-proper treatment of stellar evolutionary tracks for the WR phase and/or by not including WR spectra for surface hydrogen mass fractions higher than a certain threshold, hence missing the contribution of WNh stars to the stellar lines \citep{Molla_pop_synth_2009, EldridgeStanway2009, Hainich2014, Martins_Palacios_VMS_2022}.

We also attempted to reproduce the observed profile by fitting the spectrum with the set of empirical WR spectral templates by \cite{Crowther_WR_LMC_2023}. 
We note that such templates are provided at a coarser resolution ($10\AA$) than the actual resolution of the NIRSpec data ($\sim5-7\AA$ in the rest-frame); therefore we do not perform any convolution of the \sourceshort spectrum before fitting, noting that information that can be retrieved from the fit is limited by the templates.  
We chose the templates provided for the Large Magellanic Cloud (LMC) (in light of its similar metallicity to \sourceshort), and performed a simultaneous least-squares fit with a linear combination of stellar templates to the combined spectral regions hosting the blue (4550-4800~$\AA$) and red (5600-6000~$\AA$) bumps; the WR templates are converted into flux density units based on the luminosity distance of \sourceshort, and complementary nebular line emission from \FeIII, \HeII, and He I were included in the model by manually adding individual, unresolved Gaussian components.

Based on the apparent strength of the stellar features  \HeIIL[4686] and nitrogen, and on the insights from the stellar population synthesis models, we started by performing a fit including only WR templates of the WN type.
The best-fit model reproduces well the broad \HeIIL[4686] feature, while only partially accounting for the nitrogen stellar wind lines between 4610-4640~$\AA$, which are much weaker in the templates than observed.
The total model, shown by the red curve in the middle panels of Fig.~\ref{fig:WR_bumps}, results from a combination of early- and late-type He-burning WR stars of the nitrogen sequence (WN3-7s, WN6-8, and WN9-11).
As seen in the right-hand panel, such templates do not contribute significantly to the red bump, which remains unfit.
Although emission in the $5750-5800 \AA$ range could possibly originate from a combination of stellar \NIVp$\lambda$5737 and \SiIIIp$\lambda$5740, and nebular\NIIp$\lambda$5755, WN6-8 templates do not produce significant contributions to the stellar component, and no narrow line emission is observed at $\lambda$5755.  

Next, we ran a fit including also templates of WR stars of the early-type carbon sequence, namely WC4-5 (bottom panels of Fig.~\ref{fig:WR_bumps}): these, however, tend to produce a relatively brighter bump at $\approx5800 \AA$ than at $\approx5760 \AA$, as opposed to what observed in \sourceshort, hence still delivering a poor fit to the whole red bump.
At the same time, the inclusion of WC templates results in a strong contribution from these stars to the blue bump, which is now comparable to that of WN stars in the best-fit model, but that at the same time provides both a poorer match to the \HeII broad emission, and produces a broad feature that is partially degenerate with the narrow \FeIIIL[4658] nebular line (likely also a consequence of the coarser resolution of the templates compared to the NIRSpec data).
Finally, we performed a fit by including different types of transition WR stars: we discuss the main insights from this analysis here and show the resulting best-fit models in Appendix~\ref{sec:appendix_C}.

We conclude this section by noting that similar features have been recently observed in the LyC-emitting (LCE) cluster within the strongly lensed galaxy Sunburst Arc at a similar redshift ($z\sim2.37$) as \sourceshort \citep{Rivera_Thorsen_WR_sunburst_arc_2024}, and interpreted as possible evidence of proto-globular-cluster seeds hosting a large number of massive stars enriching the surrounding medium via the release of CNO-cycle H-burning products \citep[see also,][]{Pascale_sunburst_arc_2023, mestric_sunburst_2023, Adamo_clusters_high-z_Nat_2024}: such an enrichment mechanism is among those proposed also to explain the strong nitrogen-enrichment level observed in several $z\gtrsim6$ galaxies \citep[e.g.,][]{cameron_gnz11_2023, charbonnel_gnz11_2023, isobe_CNO_2023, topping_z6_lens_2024, curti_GSz9_2025}. 
Similarly to \cite{Rivera_Thorsen_WR_sunburst_arc_2024}, we speculate that the high star-formation rate density characterizing high-z galaxies could possibly result in an enhanced fraction of (possibly rotating) binaries evolving into WR stars with increased mass-loss rates, which compensates for the relatively low metallicity of high-redshift systems in driving the stellar winds required to enrich the ISM with nitrogen (as N-enrichment is expected to scale with the wind strength, which in turn depends on the metallicity of the stellar atmospheres \citep[e.g.,][]{EldridgeVink2006}.
However, direct signatures of WR stars in the spectra of early galaxies are extremely challenging to detect (and indeed not observed as of yet), and therefore deep spectra of galaxies at Cosmic Noon such as \sourceshort can serve as useful observational benchmarks for stellar population models.

\subsubsection{Wolf-Rayet or very massive stars?}
\label{sssec:WR_or_VMS}

An alternative scenario invokes the contribution from the class of "very massive stars" (VMS, M$_{\star}\gtrsim100$~\MSun). 
As proposed, for example,  by \cite{vink_vms_2024}, such stars are luminous WNh that evolve vertically in the Hertzsprung-Russell (HR) diagram to become classical WR stars (cWR) at later times, have large convective cores leading to chemically homogeneity, and are expected to drive strong, nitrogen-enriched stellar winds, making them robust candidates for polluting the surrounding ISM within the early phases of galaxy formation (due to their large mass-loss rates and relatively slow wind velocities). Moreover, they could also account for the strong \HeII emission, if the correct mass-loss rate is implemented at the optically thin-thick transition point  
\citep{Vink_WR_massloss_2011, Vink_WR_massloss_2012, Sabhahit_VMS_2023}.
However, identifying spectral signatures of cWR from those of VMS is not trivial, especially in galaxies characterized by extended star-formation histories.

Here, we compare our observations with the empirical classification scheme discussed in \cite{Martins_VMS_2023}, based on the relative strength of blue and red bump optical features to discriminate the possible presence of VMS in \sourceshort.
This approach is based on observing that the spectra of a prototypical VMS host, such as the young stellar cluster R136 in the LMC, showcase strong \HeIIL[4686] emission, while having negligible N III 4640 + CIII 4650 \citep{Hainich2014}, the latter being also influenced by the relative abundance of N and C (hence by the metallicity); in contrast, cWR spectra have an N III 4640 + C III 4650 complex of comparable intensity to \HeIIL[4686]. 
In Fig.~\ref{fig:VMS_vs_WR}, we report the location of \sourceshort on two of the diagnostic diagrams proposed by \cite{Martins_VMS_2023}, comparing the EW of the blue bump in the $4605-4650 \AA$ complex with that of the red bump (measured within the [5780-5850~$\AA$] wavelength interval, left-hand panel) and of \HeIIL[4686] (right-hand panel).
In both panels, we report the sample of local sources compiled by \cite{Martins_VMS_2023}, color-coded according to the spectral classification discussed in that paper.
Within these diagrams, \sourceshort occupies an intermediate region, but its features are more aligned to those observed in WR hosts than in VMS hosts, particularly in light of the comparable strength of the $4605-4650 \AA$ and \HeII features in the blue bump (where EW[\HeIIL[4686] is generally much higher in VMS candidates), and of the typical weakness of the red bump in VMS spectra.

Applying such a classification approach to the spectrum of \sourceshort is nonetheless subject to several uncertainties (beyond those pertaining to the S/N of the involved features), including the "dilution" effect that the presence of multiple stellar populations and/or of nebular continuum emission can have on the WR and/or VMS spectrum, altering the inferred EWs.
We also note that, in principle, BPASS models with an upper mass cutoff of 300~\MSun include VMS, however the spectra used for VMS are the same as for normal OB and WR stars \citep{Martins_VMS_2023}.
Furthermore, as VMS and WN reach the CNO equilibrium in their atmospheres, their nitrogen content reflects the initial carbon content, which scales with metallicity.
The relative strength of the nitrogen lines in the complex
depends also on the ionization of the stellar atmospheres, with \NVp dominating at the highest T$_{\text{eff}}$, as well as on the properties (and relative treatment in SPS models) of stellar winds and relative mass loss rates \citep{Rivero_Gonzalez_stellar_NIII_2011, Martins_Palacios_VMS_2022}.
In this sense, additional and complementary clues to the nature of the population of massive stars in this system (and more in general) might come from forthcoming observations of the rest-UV spectrum, as both stellar population synthesis models and observations suggest that the presence of strong (EW > 3~$\AA$, and up to $\sim$30~$\AA$) and broad (FWHM $\gtrsim$ 1000~km~s$^{-1}$) \HeIIL[1640] emission and multiple, strong P-Cygni profiles for \CIV$\lambda$1550, \NIVp$\lambda$1486, and \NIVp$\lambda$1720 stellar features could be indicative of the presence of VMS \citep{Martins_Palacios_VMS_2022, Martins_VMS_2023, Berg_WR_M101_2024}.

\subsection{Chemical abundances and star-formation history}
\label{ssec:SFH}

\begin{figure*}
    \centering
    \includegraphics[width=0.99\textwidth]{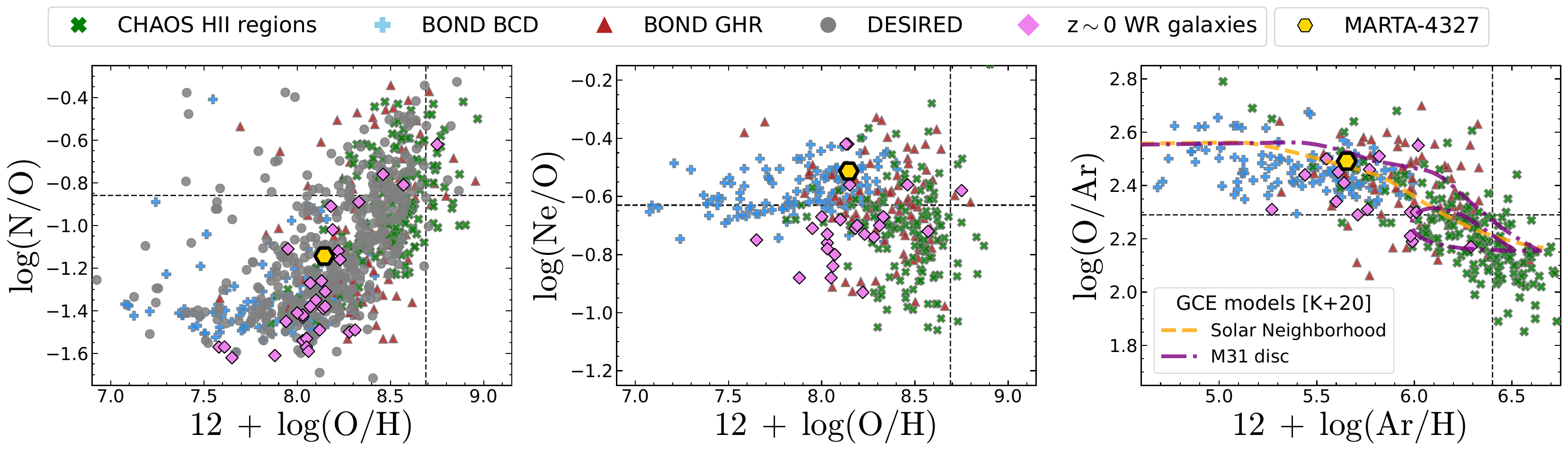}
    \caption{Chemical abundance patterns. 
    From left to right, we show the position of \sourceshort in the N/O vs. O/H, Ne/O vs. O/H, and O/Ar vs. Ar/H diagrams.
    A compilation of local galaxies and HII regions with \Te-based abundance measurements from CHAOS \citep{berg_carbon_2016}, BOND \citep{vale_asari_bond_2016}, and DESIRED \citep{Mendez-Delgado_Fe_O_2024} is shown for comparison, together with the sample of Wolf-Rayet galaxies from \cite{lopez_sanchez_WR_abundances_2010}.  
    In the O/Ar vs. Ar/H diagram, we overplot the galactic chemical evolution tracks for the MW (solar neighborhood, orange) and the disc of M31 (between 3 and 14 kpc, purple) from \cite{kobayashi_origin_2020} (see also, \citealt{Arnaboldi_M31_PNe_2022}). 
    }
    \label{fig:abundance_patterns}
\end{figure*}

\subsubsection{Abundance patterns}
\label{sssec:abundance_patterns}

In Fig.~\ref{fig:abundance_patterns} we report the location of \sourceshort in different abundance pattern diagrams, namely N/O versus O/H, Ne/O versus O/H, and O/Ar versus Ar/H (here used as a proxy of the metallicity). 
A sample of local galaxies and $\rm HII$ regions with \Te-based abundances is also shown for comparison, as compiled from CHAOS \citep{berg_chaos_2020}, BOND \citep{vale_asari_bond_2016}, and DESIRED \citep{Mendez-Delgado_Fe_O_2024} projects, and for which elemental abundances have been consistently re-computed adopting the same set of atomic parameters and assumptions as for \sourceshort; in addition, the sample of star-forming galaxies with Wolf-Rayet spectral features presented in \cite{lopez_sanchez_WR_abundances_2010} is shown by violet diamond markers.
Within these panels, \sourceshort overlaps with the distribution of local systems and aligns with the expectations of standard chemical evolution models.

In the left-hand panel of Fig.~\ref{fig:abundance_patterns} we investigate the log(N/O) vs log(O/H) pattern.
We do not observe any evidence of significant N/O-enhancement above the expected level given the metallicity of \sourceshort, which overlaps with the local distribution of galaxies \citep[as well as with most $z \sim 2$ galaxies of comparable metallicity with \Te-based N/O measurements, see][]{Cataldi_NO_2025}.
This implies that the population of massive stars in the WR phase has not significantly impacted the global ISM of the galaxy, and that the nitrogen observed in the form of N III wind lines in the blue bump is likely confined to the stellar atmospheres or their immediate surroundings.
Although some past studies have suggested that winds from WR stars can cool and mix with the ISM on relatively shorter timescales compared to CC-SNe ejecta \citep{KobulnickySkillman1997, lopez_sanchez_2007}, the patterns observed in \sourceshort align with those of the WR galaxy sample with Te-based abundance measurements from \cite{lopez_sanchez_WR_abundances_2010}. Furthermore, systematic studies of large samples of galaxies from the Sloan Digital Sky Survey (SDSS) suggest that WR galaxies of high EW(\Hbeta) (hence, systems with prevalence of young stellar populations) do not show any significant enhancement in N/O (at fixed O/H) compared to sources with the same EW(\Hbeta) but without WR features in their spectra, in contrast to what is more commonly observed at low-metallicity and low-EW(\Hbeta) \citep{Brinchmann_WR_2008}.

In the previous Section, we have seen that the presence of N III features in the blue bump (together with the absence of strong C lines) suggests that the WR population in \sourceshort is dominated by WN.
As WN stars are expected to be at (or close to) the CNO equilibrium in their atmospheres \citep{Crowther_WR_2000, Crowther_WR_review_2007}, one could also expect an elevated He/H abundance \citep{Kobulnicky_NGC5253_massive_stars_1997}, which would be, however, reflected into ISM abundances only if efficient helium mass-loss mechanisms (associated, for example, with rotational mixing) are in place and wind mixing timescales into the ISM are short enough.
The absence of any significant He/H enhancement in the ISM of \sourceshort again suggests that either WR spectral features have shown up in the spectrum before enrichment becomes detectable on global ISM scales (being associated with a young $\sim5$~Myr old burst), that the enrichment is strongly localized, and/or that such a young population includes binaries (with not fully stripped and partially CNO-processed H-envelopes) formed in "intermediate" ($\approx 0.5$~Z$_{\odot}$) metallicity environments \citep{Gotberg_binaries_2017}, as predicted, for example,  by BPASS models \citep[][see also the previous Section]{Eldridge_BPASS_2017}. 
At the same time, it is also possible that the N/O abundance inferred from the low-ionization \NII species provides only a partial representation of the full chemical structure of \sourceshort and that, in the presence of multiple emitting regions of different chemical properties ("chemical stratification"), any nitrogen enrichment due to WR stars might be visible only in high-ionization nitrogen lines (emitted in the rest-frame UV) probing the region more closely surrounding such population, in a way not dissimilar to that observed (or postulated) in some galaxies at higher redshift \citep[e.g.,][]{Pascale_sunburst_arc_2023, ji_nitrogen_AGN_z5_2024}.      

The middle panel of Fig.~\ref{fig:abundance_patterns} reports the location of \sourceshort on the Ne/O versus O/H diagram. 
This source aligns with the local distribution of galaxies,  reflecting the common nucleosynthetic origin of the two elements (since neon is an $\alpha-$element like oxygen, their ratio is not expected to vary significantly with metallicity, nor to evolve with redshift, see, e.g., \citealt{arellano-corodva_2023, Stanton_excels_2025}).  

Finally, we explore the behavior of \sourceshort in terms of its argon-to-oxygen abundance ratio. 
The argon enrichment is in fact expected to partially follow that of iron in galaxies, in virtue of a non-negligible contribution from SNe-Ia; therefore, the O/Ar abundance ratio has been identified as a good proxy of O/Fe, reflecting the level of $\alpha$-enhancement and its connection to the timescales of star-formation and chemical evolution \citep[e.g.,][]{Arnaboldi_M31_PNe_2022, Bhattacharya_argon_2025, Stanton_excels_2025}.
On the O/Ar versus Ar/H diagram (right-hand panel of Fig.~\ref{fig:abundance_patterns}, and proxy for O/Fe versus Fe/H), \sourceshort follows the trend outlined by local galaxies and chemical evolution models of the solar neighborhood (orange track) and the M31 disc (modeled between 3 and 14 kpc, purple track) \citep{kobayashi_origin_2020, Arnaboldi_M31_PNe_2022}, occupying the region around the "knee" of the relationship where the contribution of SNe-Ia starts to be significant and the abundance pattern deviates from the plateau expected for pure CC-SNe enrichment.
More specifically, the chemical evolution model track of the solar neighborhood from \citep{kobayashi_origin_2020} crosses the \sourceshort value on the O/Ar versus Ar/H diagram at a time of $\sim3.4$~Gyr after the initial burst of star-formation, closely matching (considering the uncertainties involved) the age of the Universe at the source redshift, i.e., $\sim2.96$~Gyr.  

\subsubsection{Wolf-Rayet, Fe enrichment, and dust depletion}
\label{sssec:WR_and_dust}

In Fig.~\ref{fig:Fe_O}, we have seen that the gas-phase Fe/O abundance as measured in \sourceshort is higher than the average trend followed by local nebulae, given its oxygen abundance.
Interestingly, a similar behavior can be observed in other sources with identified Wolf-Rayet spectral features, 
including the low-metallicity, nitrogen-enhanced LyC emitter in the Sunburst Arc at $z\sim2.37$ \citep{Rivera_Thorsen_WR_sunburst_arc_2024, Welch_Sunburst_arc_2025}, some extremely-metal-poor galaxies in the local Universe (e.g., J0811+4730, \citet{Izotov_J0811_2018}; J1631+4426, \citet{Kojima2021}; J1205+4551, \citet{Izotov2017, izotov_xmp_2021}, and the sample of WR galaxies from \cite{lopez_sanchez_WR_abundances_2010}; these are also shown in Fig.~\ref{fig:Fe_O}).
We can hence speculate about the possible role of WR stars in boosting Fe abundance at low-to-moderate metallicity over short star-formation timescales. 

In the atmosphere of WN stars, nitrogen is quickly enriched at the expense of oxygen, potentially leading to a localized enrichment of the ISM with nitrogen via strong winds \citep{Abbott_WR_review_1987, Maeder_WR_2014, lopez-sanchez_NGC5264_2012}.
At the same time, these processes would affect the Fe/O ratio while keeping the Fe/N ratio almost constant, thus suggesting a possible link between regions with enhanced N/O abundances, relatively high (Fe/O)$_{\text{gas}}$, and the presence of WR stars \citep{Mendez-Delgado_Fe_O_2024}. However, we have seen that the N/O abundance measured in \sourceshort from optical lines (likely tracing the global galaxy ISM as probed by the integrated NIRSpec spectrum) is not particularly elevated compared to the distribution of local galaxies. 

Instead, a plausible scenario for \sourceshort links the enhanced gas-phase Fe/O abundance with dust destruction, either via shocks or intense radiation fields possibly associated with the WR population \citep{Jones_dust_grains_1996, izotov_chemical_2006}. 
These could possibly displace and destroy dust grains, releasing the previously depleted Fe into the gas-phase, and producing a transient boost in gas-phase Fe/O.
Strong Fe enhancements (by about one dex) have been, for instance, observed in gamma-ray burst host galaxies \citep[e.g.,][]{Waxman_dust_sublimation_2000, DeCia_GRB_2012}, suggesting a very recent and isolated episode of massive star-formation that caused dust destruction.
This scenario would also be reflected in a lower dust mass and dust-to-metal ratio in the proximity of the WR population.

An alternative scenario instead invokes more exotic rapid Fe-enrichment mechanisms.
In extreme, very localized starbursts, some CC-SNe (especially from massive progenitors, $>30$ \MSun) may contribute iron group elements. This could be in the form of "Hypernovae" and/or "metal-rich fallback SNe", or "stripped-envelope SNe" \citep[e.g.,][]{Nomoto_hypernovae_2006, tominaga_2007, nomoto_chemical_evolution_2013}, locally elevating the gas-phase Fe/O before SNe Ia contributes on longer timescales, while dust formation proceeds normally and the dust mass should be consistent with the inferred metal content.
Forthcoming ALMA observations probing the dust properties in \sourceshort will help to distinguish among these scenarios.

\subsubsection{Fe/O abundance and specific star formation rate}
\label{sssec:FeO_vs_sSFR}

In virtue of their different production and release timescales, the relative iron-to-oxygen [Fe/O] abundance ratio can be used to constrain the star-formation history of galaxies.
Because of the intrinsic challenges associated with the direct measurement of stellar Fe abundances in high-$z$ systems from rest-UV stellar winds and continuum features, such an investigation is still limited to a few individual cases, relies heavily on stacking techniques, or it is based on indirect inference from SSP modelling aimed at reproducing optical line ratios
\citep[e.g.,][]{sommariva_stellar_2012, steidel_reconciling_2016, strom_measuring_2018, topping_mosdef-lris_2020_i, Cullen_NIRVANDELS_alpha_enhance_2021, calabro_UV_Slope_vs_metallicity_2021}.

Alternatively, \cite{Chruslinska_Fe_O_2024} proposed 
that the relationship between [O/Fe] and the specific star formation rate (sSFR) in galaxies (the latter probing the timescales of galaxy formation, and hence also of enrichment) could be used to indirectly guess the metallicity (in terms of Fe/H) of young stars, once the oxygen abundance is known (and modulate the systematic uncertainties on the normalization of the oxygen abundance scale due to the abundance discrepancy factor (ADF) e.g., \citealt{mendez_delgado_t_inhomogeneities_2023}). 
Such a relationship is expected to be universal (i.e., redshift invariant), with systems of different ages and chemical maturity probing different regions of the [O/Fe] versus sSFR plane, and it is also observed in good agreement with the same relationship derived for stars in the MW disk (for relatively young and metal rich stars) and MW halo (probing older and metal poor stars), once a sSFR value is assigned based on the stellar age and under a given parametrization of the MW star-formation history \citep{Chruslinska_Fe_O_2024}.
Direct measurements of both quantities in high-z systems not only help to validate such a relationship, but also help constrain the underlying physics, including the yields of core-collapse SNe (mainly responsible for oxygen enrichment) and the time-delay and progenitors' properties of type-Ia SNe (main sources of iron enrichment), as different prescriptions on such parameters have been demonstrated to produce relatively different shapes of the [O/Fe] versus sSFR relationship in cosmological simulations such as EAGLE \citep{Schaye_EAGLE_2015} and TNG \citep{nelson_TNG_2019}.

\begin{figure}
    \centering
    \includegraphics[width=\columnwidth]{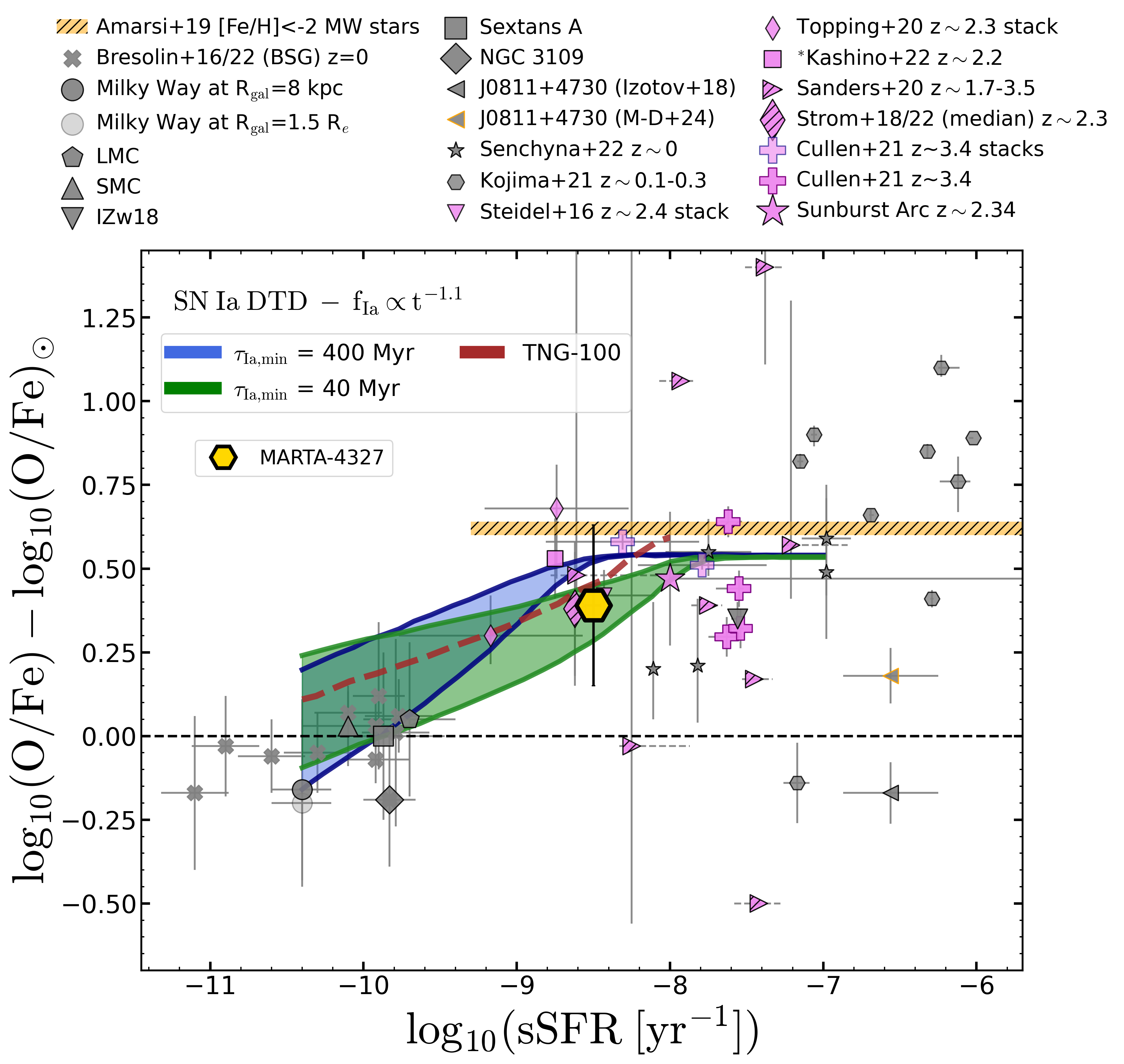}
    \caption{[O/Fe] versus sSFR relationship for galaxies.
    The location of \sourceshort is reported alongside a literature sample comprised of nearby dwarf galaxies, local galaxies with blue supergiant-based metallicity estimates, extremely metal-poor dwarf galaxies,
    and high-redshift star-forming galaxies/stacks, as presented in \citet{Chruslinska_Fe_O_2024} (see the legend, and references therein). \sourceshort represents, together with the Sunburst Arc, the only high-z galaxy with Te-based determination of both gas-phase oxygen and Fe abundances (the latter corrected to total Fe/O abundance as detailed in Sect.~\ref{sssec:gas_Fe_O}). The hatched bar indicates the average abundance ratio of the MW thick disc and/or halo dwarf stars with [Fe/H] <–2 from \cite{Amarsi_abundances_MW_2019}, whereas hatched symbols mark high-z galaxies with indirect determination of stellar metallicity (Fe/H) via  modeling of optical lines.  
    Shaded areas represent the model [O/Fe]–sSFR relation derived as detailed in \citet{Chruslinska_Fe_O_2024} (see their Eq.~1) for different assumptions on the SN Ia delay-time distribution, namely a power-law DTD with slope equal to $-1.1$ with $\tau_{Ia,min}=400$~Myr (blue) and $40$~Myr (green); upper and/or lower curves of the same color correspond to different choices of the relative iron yield of SN Ia and CCSN, namely $C_{Ia/\text{CC}}=0.74$ and $2.5$, respectively. The brown, dashed line marks the same relationship as obtained in TNG100 simulations with $f_{Ia} \propto -1.12$ and $C_{Ia/\text{CC}}=2.04$. 
    }
    \label{fig:FeO_sSFR}
\end{figure}

This is shown in Fig.~\ref{fig:FeO_sSFR}, where we populate the [O/Fe] versus sSFR diagram with a combined sample of MW stars, local galaxies, and high-z sources where both oxygen and iron abundances have been derived by means of either direct or indirect methodologies, and which have been reported to a common baseline tied to Te-based, gas-phase oxygen abundance (corrected for oxygen depletion onto dust, where needed) and \textsc{bpass} stellar population synthesis-based Fe abundance measurements, respectively \citep[see ][for a detailed discussion on the involved systematic uncertainties and on how to convert these measurements to a common abundance scale for Fe and O]{Chruslinska_Fe_O_2024}.
Once accounting for the residual fraction of Fe depleted onto dust grains (Sect.~\ref{sssec:gas_Fe_O}), \sourceshort aligns with galaxies of similar sSFR and redshift, representing one of the few high-z sources with self-consistent gas-phase probes of both quantities. 

In Fig.~\ref{fig:FeO_sSFR} we also plot different models of the [O/Fe] versus sSFR relation for galaxies, derived by means of different assumptions on the delay-time distribution of SN Ia, Fe yields of CC-SNe, and the shape of the star-formation history of galaxies. 
At the sSFR of \sourceshort (log(sSFR)=-8.5, corresponding to a characteristic timescale of star-formation $\tau_{SF}\sim 315$~Myr), enrichment from SNe Ia is expected to start contributing significantly.
Here, we model the [O/Fe] versus sSFR relationship as described in \cite{Kashino_stellar_MZR_z2_2022} and \cite{Chruslinska_Fe_O_2024} (see their Eq.~1), assuming a specific power-law parametrization for the SN Ia delay-time distribution (DTD) of the form $f_{Ia} \propto t^{-1.1}$ \citep{Maoz_Graur_SnIa_2017}, noting, however, that the exact shape is strongly sensitive to the formational channels and progenitors of SN~Ia \citep[e.g.][]{Nelemans_SNIa_DTD_2013}. Different minimum SN Ia delay times\footnote{The time at which SN Ia start to contribute to the chemical enrichment of the ISM.} $\tau_{\text{Ia},min} = 40$ and $400$~Myr correspond to green and blue tracks, respectively (this means that $f_{Ia}=0$ for $t < \tau_{\text{Ia},min}$), whereas at fixed color (hence, fixed choice of $\tau_{\text{Ia},min}$), the shaded areas encompass the relationships predicted by varying another critical parameter, i.e., the relative iron yield of SNe Ia to CC-SNE, $C_{Ia/CC}$, which is assumed equal to $C_{Ia/CC}= 0.74$ and $2.5$ for upper and lower curves, respectively.
In addition, we plot (as the dashed brown curve) the [O/Fe] versus sSFR relationship predicted within TNG-100 simulations for galaxies that obey the evolving star-forming main sequence (SFMS) from \citep{Kashino_stellar_MZR_z2_2022} up to $z\sim8$, with a SN~Ia DTD of the form $f_{Ia} \propto t^{-1.12}$\footnote{This is generally consistent with double-degenerate SN~Ia progenitor scenarios involving two merging white dwarfs, e.g., \citealt{Iben_SNIa_progenitors_1984, Webbink1984}.}, and by integrating Eq.~1 from \citealt{Chruslinska_Fe_O_2024} with the set of parameters reported in their Table~A.1 \citep[see, also,][]{Schaye_EAGLE_2015, Pillepich_TNG_2018, Naiman_TNG_2018}.
The general expected behavior is that longer $\tau_{\text{Ia},min}$ timescales produce a turnover in the relationship that occurs at the high end in sSFR, whereas the relative iron production efficiency of CC-SNe and SN Ia (the former highly uncertain in both observations, e.g.,\citealt{Anderson2019_AA_56Ni, Rodriguez_Fe_SNIa_2023}, and models, e.g., \citealt{WoosleyHeger2007, SawadaSuwa2023, Imasheva_SNIa_yields_2023}) impacts the steepness of the relationship and the saturation [O/Fe] value at the low sSFR end. In this sense, and while acknowledging the large systematic uncertainties affecting its [O/Fe] determination, if taken at face value \sourceshort overlaps with the set of curves obtained by assuming shorter values for $\tau_{\text{Ia},min} = 40$~Myr, falling in the steep region of the relationship beyond the turnover knee.

\subsection{Ionized gas outflow}
\label{ssec:discussion_outflow}

In Sect.~\ref{sec:g235h_fit} we have reported the detection of a broad and slightly blue-shifted kinematical component under the \Halpha emission line. If we interpret such a component as tracing outflowing gas powered by star-formation, we can estimate the ionized--phase mass outflow rate and mass loading factor using the standard Case B framework \citep[e.g.,][]{genzel_sins_2011, forster_schreiber_kmos3d_2018}. 
We discuss an alternative (while more speculative) origin of the broad \Halpha component (as associated to the presence of an intermediate-mass black hole) in Appendix~\ref{sec:appendix_D}.

Assuming a conical and approximately steady outflow with outer radius $R_{\mathrm out}$, constant velocity $v$, and electron density $n_{\mathrm e}$, the ionized mass and mass outflow rate are
\begin{equation}
M_{\mathrm out}^{\mathrm{(H\,II+He)}}\;=\;\frac{\mu\,L_{\mathrm{H}\alpha,\mathrm{br}}}{\gamma_{\mathrm{H}\alpha}(T_{\mathrm e})\,n_{\mathrm e}},\qquad
\dot M_{\mathrm out}\;=\;\frac{M_{\mathrm out}^{\mathrm{(H\,II+He)}}\,v}{R_{\mathrm out}},
\end{equation}
where $\gamma_{\mathrm{H}\alpha}$ is the \Halpha volume emissivity (function of \Te) and $\mu=1.36\,m_{\mathrm p}$ accounts for the mass of helium. 
In the case of \sourceshort, we adopt $R_{\mathrm out}=1\,\mathrm{kpc}$, given that the size of the outflow is unresolved in the 1D NIRSpec spectrum, \Te=\Tiii, and \Ne derived from the narrow \SII emission as our fiducial assumptions for the outflow--rate calculations.

We derived the mass loading factor referenced to the local star-formation rate derived from the \Halpha emission within the slit as
\begin{equation}
\eta \;\equiv\;\frac{\dot M_{\mathrm out}}{\mathrm{SFR}_{\mathrm{slit}}}
\;=\;\frac{\mu}{\gamma_{\mathrm{H}\alpha}(T_{\mathrm e})\,n_{\mathrm e}}\;
\frac{v}{R_{\mathrm out}}\;
\frac{L_{\mathrm{H}\alpha,\mathrm{br}}}{\mathrm{SFR}_{\mathrm{slit}}}.
\end{equation}
We compute $\mathrm{SFR}_{\mathrm{slit}}$ from the narrow H$\alpha$ luminosity adopting the
\cite{kennicutt_star_2012} calibration converted to a Chabrier IMF, and assuming the same nebular attenuation for the narrow and broad components.
We also corrected the measured FWHM of the broad component for the NIRSpec/G235H instrumental line-spread function computed at the observed \Halpha wavelength, i.e., $\sigma_{\rm inst}=56\ \mathrm{km\ s^{-1}}$, obtaining an intrinsic $\mathrm{FWHM}_{\rm int}=425\ \mathrm{km\ s^{-1}}$,
and adopted two common definitions for the outflow velocity in the literature, e.g., from \cite{Rupke_outflows_2005, fiore_agn_2017}
\begin{equation}
\begin{split}
v_{\rm out}\;\equiv\;|\Delta v_{\rm br}|+\frac{1}{2}\mathrm{FWHM}_{\rm int}\;=\;283\ \mathrm{km\ s^{-1}}, \\
v_{\rm max}\;\equiv\;|\Delta v_{\rm br}|+\mathrm{FWHM}_{\rm int}\;=\;495\ \mathrm{km\ s^{-1}}.
\end{split}
\end{equation}
We obtain $\dot M_{\rm out}(v_{\rm out})=4.38 \ M_\odot\ \mathrm{yr^{-1}}$ and $\dot M_{\rm out}(v_{\rm max})=6.69\ M_\odot\ \mathrm{yr^{-1}}$, which translates into 
$\eta(v_{\rm out})=0.155$ and $\eta(v_{\rm max})=0.271$.
These values sit within the SF-driven regime and are in line with that reported for main-sequence galaxies at similar redshift and stellar mass regimes\footnote{We note, however, that adopting, e.g., $n_{\mathrm e}=380\ \mathrm{cm^{-3}}$ as in \citealt{forster_schreiber_kmos3d_2018} would lower all quoted $\dot M_{\rm out}$ and $\eta$ by a factor $\simeq 1.9$; similarly, choosing a larger effective radius would reduce them proportionally.} \citep{forster_schreiber_kmos3d_2018}, while $\approx10\times$ higher than observed in local dwarf galaxies \citep{Marasco_dwarfs_outflows_2023}.
This implies that, although the observed outflow is injecting a nonnegligible amount of momentum and energy into the surrounding ISM, it is unlikely to impact the global star-formation rate or drive wholesale removal of the ISM from the galaxy: the momentum and energy rates we infer are comparable to those expected from stellar feedback, but fall short of the values required to unbind the total gas reservoir. The impact on gas-phase metallicity is uncertain, as even modest-velocity winds can transport metals to larger radii and enrich the circumgalactic medium \citep{TumlinsonPeeplesWerk2017}.
Cosmological simulations such as FIRE \citep{muratov_feedback_FIRE_2015} and TNG50 \citep{Nelson2019} predict total mass loading factors of the order of unity to tenths for $\sim10^9$\MSun. 
However, we note that since we are only probing the warm ionized gas-phase from H$\alpha$, our inferred mass outflow rate is a lower limit and
the true total (multiphase) outflow rate could be significantly higher if outflows are present in the cooler, molecular gas \citep{cicone_massive_2014} and/or hotter, ionized plasma \citep{StricklandHeckman2009} phases.

\subsection{Alternative excitation mechanisms: Interpreting the presence of the O\textsc{i}~$\lambda8446$ emission line}
\label{ssec:discussion_OI8446}

In Sect.~\ref{ssec:OI_detection} we reported the detection of the permitted O\textsc{i}$\lambda$8446 emission line in \sourceshort, 
representing one of the first of its kind in individual star-forming galaxies at high redshift.
From a physical point of view, such line emission can arise through multiple excitation mechanisms, including Ly$\beta$ fluorescence, collisional excitation in partially ionized regions, and pumping by stellar continuum \citep{Grandi1980}.
Given the star-forming nature of our source and the absence of AGN-like features in the nebular spectrum (as suggested by all classical line-ratio diagnostic diagrams), we consider below what other processes could plausibly explain the observed strength of O\textsc{i}$\lambda$8446.

First, we note that no emission is observed in the O\textsc{i}$\lambda$7776 and O\textsc{i}$\lambda$7990 lines; in particular, we derive a $3\sigma$ upper limit of $\approx0.4$ in both O\textsc{i}$\lambda$7776/$\lambda$8446 and O\textsc{i}$\lambda$7990/$\lambda$8446 line ratios. 
From these values, we can robustly exclude recombination as the mechanism responsible for powering O\textsc{i}$\lambda$8446 (since in such a case the O\textsc{i}$\lambda$7776/$\lambda$8446 ratio is expected to be $\approx1.7$), whereas the inferred upper limits are still consistent with expected line ratios predicted for Case B by \cite{Grandi1980} for collisional excitation (O\textsc{i}$\lambda$7776/$\lambda$8446 $\approx 0.3$) and stellar continuum fluorescence (O\textsc{i}$\lambda$7790/$\lambda$8446 $\approx 0.05$); therefore, in this sense they are not fully conclusive.

On the other hand, Ly$\beta$-pumping is another relevant mechanism that can possibly explain the presence (and relative strength) of the O\textsc{i}$\lambda$8446 emission in astrophysical contexts. In fact, accidental resonance\footnote{The frequency difference is roughly one thermal Doppler width of Ly$\beta$ at temperatures T~$\sim10^4$~K.} between 
Ly$\beta$ and the OI $2p^4~^3P_2$-$3d~^3D^0$ transition can pump the $3d~^3D^0$ level\footnote{If the \Halpha optical depth is large enough such that the H I~$3p$ level is not rapidly depopulated.}, which can then produce O\textsc{i} transitions at $\lambda$1304, $\lambda$8446, and $\lambda$11287 via recombination cascade \citep{NetzerPenston1976, Grandi1980}.  
In this sense, the absence of strong O\textsc{i}$\lambda$7774, O\textsc{i}$\lambda$7254, and O\textsc{i}$\lambda$7790 emission in the \sourceshort spectrum supports this scenario (though, as seen, our data do not allow us to fully discriminate Ly$\beta$ from stellar continuum fluorescence).

To produce O\textsc{i}$\lambda$8446 emission via Ly$\beta$ pumping, it is required to have a combination of dense neutral oxygen and intense UV radiation.
In terms of possible astrophysical sources, it has recently been suggested that analog configurations to Weigelt blobs observed in the proximity of luminous blue variable stars in the MW such as $\eta$ Carinae might explain the strong OI~$\lambda$8446 emission in the highly magnified stellar system dubbed "Godzilla" within the Sunburst Arc \citep{Choe_SArc_godzilla_2025}.    
These high density ($\sim10^7$~cm$^{-3}$) gas condensations are self-shielded from ionization and remain neutral or partially neutral in their interiors, while having an ionized surface layer facing the stars. 
The neutral O\textsc{i} is hence exposed to Ly$\beta$ photons emitted from the ionized surface of the blob, as well as from stellar winds, populating the $3d~^3D^0$ level of neutral oxygen atoms\footnote{As the decay of the lower $3s~^3S^1$ level to the ground level is faster than the decay of the upper $3p~^3P$ level, population inversion can occur, which, under specific conditions, can produce stimulated emission and a laser effect, e.g., \citealt{JohanssonLetokhov2005}.}.

Whether similar structures also exist in the proximity of young populations of massive WR and/or O stars (and what characteristic O\textsc{i}~$\lambda$8446/7774 line ratios should they produce) is not completely clear. 
Here, we just note that no evidence of WR or VMS is reported, for example in the Godzilla stellar complex \citep{Choe_SArc_godzilla_2025}, contrary to what was observed in the LyC-emitting cluster within the same Sunburst Arc \citep{mestric_sunburst_2023, Rivera_Thorsen_WR_sunburst_arc_2024}, and that evidence for Bowen fluorescence associated with WR stars is generally scarce.
Furthermore, no evidence for Ly$\alpha$-pumped Fe\textsc{ii} lines at $\lambda~4853,8490$ is found in \sourceshort, while these lines are brighter than O\textsc{i}$\lambda$8446 in Weigelt blobs due to the Fe\textsc{ii} laser mechanism \citep{JohanssonLetokhov_FeII_EtaCar_2004, DavidsonHumphreys_EtaCar_Nat_2012}.
At the same time, the absence of very high-ionization emission lines in the optical spectrum of \sourceshort and the derived nebular line ratios disfavor alternative scenarios associated with AGNs \citep{Grandi1980}, planetary nebulae \citep{Rudy_NGC7027_1992}, or supernova remnants \citep{Oliva_OI_SNe_1993}.
Observations of additional emission lines arising from the same neutral oxygen pumping cascade, such as the UV O\textsc{i}$\lambda1304$ multiplet and NIR O\textsc{i}$\lambda11287, 13165$ lines (not currently available for this system), together with diagnostics sensitive to higher density gas, might provide further compelling evidence in favor (or disfavor) of the Ly$\beta$ fluorescence scenario.

Finally, fast shocks from WR stellar winds and/or X-ray heating from high-mass X-ray binaries (HMXBs) could offer an alternative source of localized excitation. 
Shock models from \citet{allen_mappings_2008} predict enhanced \OI emission at velocities $v{_s}~\gtrsim150$~km/s, particularly when preshock densities are high. In such environments, the partially ionized zones behind the shock front are capable of producing strong collisional excitation of neutral oxygen lines, although we note that large-scale shocks in \sourceshort are ruled out by the observed global \OIL/\Halpha line ratio, suggesting that \OIL likely traces extended photoionized structures.

We can hence imagine a plausible scenario in which localized, dense gas condensations (e.g., Weigelt-blob analogs) are illuminated by nearby young stellar populations and processed by feedback, producing the WR spectral features, dominating the O\textsc{i} emission (from a combination of fluorescence and collisional excitation), and enhancing the inferred gas-phase Fe via dust grain (partial) destruction, leading to the apparent (Fe/O)$_{\text{gas}}$ enhancement (Sect.~\ref{sssec:gas_Fe_O}).
At the same time, the global nebular spectrum, probing the large-scale galaxy ISM and reflected into the inferred strong-line ratios, nebular attenuation, and chemical abundance patterns, remains consistent with the typical star-formation process in a $z\sim2$ galaxy.
Overall, these findings illustrate how the spatial and structural complexity of the ISM in compact, highly star-forming galaxies at high redshift (and, in particular, the possible presence of dense structures embedded within or near the star-forming complexes) can give rise to emission features that depart from the classical physical modeling of H\textsc{II} regions, even in galaxies that would appear otherwise unremarkable according to traditional diagnostics.

\section{Summary and outlook}
\label{sec:summary}

We have presented deep JWST/NIRSpec spectroscopy in G140M/F100LP, G235M/F170LP, and G235H/F170LP of the galaxy \source at $z=2.224$, targeted as part of the MARTA-GO JWST programme (Fig.~\ref{fig:cutout} and Fig.~\ref{fig:plot_spectra}). 
We report one of the highest-redshift detections to date of the Wolf–Rayet (WR) blue and red bumps in a non-lensed galaxy (Fig.~\ref{fig:WR_features}), and study the interplay between the presence of such massive stars, the elemental abundance patterns (derived via the $T_{\rm e}$-method thanks to the detection of multiple auroral lines), dust properties, and feedback in this system at Cosmic Noon.
Our main results can be summarized as follows.

\begin{itemize}
    \item The broad \HeIIL[4686] feature in the blue bump is well reproduced by a population of WN stars associated with a recent ($\sim$5--6 Myr) burst of star formation, whereas both stellar population synthesis models and empirical templates fall short in matching the observed strength of nitrogen stellar wind lines at $\approx 4640\AA$, suggesting additional N-enhancement in the stellar atmospheres that is not fully accounted for by models (Fig.~\ref{fig:WR_bumps}, see also Fig.~\ref{fig:alternative_WR_fits}).

    \item The relative strength of the blue and red bumps disfavor the presence of VMS in this system (which are generally characterized by much higher EW[\HeIIL[4686]] than seen in \sourceshort, Fig.~\ref{fig:VMS_vs_WR}), although a more definitive assessment will likely require joint analysis of rest-UV and rest-optical features.
     
    \item We do not find any evidence for strong deviations (e.g., N/O-enhancement) in most elemental abundance patterns compared to the global trends inferred for star-forming galaxies of similar metallicity in the local Universe, including $z\sim0$ starbursts with detected WR features (Fig.~\ref{fig:abundance_patterns}). This suggests that any ISM enrichment from WR-driven stellar winds is highly localized and confined to their immediate surroundings, having negligible impact on the global ISM enrichment properties inferred from the integrated galaxy spectrum.
    
    \item However, the gas-phase Fe/O ratio appears enhanced relative to local galaxies of comparable metallicity (Fig.~\ref{fig:Fe_O}). We interpret this as evidence of reduced Fe depletion, possibly linked to dust grain destruction in WR-driven environments.

    \item After correcting our measurement of (Fe/O)$_{\text{gas}}$ (one of the first \Te-based at high redshift) to (Fe/O)$_{\text{total}}$, 
    \sourceshort agrees well with the [O/Fe]–sSFR relationship for galaxies (Fig.~\ref{fig:FeO_sSFR}), formally matching the expected relationship for relatively short ($\sim40$~Myr) SN Ia minimum delay times.   
    This demonstrates the potential of joint O and Fe abundance measurements in constraining star-formation histories and SN Ia delay-time distributions at high redshift.    
    
    \item The detection of a broad H$\alpha$ component in the G235H spectrum, blueshifted by $\sim 70$ km s$^{-1}$ (Fig.~\ref{fig:g235h_fit}), suggests the presence of an ionized outflow with a mass loading factor $\eta\sim0.2$. Although consistent with previous studies of star-formation-driven winds in $z\sim2$ galaxies (and significantly larger than in local dwarfs), such an outflow is likely insufficient to expel the ISM from the galaxy on global scales. At the same time, we note that the total multiphase outflow rate may be larger, since molecular and hot gas phases are not probed by our data.
    
    \item We report the detection of O\,{\sc i}$\lambda8446$ emission at $\sim$8$\sigma$ significance, among the first such detections in a high-redshift star-forming galaxy. Although recombination origin can be excluded, both Ly$\beta$ fluorescence and collisional excitation in dense neutral clumps (similar to Weigelt blobs observed in $\eta$ Carinae) remain plausible (while not mutually exclusive) excitation channels, with additional contributions from localized shocks also possible.
    
\end{itemize}

Looking forward, future works on either this source and, more broadly, on galaxy populations at Cosmic Noon might benefit from the following.

\begin{itemize}
    \item Expanding the sample of galaxies with deep, medium-high resolution JWST/NIRspec spectroscopy, to search for signatures of massive stars, perform more robust (direct) Fe abundance measurements, and establish whether the enhanced (Fe/O)$_{\text{gas}}$ observed in \sourceshort is a common property of WR hosts at high redshift.
    \item High-resolution, deep rest-UV spectroscopy, which can provide additional constraints on the nature of the massive stellar populations (WR vs.\ VMS) and on localized chemical enrichment as probed by ionic species of different ionization.
    \item Probing multi-phase outflow and dust properties from ALMA observations, to provide a full census of the interplay between metal enrichment, dust, and feedback.
    \item Spatially resolved observations, providing a more robust physical connection between abundance patterns, dust properties, O\,{\sc i} emission, and the spatial regions associated with massive stars (or, possibly, with intermediate mass-black holes, see Appendix~\ref{sec:appendix_D}).
\end{itemize}

Taken together, \sourceshort highlights the unique capability of deep JWST/NIRSpec spectroscopy in unveiling the interplay between star formation and ISM properties in galaxies at Cosmic Noon. 
The opportunity to measure Te-based abundances of multiple elements, constrain the dust properties, and detect signatures of massive stars in "normal" star-forming galaxies provides a strong benchmark for testing stellar population models, chemical enrichment pathways, and feedback mechanisms in young, actively star-forming systems, shedding unprecedented light onto the life cycle of dust and metals during the epoch of peak cosmic star-formation activity. 

\begin{acknowledgements}
MC acknowledges support from ESO via the ESO Fellowship Europe.
FB and FM acknowledge support from the INAF Fundamental Astrophysics programme 2022 and 2023. FM and AM acknowledge support from the European Union with the Next Generation EU plan, Mission 4, through PRIN-MUR project ``PROMETEUS'' (202223XPZM), CUP C53D2300080-006.
AM acknowledges INAF funding through the ``Ricerca Fondamentale 2023'' program (mini-grant 1.05.23.04.01). 
WMB acknowledges support from DARK via the DARK Fellowship.
GC acknowledges financial support from INAF under the Large Grant 2022 ``The metal circle: a new sharp view of the baryon cycle up to Cosmic Dawn with the latest generation IFU facilities''. 
A.F. acknowledges the support from project ``VLT- MOONS'' CRAM 1.05.03.07.
FC acknowledges support from a UKRI Frontier Research Guarantee Grant (PI Cullen; grant reference EP/X021025/1). 
CK acknowledges funding from the UK Science and Technology Facility Council through grant ST/Y001443/1.
This work is based on observations made with the NASA/ESA/CSA James Webb Space Telescope. The data were obtained from the Mikulski Archive for Space Telescopes at the Space Telescope Science Institute, which is operated by the Association of Universities for Research in Astronomy, Inc., under NASA contract NAS 5-03127 for JWST. These observations are associated with program JWST 1879.\\ 
\end{acknowledgements}

\bibliography{biblio_merged_final}{}

@article{Aadland2022,
 author = {Aadland, E. and Massey, P. and Hillier, D. J. and Morrell, N.},
 doi = {10.3847/1538-4357/ac3426},
 journal = {ApJ},
 number = {2},
 pages = {44},
 title = {The Physical Parameters of Four WC-type Wolf–Rayet Stars in the Large Magellanic Cloud: Evidence of Evolution},
 volume = {924},
 year = {2022}
}

@article{Abbott_WR_review_1987,
 adsurl = {https://ui.adsabs.harvard.edu/abs/1987ARA&A..25..113A},
 author = {{Abbott}, David C. and {Conti}, Peter S.},
 doi = {10.1146/annurev.aa.25.090187.000553},
 journal = {\araa},
 month = {January},
 pages = {113-150},
 title = {{Wolf-rayet stars.}},
 volume = {25},
 year = {1987}
}

@article{Adamo_clusters_high-z_Nat_2024,
 adsurl = {https://ui.adsabs.harvard.edu/abs/2024Natur.632..513A},
 archiveprefix = {arXiv},
 author = {{Adamo}, Angela and {Bradley}, Larry D. and {Vanzella}, Eros and {Claeyssens}, Ad{\'e}la{\"\i}de and {Welch}, Brian and {Diego}, Jose M. and {Mahler}, Guillaume and {Oguri}, Masamune and {Sharon}, Keren and {Abdurro'uf} and {Hsiao}, Tiger Yu-Yang and {Xu}, Xinfeng and {Messa}, Matteo and {Lassen}, Augusto E. and {Zackrisson}, Erik and {Brammer}, Gabriel and {Coe}, Dan and {Kokorev}, Vasily and {Ricotti}, Massimo and {Zitrin}, Adi and {Fujimoto}, Seiji and {Inoue}, Akio K. and {Resseguier}, Tom and {Rigby}, Jane R. and {Jim{\'e}nez-Teja}, Yolanda and {Windhorst}, Rogier A. and {Hashimoto}, Takuya and {Tamura}, Yoichi},
 doi = {10.1038/s41586-024-07703-7},
 eprint = {2401.03224},
 journal = {\nat},
 month = {August},
 number = {8025},
 pages = {513-516},
 primaryclass = {astro-ph.GA},
 title = {{Bound star clusters observed in a lensed galaxy 460 Myr after the Big Bang}},
 volume = {632},
 year = {2024}
}

@article{allen_mappings_2008,
 author = {Allen, M. G. and Groves, B. A. and Dopita, M. A. and Sutherland, R. S. and Kewley, L. J.},
 doi = {10.1086/589652},
 journal = {\apjs},
 month = {September},
 pages = {20--55},
 title = {The {MAPPINGS} {III} {Library} of {Fast} {Radiative} {Shock} {Models}},
 volume = {178},
 year = {2008}
}

@article{Amarsi_abundances_MW_2019,
 adsurl = {https://ui.adsabs.harvard.edu/abs/2019A&A...630A.104A},
 archiveprefix = {arXiv},
 author = {{Amarsi}, A.~M. and {Nissen}, P.~E. and {Sk{\'u}lad{\'o}ttir}, {\'A}.},
 doi = {10.1051/0004-6361/201936265},
 eid = {A104},
 eprint = {1908.10319},
 journal = {\aap},
 month = {October},
 pages = {A104},
 primaryclass = {astro-ph.SR},
 title = {{Carbon, oxygen, and iron abundances in disk and halo stars. Implications of 3D non-LTE spectral line formation}},
 volume = {630},
 year = {2019}
}

@article{amayo_ICFs_2020,
 adsurl = {https://ui.adsabs.harvard.edu/abs/2021MNRAS.505.2361A},
 archiveprefix = {arXiv},
 author = {{Amayo}, A. and {Delgado-Inglada}, G. and {Stasi{\'n}ska}, G.},
 doi = {10.1093/mnras/stab1467},
 eprint = {2105.08891},
 journal = {\mnras},
 month = {August},
 number = {2},
 pages = {2361-2376},
 primaryclass = {astro-ph.GA},
 title = {{Ionization correction factors and dust depletion patterns in giant H II regions}},
 volume = {505},
 year = {2021}
}

@article{Anderson2019_AA_56Ni,
 author = {Anderson, J. P.},
 doi = {10.1051/0004-6361/201935027},
 eprint = {arXiv:1906.00761},
 journal = {A\&A},
 pages = {A7},
 title = {A meta-analysis of core-collapse supernova $^{56}$Ni masses},
 volume = {628},
 year = {2019}
}

@article{Arellano-Cordova_CNO_EXCELS_2024,
 adsurl = {https://ui.adsabs.harvard.edu/abs/2025MNRAS.540.2991A},
 archiveprefix = {arXiv},
 author = {{Arellano-C{\'o}rdova}, K.~Z. and {Cullen}, F. and {Carnall}, A.~C. and {Scholte}, D. and {Stanton}, T.~M. and {Kobayashi}, C. and {Martinez}, Z. and {Berg}, D.~A. and {Barrufet}, L. and {Begley}, R. and {Donnan}, C.~T. and {Dunlop}, J.~S. and {Hamadouche}, M.~L. and {McLeod}, D.~J. and {McLure}, R.~J. and {Rowlands}, K. and {Shapley}, A.~E.},
 doi = {10.1093/mnras/staf855},
 eid = {arXiv:2412.10557},
 eprint = {2412.10557},
 journal = {\mnras},
 month = {July},
 number = {4},
 pages = {2991-3007},
 primaryclass = {astro-ph.GA},
 title = {{The JWST EXCELS survey: direct estimates of C, N, and O abundances in two relatively metal-rich galaxies at z ≃ 5}},
 volume = {540},
 year = {2025}
}

@article{arellano-corodva_2023,
 adsurl = {https://ui.adsabs.harvard.edu/abs/2022ApJ...940L..23A},
 archiveprefix = {arXiv},
 author = {{Arellano-C{\'o}rdova}, Karla Z. and {Berg}, Danielle A. and {Chisholm}, John and {Arrabal Haro}, Pablo and {Dickinson}, Mark and {Finkelstein}, Steven L. and {Leclercq}, Floriane and {Rogers}, Noah S.~J. and {Simons}, Raymond C. and {Skillman}, Evan D. and {Trump}, Jonathan R. and {Kartaltepe}, Jeyhan S.},
 doi = {10.3847/2041-8213/ac9ab2},
 eid = {L23},
 eprint = {2208.02562},
 journal = {\apjl},
 month = {November},
 number = {1},
 pages = {L23},
 primaryclass = {astro-ph.GA},
 title = {{A First Look at the Abundance Pattern-O/H, C/O, and Ne/O-in z > 7 Galaxies with JWST/NIRSpec}},
 volume = {940},
 year = {2022}
}

@article{Arnaboldi_M31_PNe_2022,
 adsurl = {https://ui.adsabs.harvard.edu/abs/2022A&A...666A.109A},
 archiveprefix = {arXiv},
 author = {{Arnaboldi}, Magda and {Bhattacharya}, Souradeep and {Gerhard}, Ortwin and {Kobayashi}, Chiaki and {Freeman}, Kenneth C. and {Caldwell}, Nelson and {Hartke}, Johanna and {McConnachie}, Alan and {Guhathakurta}, Puragra},
 doi = {10.1051/0004-6361/202244258},
 eid = {A109},
 eprint = {2208.02328},
 journal = {\aap},
 month = {October},
 pages = {A109},
 primaryclass = {astro-ph.GA},
 title = {{The survey of planetary nebulae in Andromeda (M31). V. Chemical enrichment of the thin and thicker discs of Andromeda: Oxygen to argon abundance ratios for planetary nebulae and HII regions}},
 volume = {666},
 year = {2022}
}

@article{Asplund_solar_2021,
 adsurl = {https://ui.adsabs.harvard.edu/abs/2021A&A...653A.141A},
 archiveprefix = {arXiv},
 author = {{Asplund}, M. and {Amarsi}, A.~M. and {Grevesse}, N.},
 doi = {10.1051/0004-6361/202140445},
 eid = {A141},
 eprint = {2105.01661},
 journal = {\aap},
 month = {September},
 pages = {A141},
 primaryclass = {astro-ph.SR},
 title = {{The chemical make-up of the Sun: A 2020 vision}},
 volume = {653},
 year = {2021}
}

@article{Aver_He_2010,
 adsurl = {https://ui.adsabs.harvard.edu/abs/2010JCAP...05..003A},
 archiveprefix = {arXiv},
 author = {{Aver}, Erik and {Olive}, Keith A. and {Skillman}, Evan D.},
 doi = {10.1088/1475-7516/2010/05/003},
 eid = {003},
 eprint = {1001.5218},
 journal = {\jcap},
 month = {May},
 number = {5},
 pages = {003},
 primaryclass = {astro-ph.CO},
 title = {{A new approach to systematic uncertainties and self-consistency in helium abundance determinations}},
 volume = {2010},
 year = {2010}
}

@article{Aver_He_mcmc_2011,
 adsurl = {https://ui.adsabs.harvard.edu/abs/2011JCAP...03..043A},
 archiveprefix = {arXiv},
 author = {{Aver}, Erik and {Olive}, Keith A. and {Skillman}, Evan D.},
 doi = {10.1088/1475-7516/2011/03/043},
 eid = {043},
 eprint = {1012.2385},
 journal = {\jcap},
 month = {March},
 number = {3},
 pages = {043},
 primaryclass = {astro-ph.CO},
 title = {{Mapping systematic errors in helium abundance determinations using Markov Chain Monte Carlo}},
 volume = {2011},
 year = {2011}
}

@article{Aver_Helium_LeoP_2021,
 adsurl = {https://ui.adsabs.harvard.edu/abs/2021JCAP...03..027A},
 archiveprefix = {arXiv},
 author = {{Aver}, Erik and {Berg}, Danielle A. and {Olive}, Keith A. and {Pogge}, Richard W. and {Salzer}, John J. and {Skillman}, Evan D.},
 doi = {10.1088/1475-7516/2021/03/027},
 eid = {027},
 eprint = {2010.04180},
 journal = {\jcap},
 month = {March},
 number = {3},
 pages = {027},
 primaryclass = {astro-ph.CO},
 title = {{Improving helium abundance determinations with Leo P as a case study}},
 volume = {2021},
 year = {2021}
}

@article{berg_aurora_helium_2025,
 adsurl = {https://ui.adsabs.harvard.edu/abs/2026ApJ...996...68B},
 archiveprefix = {arXiv},
 author = {{Berg}, Danielle A. and {Sanders}, Ryan L. and {Shapley}, Alice E. and {Topping}, Michael W. and {Reddy}, Naveen A. and {Skillman}, Evan D. and {Aver}, Erik and {Cullen}, Fergus and {Donnan}, Callum T. and {Dunlop}, James S. and {Jones}, Tucker and {Khostovan}, Ali Ahmad and {McLeod}, Derek J. and {Narayanan}, Desika and {Oesch}, Pascal A. and {Pahl}, Anthony J. and {Pettini}, Max and {Schreiber}, N.~M. F{\"o}rster and {Stark}, Daniel P.},
 doi = {10.3847/1538-4357/ae18db},
 eid = {68},
 eprint = {2507.17057},
 journal = {\apj},
 month = {January},
 number = {1},
 pages = {68},
 primaryclass = {astro-ph.GA},
 title = {{The AURORA Survey: Robust Helium Abundances at High Redshift Reveal a Subpopulation of Helium-enhanced Galaxies in the Early Universe}},
 volume = {996},
 year = {2026}
}

@article{berg_carbon_2016,
 author = {Berg, D. A. and Skillman, E. D. and Henry, R. B. C. and Erb, D. K. and Carigi, L.},
 doi = {10.3847/0004-637X/827/2/126},
 journal = {\apj},
 month = {August},
 pages = {126},
 title = {Carbon and {Oxygen} {Abundances} in {Low} {Metallicity} {Dwarf} {Galaxies}},
 volume = {827},
 year = {2016}
}

@article{berg_chaos_2020,
 author = {Berg, Danielle A. and Pogge, Richard W. and Skillman, Evan D. and Croxall, Kevin V. and Moustakas, John and Rogers, Noah S. J. and Sun, Jiayi},
 doi = {10.3847/1538-4357/ab7eab},
 issn = {1538-4357},
 journal = {The Astrophysical Journal},
 language = {en},
 month = {April},
 number = {2},
 pages = {96},
 shorttitle = {{CHAOS} {IV}},
 title = {{CHAOS} {IV}: {Gas}-phase {Abundance} {Trends} from the {First} {Four} {CHAOS} {Galaxies}},
 url = {https://iopscience.iop.org/article/10.3847/1538-4357/ab7eab},
 urldate = {2020-06-05},
 volume = {893},
 year = {2020}
}

@article{Berg_WR_M101_2024,
 adsurl = {https://ui.adsabs.harvard.edu/abs/2024ApJ...971...87B},
 archiveprefix = {arXiv},
 author = {{Berg}, Danielle A. and {Skillman}, Evan D. and {Chisholm}, John and {Pogge}, Richard W. and {Gazagnes}, Simon and {Rogers}, Noah S.~J. and {Erb}, Dawn K. and {Arellano-C{\'o}rdova}, Karla Z. and {Leitherer}, Claus and {Appel}, Jackie and {Moustakas}, John},
 doi = {10.3847/1538-4357/ad5292},
 eid = {87},
 eprint = {2405.19477},
 journal = {\apj},
 month = {August},
 number = {1},
 pages = {87},
 primaryclass = {astro-ph.GA},
 title = {{CHAOS. VIII. Far-ultraviolet Spectra of M101 and the Impact of Wolf{\textendash}Rayet Stars}},
 volume = {971},
 year = {2024}
}

@article{Bhattacharya_argon_2025,
 adsurl = {https://ui.adsabs.harvard.edu/abs/2025ApJ...983L..30B},
 archiveprefix = {arXiv},
 author = {{Bhattacharya}, Souradeep and {Arnaboldi}, Magda and {Gerhard}, Ortwin and {Kobayashi}, Chiaki and {Saha}, Kanak},
 doi = {10.3847/2041-8213/adc735},
 eid = {L30},
 eprint = {2408.13396},
 journal = {\apjl},
 month = {April},
 number = {2},
 pages = {L30},
 primaryclass = {astro-ph.GA},
 title = {{Unveiling Galaxy Chemical Enrichment Mechanisms Out to z {\ensuremath{\sim}} 8 from Direct Determination of O and Ar Abundances from JWST/NIRSPEC Spectroscopy}},
 volume = {983},
 year = {2025}
}

@article{bian_ldquodirectrdquo_2018,
 abstract = {We study the direct gas-phase oxygen abundance using the well-detected auroral line [O III]lambda4363 in the stacked spectra of a sample of local analogs of high-redshift galaxies. These local analogs share the same location as z 2 star-forming galaxies on the [O III]lambda5007/Hbeta versus [N II]lambda6584/Halpha Baldwin–Phillips–Terlevich diagram. This type of analog has the same ionized interstellar medium (ISM) properties as high-redshift galaxies. We establish empirical metallicity calibrations between the direct gas-phase oxygen abundances (7.8{\textless} 12+log(O/H){\textless} 8.4) and the N2 (log([N II]lambda6584/Halpha))/O3N2 (log(([O III]lambda5007/Hbeta)/([N II]lambda6584/Halpha))) indices in our local analogs. We find significant systematic offsets between the metallicity calibrations for our local analogs of high-redshift galaxies and those derived from the local H II regions and a sample of local reference galaxies selected from the Sloan Digital Sky Survey (SDSS). The N2 and O3N2 metallicities will be underestimated by 0.05–0.1 dex relative to our calibration, if one simply applies the local metallicity calibration in previous studies to high-redshift galaxies. Local metallicity calibrations also cause discrepancies of metallicity measurements in high-redshift galaxies using the N2 and O3N2 indicators. In contrast, our new calibrations produce consistent metallicities between these two indicators. We also derive metallicity calibrations for R23 (log(([O III]lambdalambda4959,5007+[O II]lambdalambda3726,3729)/Hbeta)), O32(log([O III]lambdalambda4959,5007/[O II]lambdalambda3726,3729)), log([O III]lambda5007/Hbeta), and log([Ne III]lambda3869/[O II]lambda3727) indices in our local analogs, which show significant offset compared to those in the SDSS reference galaxies. By comparing with MAPPINGS photoionization models, the different empirical metallicity calibration relations in the local analogs and the SDSS reference galaxies can be shown to be primarily due to the change of ionized ISM conditions. Assuming that temperature structure variations are minimal and ISM conditions do not change dramatically from z 2 to z 5, these empirical calibrations can be used to measure relative metallicities in galaxies with redshifts up to z 5.0 in ground-based observations.},
 author = {Bian, F. and Kewley, L. J. and Dopita, M. A.},
 doi = {10.3847/1538-4357/aabd74},
 journal = {\apj},
 keywords = {galaxies: high-redshift, galaxies: ISM, galaxies: abundances},
 month = {June},
 pages = {175},
 title = {\{ldquoDirect}\rdquo {Gas}-phase {Metallicity} in {Local} {Analogs} of {High}-redshift {Galaxies}: {Empirical} {Metallicity} {Calibrations} for {High}-redshift {Star}-forming {Galaxies}}

@article{brinchmann_bpt_highz_2008,
 adsurl = {https://ui.adsabs.harvard.edu/abs/2008MNRAS.385..769B},
 archiveprefix = {arXiv},
 author = {{Brinchmann}, Jarle and {Pettini}, Max and {Charlot}, St{\'e}phane},
 doi = {10.1111/j.1365-2966.2008.12914.x},
 eprint = {0801.1678},
 journal = {\mnras},
 month = {April},
 number = {2},
 pages = {769-782},
 primaryclass = {astro-ph},
 title = {{New insights into the stellar content and physical conditions of star-forming galaxies at z = 2-3 from spectral modelling}},
 volume = {385},
 year = {2008}
}

@article{Brinchmann_WR_2008,
 adsurl = {https://ui.adsabs.harvard.edu/abs/2008A&A...485..657B},
 archiveprefix = {arXiv},
 author = {{Brinchmann}, J. and {Kunth}, D. and {Durret}, F.},
 doi = {10.1051/0004-6361:200809783},
 eprint = {0805.1073},
 journal = {\aap},
 month = {July},
 number = {3},
 pages = {657-677},
 primaryclass = {astro-ph},
 title = {{Galaxies with Wolf-Rayet signatures in the low-redshift Universe. A survey using the Sloan Digital Sky Survey}},
 volume = {485},
 year = {2008}
}

@article{Byrne_BPASS_2023,
 adsurl = {https://ui.adsabs.harvard.edu/abs/2023MNRAS.521.4995B},
 archiveprefix = {arXiv},
 author = {{Byrne}, C.~M. and {Stanway}, E.~R.},
 doi = {10.1093/mnras/stad832},
 eprint = {2303.16920},
 journal = {\mnras},
 month = {June},
 number = {4},
 pages = {4995-5012},
 primaryclass = {astro-ph.GA},
 title = {{On the impact of spectral template uncertainties in synthetic stellar populations}},
 volume = {521},
 year = {2023}
}

@article{calabro_UV_Slope_vs_metallicity_2021,
 adsurl = {https://ui.adsabs.harvard.edu/abs/2021A&A...646A..39C},
 archiveprefix = {arXiv},
 author = {{Calabr{\`o}}, A. and {Castellano}, M. and {Pentericci}, L. and {Fontanot}, F. and {Menci}, N. and {Cullen}, F. and {McLure}, R. and {Bolzonella}, M. and {Cimatti}, A. and {Marchi}, F. and {Talia}, M. and {Amor{\'\i}n}, R. and {Cresci}, G. and {De Lucia}, G. and {Fynbo}, J. and {Fontana}, A. and {Franco}, M. and {Hathi}, N.~P. and {Hibon}, P. and {Hirschmann}, M. and {Mannucci}, F. and {Santini}, P. and {Saxena}, A. and {Schaerer}, D. and {Xie}, L. and {Zamorani}, G.},
 doi = {10.1051/0004-6361/202039244},
 eid = {A39},
 eprint = {2011.06615},
 journal = {\aap},
 month = {February},
 pages = {A39},
 primaryclass = {astro-ph.GA},
 title = {{The VANDELS survey: The relation between the UV continuum slope and stellar metallicity in star-forming galaxies at z {\ensuremath{\sim}} 3}},
 volume = {646},
 year = {2021}
}

@article{cameron_gnz11_2023,
 adsurl = {https://ui.adsabs.harvard.edu/abs/2023MNRAS.523.3516C},
 archiveprefix = {arXiv},
 author = {{Cameron}, Alex J. and {Katz}, Harley and {Rey}, Martin P. and {Saxena}, Aayush},
 doi = {10.1093/mnras/stad1579},
 eprint = {2302.10142},
 journal = {\mnras},
 month = {August},
 number = {3},
 pages = {3516-3525},
 primaryclass = {astro-ph.GA},
 title = {{Nitrogen enhancements 440 Myr after the big bang: supersolar N/O, a tidal disruption event, or a dense stellar cluster in GN-z11?}},
 volume = {523},
 year = {2023}
}

@article{cappellari_improving_2017,
 author = {Cappellari, M.},
 doi = {10.1093/mnras/stw3020},
 journal = {\mnras},
 month = {April},
 pages = {798--811},
 title = {Improving the full spectrum fitting method: accurate convolution with {Gauss}-{Hermite} functions},
 volume = {466},
 year = {2017}
}

@article{cappellari_parametric_2004,
 author = {Cappellari, M. and Emsellem, E.},
 doi = {10.1086/381875},
 journal = {\pasp},
 month = {February},
 pages = {138--147},
 title = {Parametric {Recovery} of {Line}-of-{Sight} {Velocity} {Distributions} from {Absorption}-{Line} {Spectra} of {Galaxies} via {Penalized} {Likelihood}},
 volume = {116},
 year = {2004}
}

@article{cappellari_ppxf_2022,
 adsurl = {https://ui.adsabs.harvard.edu/abs/2023MNRAS.526.3273C},
 archiveprefix = {arXiv},
 author = {{Cappellari}, Michele},
 doi = {10.1093/mnras/stad2597},
 eprint = {2208.14974},
 journal = {\mnras},
 month = {December},
 number = {3},
 pages = {3273-3300},
 primaryclass = {astro-ph.GA},
 title = {{Full spectrum fitting with photometry in PPXF: stellar population versus dynamical masses, non-parametric star formation history and metallicity for 3200 LEGA-C galaxies at redshift z {\ensuremath{\approx}} 0.8}},
 volume = {526},
 year = {2023}
}

@article{cardelli_relationship_1989,
 author = {Cardelli, J. A. and Clayton, G. C. and Mathis, J. S.},
 doi = {10.1086/167900},
 journal = {\apj},
 month = {October},
 pages = {245--256},
 title = {The relationship between infrared, optical, and ultraviolet extinction},
 volume = {345},
 year = {1989}
}

@article{Carnall_EXCELS_quiescent_2024,
 adsurl = {https://ui.adsabs.harvard.edu/abs/2024MNRAS.534..325C},
 archiveprefix = {arXiv},
 author = {{Carnall}, A.~C. and {Cullen}, F. and {McLure}, R.~J. and {McLeod}, D.~J. and {Begley}, R. and {Donnan}, C.~T. and {Dunlop}, J.~S. and {Shapley}, A.~E. and {Rowlands}, K. and {Almaini}, O. and {Arellano-C{\'o}rdova}, K.~Z. and {Barrufet}, L. and {Cimatti}, A. and {Ellis}, R.~S. and {Grogin}, N.~A. and {Hamadouche}, M.~L. and {Illingworth}, G.~D. and {Koekemoer}, A.~M. and {Leung}, H. -H. and {Lovell}, C.~C. and {P{\'e}rez-Gonz{\'a}lez}, P.~G. and {Santini}, P. and {Stanton}, T.~M. and {Wild}, V.},
 doi = {10.1093/mnras/stae2092},
 eprint = {2405.02242},
 journal = {\mnras},
 month = {October},
 number = {1},
 pages = {325-348},
 primaryclass = {astro-ph.GA},
 title = {{The JWST EXCELS survey: too much, too young, too fast? Ultra-massive quiescent galaxies at 3 < z < 5}},
 volume = {534},
 year = {2024}
}

@article{Cataldi_MARTA_2025,
 adsurl = {https://ui.adsabs.harvard.edu/abs/2025A&A...703A.208C},
 archiveprefix = {arXiv},
 author = {{Cataldi}, E. and {Belfiore}, F. and {Curti}, M. and {Moreschini}, B. and {Mannucci}, F. and {D'Amato}, Q. and {Cresci}, G. and {Feltre}, A. and {Ginolfi}, M. and {Marconi}, A. and {Amiri}, A. and {Arnaboldi}, M. and {Bertola}, E. and {Bracci}, C. and {Carniani}, S. and {Ceci}, M. and {Chakraborty}, A. and {Cirasuolo}, M. and {Cullen}, F. and {Kobayashi}, C. and {Kumari}, N. and {Maiolino}, R. and {Marconcini}, C. and {Scialpi}, M. and {Ulivi}, L.},
 doi = {10.1051/0004-6361/202554843},
 eid = {A208},
 eprint = {2504.03839},
 journal = {\aap},
 month = {November},
 pages = {A208},
 primaryclass = {astro-ph.GA},
 title = {{MARTA: Temperature-temperature relationships and strong-line metallicity calibrations from multiple auroral-line detections at cosmic noon}},
 volume = {703},
 year = {2025}
}

@article{Cataldi_NO_2025,
 adsurl = {https://ui.adsabs.harvard.edu/abs/2025arXiv251207955C},
 archiveprefix = {arXiv},
 author = {{Cataldi}, E. and {Belfiore}, F. and {Curti}, M. and {Moreschini}, B. and {Marconi}, A. and {Maiolino}, R. and {Feltre}, A. and {Ginolfi}, M. and {Mannucci}, F. and {Cresci}, G. and {Ji}, X. and {Amiri}, A. and {Arnaboldi}, M. and {Bertola}, E. and {Bracci}, C. and {Ceci}, M. and {Chakraborty}, A. and {Cullen}, F. and {D'Amato}, Q. and {Kobayashi}, C. and {Lamperti}, I. and {Marconcini}, C. and {Scialpi}, M. and {Ulivi}, L. and {Zanchettin}, M.~V.},
 doi = {10.48550/arXiv.2512.07955},
 eid = {arXiv:2512.07955},
 eprint = {2512.07955},
 journal = {arXiv e-prints},
 month = {December},
 pages = {arXiv:2512.07955},
 primaryclass = {astro-ph.GA},
 title = {{Tracing Nitrogen Enrichment across Cosmic Time with JWST}},
 year = {2025}
}

@article{chabrier_galactic_2003,
 author = {Chabrier, G.},
 doi = {10.1086/376392},
 journal = {\pasp},
 month = {July},
 pages = {763--795},
 title = {Galactic {Stellar} and {Substellar} {Initial} {Mass} {Function}},
 volume = {115},
 year = {2003}
}

@article{Chakraborty_strong_lines_2024,
 adsurl = {https://ui.adsabs.harvard.edu/abs/2025ApJ...985...24C},
 archiveprefix = {arXiv},
 author = {{Chakraborty}, Priyanka and {Sarkar}, Arnab and {Smith}, Randall and {Ferland}, Gary J. and {McDonald}, Michael and {Forman}, William and {Vogelsberger}, Mark and {Torrey}, Paul and {Garcia}, Alex M. and {Bautz}, Mark and {Foster}, Adam and {Miller}, Eric and {Grant}, Catherine},
 doi = {10.3847/1538-4357/adc7b5},
 eid = {24},
 eprint = {2412.15435},
 journal = {\apj},
 month = {May},
 number = {1},
 pages = {24},
 primaryclass = {astro-ph.GA},
 title = {{Unveiling the Cosmic Chemistry. II. ``Direct'' T$_{e}$-based Metallicity of Galaxies at 3 < z < 10 with JWST/NIRSpec}},
 volume = {985},
 year = {2025}
}

@article{charbonnel_gnz11_2023,
 adsurl = {https://ui.adsabs.harvard.edu/abs/2023A&A...673L...7C},
 archiveprefix = {arXiv},
 author = {{Charbonnel}, C. and {Schaerer}, D. and {Prantzos}, N. and {Ram{\'\i}rez-Galeano}, L. and {Fragos}, T. and {Kuruvanthodi}, A. and {Marques-Chaves}, R. and {Gieles}, M.},
 doi = {10.1051/0004-6361/202346410},
 eid = {L7},
 eprint = {2303.07955},
 journal = {\aap},
 month = {May},
 pages = {L7},
 primaryclass = {astro-ph.GA},
 title = {{N-enhancement in GN-z11: First evidence for supermassive stars nucleosynthesis in proto-globular clusters-like conditions at high redshift?}},
 volume = {673},
 year = {2023}
}

@article{Choe_SArc_godzilla_2025,
 adsurl = {https://ui.adsabs.harvard.edu/abs/2025A&A...698A..16C},
 archiveprefix = {arXiv},
 author = {{Choe}, S. and {Emil Rivera-Thorsen}, T. and {Dahle}, H. and {Sharon}, K. and {Owens}, M. Riley and {Rigby}, J.~R. and {Bayliss}, M.~B. and {Hayes}, M.~J. and {Hutchison}, T. and {Welch}, B. and {Chisholm}, J. and {Gladders}, M.~D. and {Khullar}, G. and {Kim}, K.},
 doi = {10.1051/0004-6361/202450685},
 eid = {A16},
 eprint = {2405.06953},
 journal = {\aap},
 month = {June},
 pages = {A16},
 primaryclass = {astro-ph.GA},
 title = {{The Sunburst Arc with JWST: II. Observations of an Eta Carinae analog at z = 2.37}},
 volume = {698},
 year = {2025}
}

@article{christensen_gravitationally_2012,
 author = {Christensen, L. and Laursen, P. and Richard, J. and Hjorth, J. and Milvang-Jensen, B. and Dessauges-Zavadsky, M. and Limousin, M. and Grillo, C. and Ebeling, H.},
 doi = {10.1111/j.1365-2966.2012.22007.x},
 journal = {\mnras},
 month = {December},
 pages = {1973--1982},
 title = {Gravitationally lensed galaxies at 2 \$\lt\$ z \$\lt\$ 3.5: direct abundance measurements of {Ly} α emitters},
 volume = {427},
 year = {2012}
}

@article{Chruslinska_Fe_O_2024,
 adsurl = {https://ui.adsabs.harvard.edu/abs/2024A&A...686A.186C},
 archiveprefix = {arXiv},
 author = {{Chru{\'s}li{\'n}ska}, M. and {Pakmor}, R. and {Matthee}, J. and {Matsuno}, T.},
 doi = {10.1051/0004-6361/202347602},
 eid = {A186},
 eprint = {2308.00023},
 journal = {\aap},
 month = {June},
 pages = {A186},
 primaryclass = {astro-ph.GA},
 title = {{Trading oxygen for iron. I. The [O/Fe]-specific star formation rate relation of galaxies}},
 volume = {686},
 year = {2024}
}

@article{cicone_massive_2014,
 author = {Cicone, C. and Maiolino, R. and Sturm, E. and Graciá-Carpio, J. and Feruglio, C. and Neri, R. and Aalto, S. and Davies, R. and Fiore, F. and Fischer, J. and García-Burillo, S. and González-Alfonso, E. and Hailey-Dunsheath, S. and Piconcelli, E. and Veilleux, S.},
 doi = {10.1051/0004-6361/201322464},
 journal = {\aap},
 month = {February},
 pages = {A21},
 title = {Massive molecular outflows and evidence for {AGN} feedback from {CO} observations},
 volume = {562},
 year = {2014}
}

@article{Crowther_WR_2000,
 adsurl = {https://ui.adsabs.harvard.edu/abs/2000A&A...356..191C},
 archiveprefix = {arXiv},
 author = {{Crowther}, P.~A.},
 doi = {10.48550/arXiv.astro-ph/0001226},
 eprint = {astro-ph/0001226},
 journal = {\aap},
 month = {April},
 pages = {191-199},
 primaryclass = {astro-ph},
 title = {{Wind properties of Wolf-Rayet stars at low metallicity: Sk 41 (SMC)}},
 volume = {356},
 year = {2000}
}

@article{Crowther_WR_LMC_2023,
 adsurl = {https://ui.adsabs.harvard.edu/abs/2023MNRAS.521..585C},
 archiveprefix = {arXiv},
 author = {{Crowther}, Paul A. and {Rate}, G. and {Bestenlehner}, Joachim M.},
 doi = {10.1093/mnras/stad418},
 eprint = {2301.11297},
 journal = {\mnras},
 month = {May},
 number = {1},
 pages = {585-612},
 primaryclass = {astro-ph.SR},
 title = {{Line luminosities of Galactic and Magellanic Cloud Wolf-Rayet stars}},
 volume = {521},
 year = {2023}
}

@article{Crowther_WR_review_2007,
 adsurl = {https://ui.adsabs.harvard.edu/abs/2007ARA&A..45..177C},
 archiveprefix = {arXiv},
 author = {{Crowther}, Paul A.},
 doi = {10.1146/annurev.astro.45.051806.110615},
 eprint = {astro-ph/0610356},
 journal = {\araa},
 month = {September},
 number = {1},
 pages = {177-219},
 primaryclass = {astro-ph},
 title = {{Physical Properties of Wolf-Rayet Stars}},
 volume = {45},
 year = {2007}
}

@article{Cullen_NIRVANDELS_alpha_enhance_2021,
 adsurl = {https://ui.adsabs.harvard.edu/abs/2021MNRAS.505..903C},
 archiveprefix = {arXiv},
 author = {{Cullen}, F. and {Shapley}, A.~E. and {McLure}, R.~J. and {Dunlop}, J.~S. and {Sanders}, R.~L. and {Topping}, M.~W. and {Reddy}, N.~A. and {Amor{\'\i}n}, R. and {Begley}, R. and {Bolzonella}, M. and {Calabr{\`o}}, A. and {Carnall}, A.~C. and {Castellano}, M. and {Cimatti}, A. and {Cirasuolo}, M. and {Cresci}, G. and {Fontana}, A. and {Fontanot}, F. and {Garilli}, B. and {Guaita}, L. and {Hamadouche}, M. and {Hathi}, N.~P. and {Mannucci}, F. and {McLeod}, D.~J. and {Pentericci}, L. and {Saxena}, A. and {Talia}, M. and {Zamorani}, G.},
 doi = {10.1093/mnras/stab1340},
 eprint = {2103.06300},
 journal = {\mnras},
 month = {July},
 number = {1},
 pages = {903-920},
 primaryclass = {astro-ph.GA},
 title = {{The NIRVANDELS Survey: a robust detection of {\ensuremath{\alpha}}-enhancement in star-forming galaxies at z ≃ 3.4}},
 volume = {505},
 year = {2021}
}

@article{curti_GSz9_2025,
 adsurl = {https://ui.adsabs.harvard.edu/abs/2025A&A...697A..89C},
 archiveprefix = {arXiv},
 author = {{Curti}, Mirko and {Witstok}, Joris and {Jakobsen}, Peter and {Kobayashi}, Chiaki and {Curtis-Lake}, Emma and {Hainline}, Kevin and {Ji}, Xihan and {D'Eugenio}, Francesco and {Chevallard}, Jacopo and {Maiolino}, Roberto and {Scholtz}, Jan and {Carniani}, Stefano and {Arribas}, Santiago and {Baker}, William M. and {Bhatawdekar}, Rachana and {Boyett}, Kristan and {Bunker}, Andrew J. and {Cameron}, Alex and {Cargile}, Phillip A. and {Charlot}, St{\'e}phane and {Eisenstein}, Daniel J. and {Ji}, Zhiyuan and {Johnson}, Benjamin D. and {Kumari}, Nimisha and {Maseda}, Michael V. and {Robertson}, Brant and {Silcock}, Maddie S. and {Tacchella}, Sandro and {{\"U}bler}, Hannah and {Venturi}, Giacomo and {Williams}, Christina C. and {Willmer}, Christopher N.~A. and {Willott}, Chris},
 doi = {10.1051/0004-6361/202451410},
 eid = {A89},
 eprint = {2407.02575},
 journal = {\aap},
 month = {May},
 pages = {A89},
 primaryclass = {astro-ph.GA},
 title = {{JADES: The star formation and chemical enrichment history of a luminous galaxy at z {\ensuremath{\sim}} 9.43 probed by ultra-deep JWST/NIRSpec spectroscopy}},
 volume = {697},
 year = {2025}
}

@article{curti_new_2017,
 author = {Curti, M. and Cresci, G. and Mannucci, F. and Marconi, A. and Maiolino, R. and Esposito, S.},
 doi = {10.1093/mnras/stw2766},
 journal = {\mnras},
 month = {February},
 pages = {1384--1400},
 title = {New fully empirical calibrations of strong-line metallicity indicators in star-forming galaxies},
 volume = {465},
 year = {2017}
}

@article{curti_smacs_2023,
 adsurl = {https://ui.adsabs.harvard.edu/abs/2023MNRAS.518..425C},
 archiveprefix = {arXiv},
 author = {{Curti}, Mirko and {D'Eugenio}, Francesco and {Carniani}, Stefano and {Maiolino}, Roberto and {Sandles}, Lester and {Witstok}, Joris and {Baker}, William M. and {Bennett}, Jake S. and {Piotrowska}, Joanna M. and {Tacchella}, Sandro and {Charlot}, Stephane and {Nakajima}, Kimihiko and {Maheson}, Gabriel and {Mannucci}, Filippo and {Amiri}, Amirnezam and {Arribas}, Santiago and {Belfiore}, Francesco and {Bonaventura}, Nina R. and {Bunker}, Andrew J. and {Chevallard}, Jacopo and {Cresci}, Giovanni and {Curtis-Lake}, Emma and {Hayden-Pawson}, Connor and {Jones}, Gareth C. and {Kumari}, Nimisha and {Laseter}, Isaac and {Looser}, Tobias J. and {Marconi}, Alessandro and {Maseda}, Michael V. and {Scholtz}, Jan and {Smit}, Renske and {{\"U}bler}, Hannah and {Wallace}, Imaan E.~B.},
 doi = {10.1093/mnras/stac2737},
 eprint = {2207.12375},
 journal = {\mnras},
 month = {January},
 number = {1},
 pages = {425-438},
 primaryclass = {astro-ph.GA},
 title = {{The chemical enrichment in the early Universe as probed by JWST via direct metallicity measurements at z   8}},
 volume = {518},
 year = {2023}
}

@article{DavidsonHumphreys_EtaCar_Nat_2012,
 adsurl = {https://ui.adsabs.harvard.edu/abs/2012Natur.486E...1D},
 archiveprefix = {arXiv},
 author = {{Davidson}, Kris and {Humphreys}, Roberta M.},
 doi = {10.1038/nature11166},
 eprint = {1205.2010},
 journal = {\nat},
 month = {June},
 number = {7403},
 pages = {E1},
 primaryclass = {astro-ph.SR},
 title = {{The Great Eruption of {\ensuremath{\eta}} Carinae}},
 volume = {486},
 year = {2012}
}

@article{DeCia_dust_LyA_absorbers_2016,
 adsurl = {https://ui.adsabs.harvard.edu/abs/2016A&A...596A..97D},
 archiveprefix = {arXiv},
 author = {{De Cia}, A. and {Ledoux}, C. and {Mattsson}, L. and {Petitjean}, P. and {Srianand}, R. and {Gavignaud}, I. and {Jenkins}, E.~B.},
 doi = {10.1051/0004-6361/201527895},
 eid = {A97},
 eprint = {1608.08621},
 journal = {\aap},
 month = {December},
 pages = {A97},
 primaryclass = {astro-ph.GA},
 title = {{Dust-depletion sequences in damped Lyman-{\ensuremath{\alpha}} absorbers. A unified picture from low-metallicity systems to the Galaxy}},
 volume = {596},
 year = {2016}
}

@article{DeCia_GRB_2012,
 adsurl = {https://ui.adsabs.harvard.edu/abs/2012A&A...545A..64D},
 archiveprefix = {arXiv},
 author = {{De Cia}, A. and {Ledoux}, C. and {Fox}, A.~J. and {Vreeswijk}, P.~M. and {Smette}, A. and {Petitjean}, P. and {Bj{\"o}rnsson}, G. and {Fynbo}, J.~P.~U. and {Hjorth}, J. and {Jakobsson}, P.},
 doi = {10.1051/0004-6361/201218884},
 eid = {A64},
 eprint = {1207.6102},
 journal = {\aap},
 month = {September},
 pages = {A64},
 primaryclass = {astro-ph.CO},
 title = {{Rapid-response mode VLT/UVES spectroscopy of super iron-rich gas exposed to GRB 080310. Evidence of ionization in action and episodic star formation in the host}},
 volume = {545},
 year = {2012}
}

@misc{Dunlop_Primer_proposal_2021,
 adsurl = {https://ui.adsabs.harvard.edu/abs/2021jwst.prop.1837D},
 author = {{Dunlop}, James S. and {Abraham}, Roberto G. and {Ashby}, Matthew L.~N. and {Bagley}, Micaela and {Best}, Philip N. and {Bongiorno}, Angela and {Bouwens}, Rychard and {Bowler}, Rebecca A.~A. and {Brammer}, Gabriel and {Bremer}, Malcolm and {Calabro'}, Antonello and {Carnall}, Adam and {Castellano}, Marco and {Cirasuolo}, Michele and {Conselice}, Christopher and {Cullen}, Fergus and {Dave}, Romeel and {Dayal}, Pratika and {Dekel}, Avishai and {Dickinson}, Mark and {Duncan}, Kenneth James and {Elbaz}, David and {Ellis}, Richard S. and {Ferguson}, Harry C. and {Ferrara}, Andrea and {Finkelstein}, Steven L. and {Fontana}, Adriano and {Furlanetto}, Steven and {Fynbo}, Johan P.~U. and {Gallerani}, Simona and {Gardner}, Jonathan P. and {Giavalisco}, Mauro and {Grazian}, Andrea and {Grogin}, Norman and {Harikane}, Yuichi and {Hopkins}, Philip F. and {Ilbert}, Olivier and {Illingworth}, Garth D. and {Juneau}, Stephanie and {Jung}, Intae and {Kartaltepe}, Jeyhan and {Kassin}, Susan and {Kauffmann}, Olivier Benjamin and {Khochfar}, Sadegh and {Kirkpatrick}, Allison and {Kocevski}, Dale D. and {Koekemoer}, Anton M. and {Labbe}, Ivo and {Laporte}, Nicolas and {Larson}, Rebecca L. and {Lucas}, Ray A. and {Magee}, Daniel K. and {Mason}, Charlotte and {McCracken}, Henry Joy and {McLeod}, Derek and {McLure}, Ross and {Merlin}, Emiliano and {Mesinger}, Andrei and {Milvang-Jensen}, Bo and {Newman}, Jeffrey Allen and {Oesch}, Pascal and {Ouchi}, Masami and {Pacifici}, Camilla and {Papovich}, Casey and {Peacock}, John and {Peeples}, Molly and {Pentericci}, Laura and {Perez-Gonzalez}, Pablo G. and {Pirzkal}, Norbert and {Pope}, Alexandra and {Pye}, John P. and {Reddy}, Naveen A. and {Robertson}, Brant and {Salvato}, Mara and {Santini}, Paola and {Schaerer}, Daniel and {Shapley}, Alice E. and {Simons}, Raymond and {Smit}, Renske and {Smith}, Britton D. and {Snyder}, Greg and {Somerville}, Rachel S. and {Stanway}, Elizabeth R. and {Stefanon}, Mauro and {Tasca}, Lidia and {Tikkanen}, Tuomo and {Tresse}, Laurence and {Trump}, Jonathan R. and {Whitaker}, Katherine E. and {Wilkins}, Stephen Matthew and {Wright}, Gillian and {Wyithe}, J. Stuart B. and {van Dokkum}, Pieter and {van der Werf}, Paul},
 howpublished = {JWST Proposal. Cycle 1, ID. \#1837},
 month = {March},
 pages = {1837},
 title = {{PRIMER: Public Release IMaging for Extragalactic Research}},
 year = {2021}
}

@article{Eldridge_BPASS_2017,
 adsurl = {https://ui.adsabs.harvard.edu/abs/2017PASA...34...58E},
 archiveprefix = {arXiv},
 author = {{Eldridge}, J.~J. and {Stanway}, E.~R. and {Xiao}, L. and {McClelland}, L.~A.~S. and {Taylor}, G. and {Ng}, M. and {Greis}, S.~M.~L. and {Bray}, J.~C.},
 doi = {10.1017/pasa.2017.51},
 eid = {e058},
 eprint = {1710.02154},
 journal = {\pasa},
 month = {November},
 pages = {e058},
 primaryclass = {astro-ph.SR},
 title = {{Binary Population and Spectral Synthesis Version 2.1: Construction, Observational Verification, and New Results}},
 volume = {34},
 year = {2017}
}

@article{EldridgeStanway2009,
 author = {Eldridge, J. J. and Stanway, E. R.},
 doi = {10.1111/j.1365-2966.2009.15566.x},
 eprint = {arXiv:0908.1386},
 journal = {MNRAS},
 number = {2},
 pages = {1019--1028},
 title = {Spectral synthesis including massive binaries},
 volume = {400},
 year = {2009}
}

@article{EldridgeVink2006,
 author = {Eldridge, J. J. and Vink, Jorick S.},
 doi = {10.1051/0004-6361:20064827},
 eprint = {astro-ph/0603188},
 journal = {A\&A},
 pages = {295–301},
 title = {Implications of the metallicity dependence of Wolf–Rayet winds},
 volume = {452},
 year = {2006}
}

@article{emcee_2013,
 adsurl = {https://ui.adsabs.harvard.edu/abs/2013PASP..125..306F},
 archiveprefix = {arXiv},
 author = {{Foreman-Mackey}, Daniel and {Hogg}, David W. and {Lang}, Dustin and {Goodman}, Jonathan},
 doi = {10.1086/670067},
 eprint = {1202.3665},
 journal = {\pasp},
 month = {March},
 number = {925},
 pages = {306},
 primaryclass = {astro-ph.IM},
 title = {{emcee: The MCMC Hammer}},
 volume = {125},
 year = {2013}
}

@article{erb_h_2006,
 author = {Erb, D. K. and Steidel, C. C. and Shapley, A. E. and Pettini, M. and Reddy, N. A. and Adelberger, K. L.},
 doi = {10.1086/505341},
 journal = {\apj},
 month = {August},
 pages = {128--139},
 title = {Hα {Observations} of a {Large} {Sample} of {Galaxies} at z ∼ 2: {Implications} for {Star} {Formation} in {High}-{Redshift} {Galaxies}},
 volume = {647},
 year = {2006}
}

@article{fiore_agn_2017,
 author = {Fiore, F. and Feruglio, C. and Shankar, F. and Bischetti, M. and Bongiorno, A. and Brusa, M. and Carniani, S. and Cicone, C. and Duras, F. and Lamastra, A. and Mainieri, V. and Marconi, A. and Menci, N. and Maiolino, R. and Piconcelli, E. and Vietri, G. and Zappacosta, L.},
 doi = {10.1051/0004-6361/201629478},
 journal = {\aap},
 month = {May},
 pages = {A143},
 title = {{AGN} wind scaling relations and the co-evolution of black holes and galaxies},
 volume = {601},
 year = {2017}
}

@article{forster_schreiber_kmos3d_2018,
 author = {Förster Schreiber, N. M. and Übler, H. and Davies, R. L. and Genzel, R. and Wisnioski, E. and Belli, S. and Shimizu, T. and Lutz, D. and Fossati, M. and Herrera-Camus, R. and Mendel, J. T. and Tacconi, L. J. and Wilman, D. and Beifiori, A. and Brammer, G. and Burkert, A. and Carollo, C. M. and Davies, R. I. and Eisenhauer, F. and Fabricius, M. and Lilly, S. J. and Momcheva, I. and Naab, T. and Nelson, E. J. and Price, S. and Renzini, A. and Saglia, R. and Sternberg, A. and van Dokkum, P. and Wuyts, S.},
 journal = {arXiv e-prints},
 month = {July},
 title = {The {KMOS}{\textasciicircum}{3D} {Survey}: {Demographics} and {Properties} of {Galactic} {Outflows} at z = 0.6 - 2.7},
 year = {2018}
}

@article{genzel_sins_2011,
 author = {Genzel, R. and Newman, S. and Jones, T. and Förster Schreiber, N. M. and Shapiro, K. and Genel, S. and Lilly, S. J. and Renzini, A. and Tacconi, L. J. and Bouché, N. and Burkert, A. and Cresci, G. and Buschkamp, P. and Carollo, C. M. and Ceverino, D. and Davies, R. and Dekel, A. and Eisenhauer, F. and Hicks, E. and Kurk, J. and Lutz, D. and Mancini, C. and Naab, T. and Peng, Y. and Sternberg, A. and Vergani, D. and Zamorani, G.},
 doi = {10.1088/0004-637X/733/2/101},
 journal = {\apj},
 month = {June},
 pages = {101},
 title = {The {Sins} {Survey} of z ∼ 2 {Galaxy} {Kinematics}: {Properties} of the {Giant} {Star}-forming {Clumps}},
 volume = {733},
 year = {2011}
}

@article{gordon_LMC_attenuation_2003,
 adsurl = {https://ui.adsabs.harvard.edu/abs/2003ApJ...594..279G},
 archiveprefix = {arXiv},
 author = {{Gordon}, Karl D. and {Clayton}, Geoffrey C. and {Misselt}, K.~A. and {Landolt}, Arlo U. and {Wolff}, Michael J.},
 doi = {10.1086/376774},
 eprint = {astro-ph/0305257},
 journal = {\apj},
 month = {September},
 number = {1},
 pages = {279-293},
 primaryclass = {astro-ph},
 title = {{A Quantitative Comparison of the Small Magellanic Cloud, Large Magellanic Cloud, and Milky Way Ultraviolet to Near-Infrared Extinction Curves}},
 volume = {594},
 year = {2003}
}

@article{Gotberg_binaries_2017,
 adsurl = {https://ui.adsabs.harvard.edu/abs/2017A&A...608A..11G},
 archiveprefix = {arXiv},
 author = {{G{\"o}tberg}, Y. and {de Mink}, S.~E. and {Groh}, J.~H.},
 doi = {10.1051/0004-6361/201730472},
 eid = {A11},
 eprint = {1701.07439},
 journal = {\aap},
 month = {November},
 pages = {A11},
 primaryclass = {astro-ph.SR},
 title = {{Ionizing spectra of stars that lose their envelope through interaction with a binary companion: role of metallicity}},
 volume = {608},
 year = {2017}
}

@article{Grandi1980,
 author = {Grandi, S. A.},
 doi = {10.1086/157961},
 journal = {ApJ},
 pages = {10-17},
 title = {OI \lambda8446 emission in Seyfert 1 galaxies},
 volume = {238},
 year = {1980}
}

@article{Greene_AGN_mass_function_2007,
 adsurl = {https://ui.adsabs.harvard.edu/abs/2007ApJ...667..131G},
 archiveprefix = {arXiv},
 author = {{Greene}, Jenny E. and {Ho}, Luis C.},
 doi = {10.1086/520497},
 eprint = {0705.0020},
 journal = {\apj},
 month = {September},
 number = {1},
 pages = {131-148},
 primaryclass = {astro-ph},
 title = {{The Mass Function of Active Black Holes in the Local Universe}},
 volume = {667},
 year = {2007}
}

@article{Greene_Ho_BHmass_Ha_2005,
 adsurl = {https://ui.adsabs.harvard.edu/abs/2005ApJ...630..122G},
 archiveprefix = {arXiv},
 author = {{Greene}, Jenny E. and {Ho}, Luis C.},
 doi = {10.1086/431897},
 eprint = {astro-ph/0508335},
 journal = {\apj},
 month = {September},
 number = {1},
 pages = {122-129},
 primaryclass = {astro-ph},
 title = {{Estimating Black Hole Masses in Active Galaxies Using the H{\ensuremath{\alpha}} Emission Line}},
 volume = {630},
 year = {2005}
}

@article{Gunawardhana_WR_GNz11_2025,
 adsurl = {https://ui.adsabs.harvard.edu/abs/2025MNRAS.543.3172G},
 archiveprefix = {arXiv},
 author = {{Gunawardhana}, M.~L.~P. and {Brinchmann}, J. and {Croom}, S. and {Bunker}, A.~J. and {Bryant}, J. and {Oh}, S.},
 doi = {10.1093/mnras/staf1620},
 eid = {arXiv:2504.12584},
 eprint = {2504.12584},
 journal = {\mnras},
 month = {November},
 number = {4},
 pages = {3172-3195},
 primaryclass = {astro-ph.GA},
 title = {{JADES NIRSpec spectroscopy of GN-z11: evidence for Wolf─Rayet contribution to stellar populations at 430 Myr after big bang?}},
 volume = {543},
 year = {2025}
}

@article{Hainich2014,
 author = {Hainich, R. and R{\"u}hling, U. and Todt, H. and others},
 doi = {10.1051/0004-6361/201322696},
 journal = {A\&A},
 pages = {A27},
 title = {The Wolf–Rayet stars in the Large Magellanic Cloud. A comprehensive analysis of the WN class},
 volume = {565},
 year = {2014}
}

@article{HillierMiller1999,
 author = {Hillier, D. J. and Miller, D. L.},
 doi = {10.1086/307356},
 journal = {ApJ},
 pages = {354-371},
 title = {The Effects of Line Blanketing in Spherically Expanding Outflows},
 volume = {519},
 year = {1999}
}

@article{Hsyu_PLEK_2020,
 adsurl = {https://ui.adsabs.harvard.edu/abs/2020ApJ...896...77H},
 archiveprefix = {arXiv},
 author = {{Hsyu}, Tiffany and {Cooke}, Ryan J. and {Prochaska}, J. Xavier and {Bolte}, Michael},
 doi = {10.3847/1538-4357/ab91af},
 eid = {77},
 eprint = {2005.12290},
 journal = {\apj},
 month = {June},
 number = {1},
 pages = {77},
 primaryclass = {astro-ph.GA},
 title = {{The PHLEK Survey: A New Determination of the Primordial Helium Abundance}},
 volume = {896},
 year = {2020}
}

@article{Iben_SNIa_progenitors_1984,
 adsurl = {https://ui.adsabs.harvard.edu/abs/1984ApJS...54..335I},
 author = {{Iben}, Jr., I. and {Tutukov}, A.~V.},
 doi = {10.1086/190932},
 journal = {\apjs},
 month = {February},
 pages = {335-372},
 title = {{Supernovae of type I as end products of the evolution of binaries with components of moderate initial mass.}},
 volume = {54},
 year = {1984}
}

@article{Imasheva_SNIa_yields_2023,
 adsurl = {https://ui.adsabs.harvard.edu/abs/2023MNRAS.518.1818I},
 archiveprefix = {arXiv},
 author = {{Imasheva}, Liliya and {Janka}, Hans-Thomas and {Weiss}, Achim},
 doi = {10.1093/mnras/stac3239},
 eprint = {2209.10989},
 journal = {\mnras},
 month = {January},
 number = {2},
 pages = {1818-1839},
 primaryclass = {astro-ph.HE},
 title = {{Parametrizations of thermal bomb explosions for core-collapse supernovae and $^{56}$Ni production}},
 volume = {518},
 year = {2023}
}

@article{isobe_CNO_2023,
 adsurl = {https://ui.adsabs.harvard.edu/abs/2023ApJ...959..100I},
 archiveprefix = {arXiv},
 author = {{Isobe}, Yuki and {Ouchi}, Masami and {Tominaga}, Nozomu and {Watanabe}, Kuria and {Nakajima}, Kimihiko and {Umeda}, Hiroya and {Yajima}, Hidenobu and {Harikane}, Yuichi and {Fukushima}, Hajime and {Xu}, Yi and {Ono}, Yoshiaki and {Zhang}, Yechi},
 doi = {10.3847/1538-4357/ad09be},
 eid = {100},
 eprint = {2307.00710},
 journal = {\apj},
 month = {December},
 number = {2},
 pages = {100},
 primaryclass = {astro-ph.GA},
 title = {{JWST Identification of Extremely Low C/N Galaxies with [N/O] {\ensuremath{\gtrsim}} 0.5 at z 6-10 Evidencing the Early CNO-cycle Enrichment and a Connection with Globular Cluster Formation}},
 volume = {959},
 year = {2023}
}

@article{Izotov2017,
 author = {Izotov, Y. I. and Thuan, T. X. and Guseva, N. G.},
 doi = {10.1093/mnras/stx1491},
 journal = {MNRAS},
 pages = {548-567},
 title = {He I emission-line ratios as diagnostics of Lyman continuum leakage},
 volume = {471},
 year = {2017}
}

@article{izotov_chemical_2006,
 author = {Izotov, Y. I. and Stasińska, G. and Meynet, G. and Guseva, N. G. and Thuan, T. X.},
 doi = {10.1051/0004-6361:20053763},
 journal = {\aap},
 month = {March},
 pages = {955--970},
 title = {The chemical composition of metal-poor emission-line galaxies in the {Data} {Release} 3 of the {Sloan} {Digital} {Sky} {Survey}},
 volume = {448},
 year = {2006}
}

@article{Izotov_J0811_2018,
 adsurl = {https://ui.adsabs.harvard.edu/abs/2018MNRAS.473.1956I},
 archiveprefix = {arXiv},
 author = {{Izotov}, Y.~I. and {Thuan}, T.~X. and {Guseva}, N.~G. and {Liss}, S.~E.},
 doi = {10.1093/mnras/stx2478},
 eprint = {1709.00202},
 journal = {\mnras},
 month = {January},
 number = {2},
 pages = {1956-1966},
 primaryclass = {astro-ph.GA},
 title = {{J0811+4730: the most metal-poor star-forming dwarf galaxy known}},
 volume = {473},
 year = {2018}
}

@article{izotov_xmp_2021,
 adsurl = {https://ui.adsabs.harvard.edu/abs/2021MNRAS.504.3996I},
 archiveprefix = {arXiv},
 author = {{Izotov}, Y.~I. and {Thuan}, T.~X. and {Guseva}, N.~G.},
 doi = {10.1093/mnras/stab1099},
 eprint = {2104.08035},
 journal = {\mnras},
 month = {July},
 number = {3},
 pages = {3996-4004},
 primaryclass = {astro-ph.GA},
 title = {{J2229+2725: an extremely low metallicity dwarf compact star-forming galaxy with an exceptionally high [O III]{\ensuremath{\lambda}}5007/[O II]{\ensuremath{\lambda}}3727 flux ratio of 53}},
 volume = {504},
 year = {2021}
}

@article{jenkins_depletion_2009,
 adsurl = {https://ui.adsabs.harvard.edu/abs/2009ApJ...700.1299J},
 archiveprefix = {arXiv},
 author = {{Jenkins}, Edward B.},
 doi = {10.1088/0004-637X/700/2/1299},
 eprint = {0905.3173},
 journal = {\apj},
 month = {August},
 number = {2},
 pages = {1299-1348},
 primaryclass = {astro-ph.GA},
 title = {{A Unified Representation of Gas-Phase Element Depletions in the Interstellar Medium}},
 volume = {700},
 year = {2009}
}

@article{ji_nitrogen_AGN_z5_2024,
 adsurl = {https://ui.adsabs.harvard.edu/abs/2024MNRAS.535..881J},
 archiveprefix = {arXiv},
 author = {{Ji}, Xihan and {{\"U}bler}, Hannah and {Maiolino}, Roberto and {D'Eugenio}, Francesco and {Arribas}, Santiago and {Bunker}, Andrew J. and {Charlot}, St{\'e}phane and {Perna}, Michele and {Rodr{\'\i}guez Del Pino}, Bruno and {B{\"o}ker}, Torsten and {Cresci}, Giovanni and {Curti}, Mirko and {Kumari}, Nimisha and {Lamperti}, Isabella},
 doi = {10.1093/mnras/stae2375},
 eid = {arXiv:2404.04148},
 eprint = {2404.04148},
 journal = {\mnras},
 month = {November},
 number = {1},
 pages = {881-908},
 primaryclass = {astro-ph.GA},
 title = {{GA-NIFS: an extremely nitrogen-loud and chemically stratified galaxy at z   5.55}},
 volume = {535},
 year = {2024}
}

@article{JohanssonLetokhov2005,
 author = {Johansson, S. and Letokhov, V. S.},
 doi = {10.1111/j.1365-2966.2005.09596.x},
 journal = {MNRAS},
 pages = {731--735},
 title = {Astrophysical laser operating in the OI $\lambda$8446-Å line in the Weigelt blobs of $\eta$ Carinae},
 volume = {364},
 year = {2005}
}

@article{JohanssonLetokhov_FeII_EtaCar_2004,
 adsurl = {https://ui.adsabs.harvard.edu/abs/2004A&A...428..497J},
 archiveprefix = {arXiv},
 author = {{Johansson}, S. and {Letokhov}, V.~S.},
 doi = {10.1051/0004-6361:20040379},
 eprint = {astro-ph/0409069},
 journal = {\aap},
 month = {December},
 pages = {497-509},
 primaryclass = {astro-ph},
 title = {{Astrophysical lasers operating in optical Fe II lines in stellar ejecta of {\ensuremath{\eta}} Carinae}},
 volume = {428},
 year = {2004}
}

@article{Jones_dust_grains_1996,
 adsurl = {https://ui.adsabs.harvard.edu/abs/1996ApJ...469..740J},
 author = {{Jones}, A.~P. and {Tielens}, A.~G.~G.~M. and {Hollenbach}, D.~J.},
 doi = {10.1086/177823},
 journal = {\apj},
 month = {October},
 pages = {740},
 title = {{Grain Shattering in Shocks: The Interstellar Grain Size Distribution}},
 volume = {469},
 year = {1996}
}

@article{kashino_fmos-cosmos_2019,
 author = {Kashino, D. and Silverman, J. D. and Sanders, D. and Kartaltepe, J. and Daddi, E. and Renzini, A. and Rodighiero, G. and Puglisi, A. and Valentino, F. and Juneau, S. and Arimoto, N. and Nagao, T. and Ilbert, O. and Le Fèvre, O. and Koekemoer, A. M.},
 doi = {10.3847/1538-4365/ab06c4},
 journal = {\apjs},
 month = {March},
 pages = {10},
 title = {The {FMOS}-{COSMOS} {Survey} of {Star}-forming {Galaxies} at z\nbsp\sim\nbsp1.6. {VI}. {Redshift} and {Emission}-line {Catalog} and {Basic} {Properties} of {Star}-forming {Galaxies}},
 volume = {241},
 year = {2019}
}

@article{Kashino_stellar_MZR_z2_2022,
 adsurl = {https://ui.adsabs.harvard.edu/abs/2022ApJ...925...82K},
 archiveprefix = {arXiv},
 author = {{Kashino}, Daichi and {Lilly}, Simon J. and {Renzini}, Alvio and {Daddi}, Emanuele and {Zamorani}, Giovanni and {Silverman}, John D. and {Ilbert}, Olivier and {Peng}, Ying-jie and {Mainieri}, Vincenzo and {Bardelli}, Sandro and {Zucca}, Elena and {Kartaltepe}, Jeyhan S. and {Sanders}, David B.},
 doi = {10.3847/1538-4357/ac399e},
 eid = {82},
 eprint = {2109.06044},
 journal = {\apj},
 month = {January},
 number = {1},
 pages = {82},
 primaryclass = {astro-ph.GA},
 title = {{The Stellar Mass versus Stellar Metallicity Relation of Star-forming Galaxies at 1.6 {\ensuremath{\leq}} z {\ensuremath{\leq}} 3.0 and Implications for the Evolution of the {\ensuremath{\alpha}}-enhancement}},
 volume = {925},
 year = {2022}
}

@article{kennicutt_star_2012,
 author = {Kennicutt, R. C. and Evans, N. J.},
 doi = {10.1146/annurev-astro-081811-125610},
 journal = {\araa},
 month = {September},
 pages = {531--608},
 title = {Star {Formation} in the {Milky} {Way} and {Nearby} {Galaxies}},
 volume = {50},
 year = {2012}
}

@article{kobayashi_origin_2020,
 adsurl = {https://ui.adsabs.harvard.edu/abs/2020ApJ...900..179K},
 archiveprefix = {arXiv},
 author = {{Kobayashi}, Chiaki and {Karakas}, Amanda I. and {Lugaro}, Maria},
 doi = {10.3847/1538-4357/abae65},
 eid = {179},
 eprint = {2008.04660},
 journal = {\apj},
 month = {September},
 number = {2},
 pages = {179},
 primaryclass = {astro-ph.GA},
 title = {{The Origin of Elements from Carbon to Uranium}},
 volume = {900},
 year = {2020}
}

@article{Kobulnicky_NGC5253_massive_stars_1997,
 adsurl = {https://ui.adsabs.harvard.edu/abs/1997ApJ...477..679K},
 author = {{Kobulnicky}, Henry A. and {Skillman}, Evan D. and {Roy}, Jean-Ren{\'e} and {Walsh}, J.~R. and {Rosa}, Michael R.},
 doi = {10.1086/303742},
 journal = {\apj},
 month = {March},
 number = {2},
 pages = {679-692},
 title = {{Hubble Space Telescope Faint Object Spectroscope Spectroscopy of Localized Chemical Enrichment from Massive Stars in NGC 5253}},
 volume = {477},
 year = {1997}
}

@article{KobulnickySkillman1997,
 author = {Kobulnicky, H. A. and Skillman, E. D.},
 doi = {10.1086/304803},
 journal = {ApJ},
 pages = {636-660},
 title = {Oxygen Abundances in Nearby Dwarf Irregular Galaxies},
 volume = {489},
 year = {1997}
}

@article{Kojima2021,
 author = {Kojima, T. and Ouchi, M. and Rauch, M. and others},
 doi = {10.3847/1538-4357/abf0a0},
 journal = {ApJ},
 pages = {22},
 title = {EMPRESS II: Highly Fe-enriched Metal-poor Galaxies with $\sim$1.0 (Fe/O)$_\odot$ and 0.02 (O/H)$_\odot$},
 volume = {913},
 year = {2021}
}

@article{laseter_auroral_jades_2023,
 adsurl = {https://ui.adsabs.harvard.edu/abs/2024A&A...681A..70L},
 archiveprefix = {arXiv},
 author = {{Laseter}, Isaac H. and {Maseda}, Michael V. and {Curti}, Mirko and {Maiolino}, Roberto and {D'Eugenio}, Francesco and {Cameron}, Alex J. and {Looser}, Tobias J. and {Arribas}, Santiago and {Baker}, William M. and {Bhatawdekar}, Rachana and {Boyett}, Kristan and {Bunker}, Andrew J. and {Carniani}, Stefano and {Charlot}, Stephane and {Chevallard}, Jacopo and {Curtis-lake}, Emma and {Egami}, Eiichi and {Eisenstein}, Daniel J. and {Hainline}, Kevin and {Hausen}, Ryan and {Ji}, Zhiyuan and {Kumari}, Nimisha and {Perna}, Michele and {Rawle}, Tim and {Rix}, Hans-Walter and {Robertson}, Brant and {Rodr{\'\i}guez Del Pino}, Bruno and {Sandles}, Lester and {Scholtz}, Jan and {Smit}, Renske and {Tacchella}, Sandro and {{\"U}bler}, Hannah and {Williams}, Christina C. and {Willott}, Chris and {Witstok}, Joris},
 doi = {10.1051/0004-6361/202347133},
 eid = {A70},
 eprint = {2306.03120},
 journal = {\aap},
 month = {January},
 pages = {A70},
 primaryclass = {astro-ph.GA},
 title = {{JADES: Detecting [OIII]{\ensuremath{\lambda}}4363 emitters and testing strong line calibrations in the high-z Universe with ultra-deep JWST/NIRSpec spectroscopy up to z {\ensuremath{\sim}} 9.5}},
 volume = {681},
 year = {2024}
}

@article{lopez-sanchez_NGC5264_2012,
 adsurl = {https://ui.adsabs.harvard.edu/abs/2012MNRAS.419.1051L},
 archiveprefix = {arXiv},
 author = {{L{\'o}pez-S{\'a}nchez}, {\'A}. R. and {Koribalski}, B.~S. and {van Eymeren}, J. and {Esteban}, C. and {Kirby}, E. and {Jerjen}, H. and {Lonsdale}, N.},
 doi = {10.1111/j.1365-2966.2011.19762.x},
 eprint = {1109.0806},
 journal = {\mnras},
 month = {January},
 number = {2},
 pages = {1051-1069},
 primaryclass = {astro-ph.CO},
 title = {{The intriguing H I gas in NGC 5253: an infall of a diffuse, low-metallicity H I cloud?}},
 volume = {419},
 year = {2012}
}

@article{lopez_sanchez_2007,
 adsurl = {https://ui.adsabs.harvard.edu/abs/2007ApJ...656..168L},
 archiveprefix = {arXiv},
 author = {{L{\'o}pez-S{\'a}nchez}, {\'A}ngel R. and {Esteban}, C{\'e}sar and {Garc{\'\i}a-Rojas}, Jorge and {Peimbert}, Manuel and {Rodr{\'\i}guez}, M{\'o}nica},
 doi = {10.1086/510112},
 eprint = {astro-ph/0609498},
 journal = {\apj},
 month = {February},
 number = {1},
 pages = {168-185},
 primaryclass = {astro-ph},
 title = {{The Localized Chemical Pollution in NGC 5253 Revisited: Results from Deep Echelle Spectrophotometry}},
 volume = {656},
 year = {2007}
}

@article{lopez_sanchez_WR_abundances_2010,
 adsurl = {https://ui.adsabs.harvard.edu/abs/2010A&A...517A..85L},
 archiveprefix = {arXiv},
 author = {{L{\'o}pez-S{\'a}nchez}, {\'A}. R. and {Esteban}, C.},
 doi = {10.1051/0004-6361/201014156},
 eid = {A85},
 eprint = {1004.0626},
 journal = {\aap},
 month = {July},
 pages = {A85},
 primaryclass = {astro-ph.CO},
 title = {{Massive star formation in Wolf-Rayet galaxies. IV. Colours, chemical-composition analysis and metallicity-luminosity relations}},
 volume = {517},
 year = {2010}
}

@article{Maeder_WR_2014,
 adsurl = {https://ui.adsabs.harvard.edu/abs/2014A&A...565A..39M},
 archiveprefix = {arXiv},
 author = {{Maeder}, Andr{\'e} and {Przybilla}, Norbert and {Nieva}, Mar{\'\i}a-Fernanda and {Georgy}, Cyril and {Meynet}, Georges and {Ekstr{\"o}m}, Sylvia and {Eggenberger}, Patrick},
 doi = {10.1051/0004-6361/201220602},
 eid = {A39},
 eprint = {1404.1020},
 journal = {\aap},
 month = {May},
 pages = {A39},
 primaryclass = {astro-ph.SR},
 title = {{Evolution of surface CNO abundances in massive stars}},
 volume = {565},
 year = {2014}
}

@article{maiolino_amaze_2008,
 author = {Maiolino, R. and Nagao, T. and Grazian, A. and Cocchia, F. and Marconi, A. and Mannucci, F. and Cimatti, A. and Pipino, A. and Ballero, S. and Calura, F. and Chiappini, C. and Fontana, A. and Granato, G. L. and Matteucci, F. and Pastorini, G. and Pentericci, L. and Risaliti, G. and Salvati, M. and Silva, L.},
 doi = {10.1051/0004-6361:200809678},
 journal = {\aap},
 month = {September},
 pages = {463--479},
 title = {{AMAZE}. {I}. {The} evolution of the mass-metallicity relation at z \$\gt\$ 3},
 volume = {488},
 year = {2008}
}

@article{mannucci_lsd_2009,
 author = {Mannucci, F. and Cresci, G. and Maiolino, R. and Marconi, A. and Pastorini, G. and Pozzetti, L. and Gnerucci, A. and Risaliti, G. and Schneider, R. and Lehnert, M. and Salvati, M.},
 doi = {10.1111/j.1365-2966.2009.15185.x},
 journal = {\mnras},
 month = {October},
 pages = {1915--1931},
 title = {{LSD}: {Lyman}-break galaxies {Stellar} populations and {Dynamics} - {I}. {Mass}, metallicity and gas at z ∼ 3.1},
 volume = {398},
 year = {2009}
}

@article{Maoz_Graur_SnIa_2017,
 adsurl = {https://ui.adsabs.harvard.edu/abs/2017ApJ...848...25M},
 archiveprefix = {arXiv},
 author = {{Maoz}, Dan and {Graur}, Or},
 doi = {10.3847/1538-4357/aa8b6e},
 eid = {25},
 eprint = {1703.04540},
 journal = {\apj},
 month = {October},
 number = {1},
 pages = {25},
 primaryclass = {astro-ph.HE},
 title = {{Star Formation, Supernovae, Iron, and {\ensuremath{\alpha}}: Consistent Cosmic and Galactic Histories}},
 volume = {848},
 year = {2017}
}

@article{Marasco_dwarfs_outflows_2023,
 adsurl = {https://ui.adsabs.harvard.edu/abs/2023A&A...670A..92M},
 archiveprefix = {arXiv},
 author = {{Marasco}, A. and {Belfiore}, F. and {Cresci}, G. and {Lelli}, F. and {Venturi}, G. and {Hunt}, L.~K. and {Concas}, A. and {Marconi}, A. and {Mannucci}, F. and {Mingozzi}, M. and {McLeod}, A.~F. and {Kumari}, N. and {Carniani}, S. and {Vanzi}, L. and {Ginolfi}, M.},
 doi = {10.1051/0004-6361/202244895},
 eid = {A92},
 eprint = {2209.02726},
 journal = {\aap},
 month = {February},
 pages = {A92},
 primaryclass = {astro-ph.GA},
 title = {{Shaken, but not expelled: Gentle baryonic feedback from nearby starburst dwarf galaxies}},
 volume = {670},
 year = {2023}
}

@article{Martins_Palacios_VMS_2022,
 adsurl = {https://ui.adsabs.harvard.edu/abs/2022A&A...659A.163M},
 archiveprefix = {arXiv},
 author = {{Martins}, F. and {Palacios}, A.},
 doi = {10.1051/0004-6361/202243048},
 eid = {A163},
 eprint = {2202.13703},
 journal = {\aap},
 month = {March},
 pages = {A163},
 primaryclass = {astro-ph.SR},
 title = {{Spectroscopic evolution of very massive stars at Z = 1/2.5 Z$_{{\ensuremath{\odot}}}$}},
 volume = {659},
 year = {2022}
}

@article{Martins_VMS_2023,
 adsurl = {https://ui.adsabs.harvard.edu/abs/2023A&A...678A.159M},
 archiveprefix = {arXiv},
 author = {{Martins}, F. and {Schaerer}, D. and {Marques-Chaves}, R. and {Upadhyaya}, A.},
 doi = {10.1051/0004-6361/202346732},
 eid = {A159},
 eprint = {2308.14489},
 journal = {\aap},
 month = {October},
 pages = {A159},
 primaryclass = {astro-ph.GA},
 title = {{Inferring the presence of very massive stars in local star-forming regions}},
 volume = {678},
 year = {2023}
}

@article{Matsuoka_OI_AGN_2007,
 adsurl = {https://ui.adsabs.harvard.edu/abs/2007ApJ...663..781M},
 archiveprefix = {arXiv},
 author = {{Matsuoka}, Y. and {Oyabu}, S. and {Tsuzuki}, Y. and {Kawara}, K.},
 doi = {10.1086/518399},
 eprint = {astro-ph/0703659},
 journal = {\apj},
 month = {July},
 number = {2},
 pages = {781-798},
 primaryclass = {astro-ph},
 title = {{Observations of O I and Ca II Emission Lines in Quasars: Implications for the Site of Fe II Line Emission}},
 volume = {663},
 year = {2007}
}

@article{Mendez-Delgado_Fe_O_2024,
 adsurl = {https://ui.adsabs.harvard.edu/abs/2024A&A...690A.248M},
 archiveprefix = {arXiv},
 author = {{M{\'e}ndez-Delgado}, J.~E. and {Kreckel}, K. and {Esteban}, C. and {Garc{\'\i}a-Rojas}, J. and {Carigi}, L. and {Sander}, A.~A.~C. and {Palla}, M. and {Chru{\'s}li{\'n}ska}, M. and {De Looze}, I. and {Rela{\~n}o}, M. and {van der Giessen}, S.~A. and {Reyes-Rodr{\'\i}guez}, E. and {S{\'a}nchez}, S.~F.},
 doi = {10.1051/0004-6361/202450928},
 eid = {A248},
 eprint = {2408.06215},
 journal = {\aap},
 month = {October},
 pages = {A248},
 primaryclass = {astro-ph.GA},
 title = {{Gas-phase Fe/O and Fe/N abundances in star-forming regions: Relations between nucleosynthesis, metallicity, and dust}},
 volume = {690},
 year = {2024}
}

@article{Mendez-Delgado_Helium_temp_2024,
 adsurl = {https://ui.adsabs.harvard.edu/abs/2025ApJ...986...74M},
 archiveprefix = {arXiv},
 author = {{M{\'e}ndez-Delgado}, J.~E. and {Skillman}, E.~D. and {Aver}, E. and {Morisset}, C. and {Esteban}, C. and {Garc{\'\i}a-Rojas}, J. and {Kreckel}, K. and {Rogers}, N.~S.~J. and {Rosales-Ortega}, F.~F. and {Arellano-C{\'o}rdova}, K.~Z. and {Flury}, S.~R. and {Reyes-Rodr{\'\i}guez}, E. and {Orte-Garc{\'\i}a}, M. and {Tan}, S.},
 doi = {10.3847/1538-4357/adc67a},
 eid = {74},
 eprint = {2410.17381},
 journal = {\apj},
 month = {June},
 number = {1},
 pages = {74},
 primaryclass = {astro-ph.GA},
 title = {{Generalized T$_{e}$([O III]){\textendash}T$_{e}$(He I) Discrepancies in Ionized Nebulae: Possible Evidence of Case B Deviations and Temperature Inhomogeneities}},
 volume = {986},
 year = {2025}
}

@article{mendez_delgado_t_inhomogeneities_2023,
 adsurl = {https://ui.adsabs.harvard.edu/abs/2023Natur.618..249M},
 archiveprefix = {arXiv},
 author = {{M{\'e}ndez-Delgado}, J. Eduardo and {Esteban}, C{\'e}sar and {Garc{\'\i}a-Rojas}, Jorge and {Kreckel}, Kathryn and {Peimbert}, Manuel},
 doi = {10.1038/s41586-023-05956-2},
 eprint = {2305.11578},
 journal = {\nat},
 month = {June},
 number = {7964},
 pages = {249-251},
 primaryclass = {astro-ph.GA},
 title = {{Temperature inhomogeneities cause the abundance discrepancy in H II regions}},
 volume = {618},
 year = {2023}
}

@article{Mesa-Delgado_HH202_2009,
 adsurl = {https://ui.adsabs.harvard.edu/abs/2009MNRAS.395..855M},
 archiveprefix = {arXiv},
 author = {{Mesa-Delgado}, A. and {Esteban}, C. and {Garc{\'\i}a-Rojas}, J. and {Luridiana}, V. and {Bautista}, M. and {Rodr{\'\i}guez}, M. and {L{\'o}pez-Mart{\'\i}n}, L. and {Peimbert}, M.},
 doi = {10.1111/j.1365-2966.2009.14554.x},
 eprint = {0901.4311},
 journal = {\mnras},
 month = {May},
 number = {2},
 pages = {855-876},
 primaryclass = {astro-ph.GA},
 title = {{Properties of the ionized gas in HH 202 - II. Results from echelle spectrophotometry with Ultraviolet Visual Echelle Spectrograph}},
 volume = {395},
 year = {2009}
}

@article{mestric_sunburst_2023,
 adsurl = {https://ui.adsabs.harvard.edu/abs/2023A&A...673A..50M},
 archiveprefix = {arXiv},
 author = {{Me{\v{s}}tri{\'c}}, U. and {Vanzella}, E. and {Upadhyaya}, A. and {Martins}, F. and {Marques-Chaves}, R. and {Schaerer}, D. and {Guibert}, J. and {Zanella}, A. and {Grillo}, C. and {Rosati}, P. and {Calura}, F. and {Caminha}, G.~B. and {Bolamperti}, A. and {Meneghetti}, M. and {Bergamini}, P. and {Mercurio}, A. and {Nonino}, M. and {Pascale}, R.},
 doi = {10.1051/0004-6361/202345895},
 eid = {A50},
 eprint = {2301.04672},
 journal = {\aap},
 month = {May},
 pages = {A50},
 primaryclass = {astro-ph.GA},
 title = {{Clues on the presence and segregation of very massive stars in the Sunburst Lyman-continuum cluster at z = 2.37}},
 volume = {673},
 year = {2023}
}

@article{Molla_pop_synth_2009,
 adsurl = {https://ui.adsabs.harvard.edu/abs/2009MNRAS.398..451M},
 archiveprefix = {arXiv},
 author = {{Moll{\'a}}, M. and {Garc{\'\i}a-Vargas}, M.~L. and {Bressan}, A.},
 doi = {10.1111/j.1365-2966.2009.15160.x},
 eprint = {0905.3664},
 journal = {\mnras},
 month = {September},
 number = {1},
 pages = {451-470},
 primaryclass = {astro-ph.CO},
 title = {{PopStar I: evolutionary synthesis model description}},
 volume = {398},
 year = {2009}
}

@article{muratov_feedback_FIRE_2015,
 adsurl = {https://ui.adsabs.harvard.edu/abs/2015MNRAS.454.2691M},
 archiveprefix = {arXiv},
 author = {{Muratov}, Alexander L. and {Kere{\v{s}}}, Du{\v{s}}an and {Faucher-Gigu{\`e}re}, Claude-Andr{\'e} and {Hopkins}, Philip F. and {Quataert}, Eliot and {Murray}, Norman},
 doi = {10.1093/mnras/stv2126},
 eprint = {1501.03155},
 journal = {\mnras},
 month = {December},
 number = {3},
 pages = {2691-2713},
 primaryclass = {astro-ph.GA},
 title = {{Gusty, gaseous flows of FIRE: galactic winds in cosmological simulations with explicit stellar feedback}},
 volume = {454},
 year = {2015}
}

@article{Naiman_TNG_2018,
 adsurl = {https://ui.adsabs.harvard.edu/abs/2018MNRAS.477.1206N},
 archiveprefix = {arXiv},
 author = {{Naiman}, Jill P. and {Pillepich}, Annalisa and {Springel}, Volker and {Ramirez-Ruiz}, Enrico and {Torrey}, Paul and {Vogelsberger}, Mark and {Pakmor}, R{\"u}diger and {Nelson}, Dylan and {Marinacci}, Federico and {Hernquist}, Lars and {Weinberger}, Rainer and {Genel}, Shy},
 doi = {10.1093/mnras/sty618},
 eprint = {1707.03401},
 journal = {\mnras},
 month = {June},
 number = {1},
 pages = {1206-1224},
 primaryclass = {astro-ph.GA},
 title = {{First results from the IllustrisTNG simulations: a tale of two elements - chemical evolution of magnesium and europium}},
 volume = {477},
 year = {2018}
}

@article{nakajima_mzr_ceers_2023,
 adsurl = {https://ui.adsabs.harvard.edu/abs/2023ApJS..269...33N},
 archiveprefix = {arXiv},
 author = {{Nakajima}, Kimihiko and {Ouchi}, Masami and {Isobe}, Yuki and {Harikane}, Yuichi and {Zhang}, Yechi and {Ono}, Yoshiaki and {Umeda}, Hiroya and {Oguri}, Masamune},
 doi = {10.3847/1538-4365/acd556},
 eid = {33},
 eprint = {2301.12825},
 journal = {\apjs},
 month = {December},
 number = {2},
 pages = {33},
 primaryclass = {astro-ph.GA},
 title = {{JWST Census for the Mass-Metallicity Star Formation Relations at z = 4-10 with Self-consistent Flux Calibration and Proper Metallicity Calibrators}},
 volume = {269},
 year = {2023}
}

@inproceedings{Nelemans_SNIa_DTD_2013,
 adsurl = {https://ui.adsabs.harvard.edu/abs/2013IAUS..281..225N},
 archiveprefix = {arXiv},
 author = {{Nelemans}, Gijs and {Toonen}, Silvia and {Bours}, Madelon},
 booktitle = {Binary Paths to Type Ia Supernovae Explosions},
 doi = {10.1017/S1743921312015098},
 editor = {{Di Stefano}, Rosanne and {Orio}, Marina and {Moe}, Maxwell},
 eprint = {1204.2960},
 month = {January},
 pages = {225-231},
 primaryclass = {astro-ph.HE},
 series = {IAU Symposium},
 title = {{Theoretical Delay Time Distributions}},
 volume = {281},
 year = {2013}
}

@article{Nelson2019,
 author = {Nelson, D. and Pillepich, A. and Springel, V. and others},
 doi = {10.1093/mnras/stz2306},
 journal = {MNRAS},
 pages = {3234-3261},
 title = {First results from the TNG50 simulation: Galactic outflows driven by supernovae and black hole feedback},
 volume = {490},
 year = {2019}
}

@article{nelson_TNG_2019,
 adsurl = {https://ui.adsabs.harvard.edu/abs/2019ComAC...6....2N},
 archiveprefix = {arXiv},
 author = {{Nelson}, Dylan and {Springel}, Volker and {Pillepich}, Annalisa and {Rodriguez-Gomez}, Vicente and {Torrey}, Paul and {Genel}, Shy and {Vogelsberger}, Mark and {Pakmor}, Ruediger and {Marinacci}, Federico and {Weinberger}, Rainer and {Kelley}, Luke and {Lovell}, Mark and {Diemer}, Benedikt and {Hernquist}, Lars},
 doi = {10.1186/s40668-019-0028-x},
 eid = {2},
 eprint = {1812.05609},
 journal = {Computational Astrophysics and Cosmology},
 month = {May},
 number = {1},
 pages = {2},
 primaryclass = {astro-ph.GA},
 title = {{The IllustrisTNG simulations: public data release}},
 volume = {6},
 year = {2019}
}

@article{NetzerPenston1976,
 author = {Netzer, H. and Penston, M. V.},
 journal = {MNRAS},
 pages = {319--325},
 title = {Line fluorescence in astrophysics: O\,{\sc i}\,λ8446 emission mechanisms},
 volume = {174},
 year = {1976}
}

@article{nomoto_chemical_evolution_2013,
 adsurl = {https://ui.adsabs.harvard.edu/abs/2013ARA&A..51..457N},
 author = {{Nomoto}, Ken'ichi and {Kobayashi}, Chiaki and {Tominaga}, Nozomu},
 doi = {10.1146/annurev-astro-082812-140956},
 journal = {\araa},
 month = {August},
 number = {1},
 pages = {457-509},
 title = {{Nucleosynthesis in Stars and the Chemical Enrichment of Galaxies}},
 volume = {51},
 year = {2013}
}

@article{Nomoto_hypernovae_2006,
 adsurl = {https://ui.adsabs.harvard.edu/abs/2006NuPhA.777..424N},
 archiveprefix = {arXiv},
 author = {{Nomoto}, Ken'ichi and {Tominaga}, Nozomu and {Umeda}, Hideyuki and {Kobayashi}, Chiaki and {Maeda}, Keiichi},
 doi = {10.1016/j.nuclphysa.2006.05.008},
 eprint = {astro-ph/0605725},
 journal = {\nphysa},
 month = {October},
 pages = {424-458},
 primaryclass = {astro-ph},
 title = {{Nucleosynthesis yields of core-collapse supernovae and hypernovae, and galactic chemical evolution}},
 volume = {777},
 year = {2006}
}

@article{Oliva_OI_SNe_1993,
 adsurl = {https://ui.adsabs.harvard.edu/abs/1993A&A...276..415O},
 author = {{Oliva}, E.},
 journal = {\aap},
 month = {September},
 pages = {415-431},
 title = {{The O I-Ly-beta fluorescence revisited and its implications on the clumping of hydrogen, O/H mixing and the pre-SN oxygen abundance in SN 1987A.}},
 volume = {276},
 year = {1993}
}

@article{Olive_Skillman_prim_He_2004,
 adsurl = {https://ui.adsabs.harvard.edu/abs/2004ApJ...617...29O},
 archiveprefix = {arXiv},
 author = {{Olive}, Keith A. and {Skillman}, Evan D.},
 doi = {10.1086/425170},
 eprint = {astro-ph/0405588},
 journal = {\apj},
 month = {December},
 number = {1},
 pages = {29-49},
 primaryclass = {astro-ph},
 title = {{A Realistic Determination of the Error on the Primordial Helium Abundance: Steps toward Nonparametric Nebular Helium Abundances}},
 volume = {617},
 year = {2004}
}

@article{palay_improved_2012,
 author = {Palay, E. and Nahar, S. N. and Pradhan, A. K. and Eissner, W.},
 doi = {10.1111/j.1745-3933.2012.01252.x},
 journal = {\mnras},
 month = {June},
 pages = {L35--L39},
 title = {Improved collision strengths and line ratios for forbidden [{O} {III}] far-infrared and optical lines},
 volume = {423},
 year = {2012}
}

@article{Pascale_sunburst_arc_2023,
 adsurl = {https://ui.adsabs.harvard.edu/abs/2023ApJ...957...77P},
 archiveprefix = {arXiv},
 author = {{Pascale}, Massimo and {Dai}, Liang and {McKee}, Christopher F. and {Tsang}, Benny T. -H.},
 doi = {10.3847/1538-4357/acf75c},
 eid = {77},
 eprint = {2301.10790},
 journal = {\apj},
 month = {November},
 number = {2},
 pages = {77},
 primaryclass = {astro-ph.GA},
 title = {{Nitrogen-enriched, Highly Pressurized Nebular Clouds Surrounding a Super Star Cluster at Cosmic Noon}},
 volume = {957},
 year = {2023}
}

@article{patricio_testing_2018,
 author = {Patrício, V. and Christensen, L. and Rhodin, H. and Cañameras, R. and Lara-López, M. A.},
 doi = {10.1093/mnras/sty2508},
 journal = {\mnras},
 month = {December},
 pages = {3520--3533},
 title = {Testing strong line metallicity diagnostics at z \tilde 2},
 volume = {481},
 year = {2018}
}

@article{Pettini_LBGs_2001,
 adsurl = {https://ui.adsabs.harvard.edu/abs/2001ApJ...554..981P},
 archiveprefix = {arXiv},
 author = {{Pettini}, Max and {Shapley}, Alice E. and {Steidel}, Charles C. and {Cuby}, Jean-Gabriel and {Dickinson}, Mark and {Moorwood}, Alan F.~M. and {Adelberger}, Kurt L. and {Giavalisco}, Mauro},
 doi = {10.1086/321403},
 eprint = {astro-ph/0102456},
 journal = {\apj},
 month = {June},
 number = {2},
 pages = {981-1000},
 primaryclass = {astro-ph},
 title = {{The Rest-Frame Optical Spectra of Lyman Break Galaxies: Star Formation, Extinction, Abundances, and Kinematics}},
 volume = {554},
 year = {2001}
}

@article{pettini_oiiinii_2004,
 author = {Pettini, M. and Pagel, B. E. J.},
 doi = {10.1111/j.1365-2966.2004.07591.x},
 journal = {\mnras},
 month = {March},
 pages = {L59--L63},
 title = {[{OIII}]/[{NII}] as an abundance indicator at high redshift},
 volume = {348},
 year = {2004}
}

@article{Pillepich_TNG_2018,
 adsurl = {https://ui.adsabs.harvard.edu/abs/2018MNRAS.473.4077P},
 archiveprefix = {arXiv},
 author = {{Pillepich}, Annalisa and {Springel}, Volker and {Nelson}, Dylan and {Genel}, Shy and {Naiman}, Jill and {Pakmor}, R{\"u}diger and {Hernquist}, Lars and {Torrey}, Paul and {Vogelsberger}, Mark and {Weinberger}, Rainer and {Marinacci}, Federico},
 doi = {10.1093/mnras/stx2656},
 eprint = {1703.02970},
 journal = {\mnras},
 month = {January},
 number = {3},
 pages = {4077-4106},
 primaryclass = {astro-ph.GA},
 title = {{Simulating galaxy formation with the IllustrisTNG model}},
 volume = {473},
 year = {2018}
}

@article{planck_2020,
 adsurl = {https://ui.adsabs.harvard.edu/abs/2020A&A...641A...6P},
 archiveprefix = {arXiv},
 author = {{Planck Collaboration} and {Aghanim}, N. and {Akrami}, Y. and {Ashdown}, M. and {Aumont}, J. and {Baccigalupi}, C. and {Ballardini}, M. and {Banday}, A.~J. and {Barreiro}, R.~B. and {Bartolo}, N. and {Basak}, S. and {Battye}, R. and {Benabed}, K. and {Bernard}, J. -P. and {Bersanelli}, M. and {Bielewicz}, P. and {Bock}, J.~J. and {Bond}, J.~R. and {Borrill}, J. and {Bouchet}, F.~R. and {Boulanger}, F. and {Bucher}, M. and {Burigana}, C. and {Butler}, R.~C. and {Calabrese}, E. and {Cardoso}, J. -F. and {Carron}, J. and {Challinor}, A. and {Chiang}, H.~C. and {Chluba}, J. and {Colombo}, L.~P.~L. and {Combet}, C. and {Contreras}, D. and {Crill}, B.~P. and {Cuttaia}, F. and {de Bernardis}, P. and {de Zotti}, G. and {Delabrouille}, J. and {Delouis}, J. -M. and {Di Valentino}, E. and {Diego}, J.~M. and {Dor{\'e}}, O. and {Douspis}, M. and {Ducout}, A. and {Dupac}, X. and {Dusini}, S. and {Efstathiou}, G. and {Elsner}, F. and {En{\ss}lin}, T.~A. and {Eriksen}, H.~K. and {Fantaye}, Y. and {Farhang}, M. and {Fergusson}, J. and {Fernandez-Cobos}, R. and {Finelli}, F. and {Forastieri}, F. and {Frailis}, M. and {Fraisse}, A.~A. and {Franceschi}, E. and {Frolov}, A. and {Galeotta}, S. and {Galli}, S. and {Ganga}, K. and {G{\'e}nova-Santos}, R.~T. and {Gerbino}, M. and {Ghosh}, T. and {Gonz{\'a}lez-Nuevo}, J. and {G{\'o}rski}, K.~M. and {Gratton}, S. and {Gruppuso}, A. and {Gudmundsson}, J.~E. and {Hamann}, J. and {Handley}, W. and {Hansen}, F.~K. and {Herranz}, D. and {Hildebrandt}, S.~R. and {Hivon}, E. and {Huang}, Z. and {Jaffe}, A.~H. and {Jones}, W.~C. and {Karakci}, A. and {Keih{\"a}nen}, E. and {Keskitalo}, R. and {Kiiveri}, K. and {Kim}, J. and {Kisner}, T.~S. and {Knox}, L. and {Krachmalnicoff}, N. and {Kunz}, M. and {Kurki-Suonio}, H. and {Lagache}, G. and {Lamarre}, J. -M. and {Lasenby}, A. and {Lattanzi}, M. and {Lawrence}, C.~R. and {Le Jeune}, M. and {Lemos}, P. and {Lesgourgues}, J. and {Levrier}, F. and {Lewis}, A. and {Liguori}, M. and {Lilje}, P.~B. and {Lilley}, M. and {Lindholm}, V. and {L{\'o}pez-Caniego}, M. and {Lubin}, P.~M. and {Ma}, Y. -Z. and {Mac{\'\i}as-P{\'e}rez}, J.~F. and {Maggio}, G. and {Maino}, D. and {Mandolesi}, N. and {Mangilli}, A. and {Marcos-Caballero}, A. and {Maris}, M. and {Martin}, P.~G. and {Martinelli}, M. and {Mart{\'\i}nez-Gonz{\'a}lez}, E. and {Matarrese}, S. and {Mauri}, N. and {McEwen}, J.~D. and {Meinhold}, P.~R. and {Melchiorri}, A. and {Mennella}, A. and {Migliaccio}, M. and {Millea}, M. and {Mitra}, S. and {Miville-Desch{\^e}nes}, M. -A. and {Molinari}, D. and {Montier}, L. and {Morgante}, G. and {Moss}, A. and {Natoli}, P. and {N{\o}rgaard-Nielsen}, H.~U. and {Pagano}, L. and {Paoletti}, D. and {Partridge}, B. and {Patanchon}, G. and {Peiris}, H.~V. and {Perrotta}, F. and {Pettorino}, V. and {Piacentini}, F. and {Polastri}, L. and {Polenta}, G. and {Puget}, J. -L. and {Rachen}, J.~P. and {Reinecke}, M. and {Remazeilles}, M. and {Renzi}, A. and {Rocha}, G. and {Rosset}, C. and {Roudier}, G. and {Rubi{\~n}o-Mart{\'\i}n}, J.~A. and {Ruiz-Granados}, B. and {Salvati}, L. and {Sandri}, M. and {Savelainen}, M. and {Scott}, D. and {Shellard}, E.~P.~S. and {Sirignano}, C. and {Sirri}, G. and {Spencer}, L.~D. and {Sunyaev}, R. and {Suur-Uski}, A. -S. and {Tauber}, J.~A. and {Tavagnacco}, D. and {Tenti}, M. and {Toffolatti}, L. and {Tomasi}, M. and {Trombetti}, T. and {Valenziano}, L. and {Valiviita}, J. and {Van Tent}, B. and {Vibert}, L. and {Vielva}, P. and {Villa}, F. and {Vittorio}, N. and {Wandelt}, B.~D. and {Wehus}, I.~K. and {White}, M. and {White}, S.~D.~M. and {Zacchei}, A. and {Zonca}, A.},
 doi = {10.1051/0004-6361/201833910},
 eid = {A6},
 eprint = {1807.06209},
 journal = {\aap},
 month = {September},
 pages = {A6},
 primaryclass = {astro-ph.CO},
 title = {{Planck 2018 results. VI. Cosmological parameters}},
 volume = {641},
 year = {2020}
}

@article{Porter_He_emiss_2013,
 adsurl = {https://ui.adsabs.harvard.edu/abs/2013MNRAS.433L..89P},
 archiveprefix = {arXiv},
 author = {{Porter}, R.~L. and {Ferland}, G.~J. and {Storey}, P.~J. and {Detisch}, M.~J.},
 doi = {10.1093/mnrasl/slt049},
 eprint = {1303.5115},
 journal = {\mnras},
 month = {July},
 number = {1},
 pages = {L89-L90},
 primaryclass = {astro-ph.CO},
 title = {{Erratum: `Improved He I emissivities in the Case B approximation'}},
 volume = {433},
 year = {2013}
}

@article{Reddy_AURORA_dust_fcov_2025,
 adsurl = {https://ui.adsabs.harvard.edu/abs/2025arXiv250617396R},
 archiveprefix = {arXiv},
 author = {{Reddy}, Naveen A. and {Shapley}, Alice E. and {Sanders}, Ryan L. and {Topping}, Michael W. and {Ellis}, Richard S. and {Pettini}, Max and {Brammer}, Gabriel and {Cullen}, Fergus and {Forster Schreiber}, Natascha M. and {Khostovan}, Ali A. and {McLeod}, Derek J. and {McLure}, Ross J. and {Narayanan}, Desika and {Oesch}, Pascal A. and {Pahl}, Anthony J. and {Steidel}, Charles C. and {Berg}, Danielle A.},
 doi = {10.48550/arXiv.2506.17396},
 eid = {arXiv:2506.17396},
 eprint = {2506.17396},
 journal = {arXiv e-prints},
 month = {June},
 pages = {arXiv:2506.17396},
 primaryclass = {astro-ph.GA},
 title = {{The JWST/AURORA Survey: Multiple Balmer and Paschen Emission Lines for Individual Star-forming Galaxies at z=1.5-4.4. I. A Diversity of Nebular Attenuation Curves and Evidence for Non-Unity Dust Covering Fractions}},
 year = {2025}
}

@article{Reines_BH_mass_2013,
 adsurl = {https://ui.adsabs.harvard.edu/abs/2013ApJ...775..116R},
 archiveprefix = {arXiv},
 author = {{Reines}, Amy E. and {Greene}, Jenny E. and {Geha}, Marla},
 doi = {10.1088/0004-637X/775/2/116},
 eid = {116},
 eprint = {1308.0328},
 journal = {\apj},
 month = {October},
 number = {2},
 pages = {116},
 primaryclass = {astro-ph.CO},
 title = {{Dwarf Galaxies with Optical Signatures of Active Massive Black Holes}},
 volume = {775},
 year = {2013}
}

@article{Rivera_Thorsen_WR_sunburst_arc_2024,
 adsurl = {https://ui.adsabs.harvard.edu/abs/2024A&A...690A.269R},
 archiveprefix = {arXiv},
 author = {{Rivera-Thorsen}, T. Emil and {Chisholm}, J. and {Welch}, B. and {Rigby}, J.~R. and {Hutchison}, T. and {Florian}, M. and {Sharon}, K. and {Choe}, S. and {Dahle}, H. and {Bayliss}, M.~B. and {Khullar}, G. and {Gladders}, M. and {Hayes}, M. and {Adamo}, A. and {Owens}, M.~R. and {Kim}, K.},
 doi = {10.1051/0004-6361/202450359},
 eid = {A269},
 eprint = {2404.08884},
 journal = {\aap},
 month = {October},
 pages = {A269},
 primaryclass = {astro-ph.GA},
 title = {{The Sunburst Arc with JWST: I. Detection of Wolf-Rayet stars injecting nitrogen into a low-metallicity, z = 2.37 proto-globular cluster leaking ionizing photons}},
 volume = {690},
 year = {2024}
}

@article{Rivero_Gonzalez_stellar_NIII_2011,
 adsurl = {https://ui.adsabs.harvard.edu/abs/2011A&A...536A..58R},
 archiveprefix = {arXiv},
 author = {{Rivero Gonz{\'a}lez}, J.~G. and {Puls}, J. and {Najarro}, F.},
 doi = {10.1051/0004-6361/201117101},
 eid = {A58},
 eprint = {1109.3595},
 journal = {\aap},
 month = {December},
 pages = {A58},
 primaryclass = {astro-ph.SR},
 title = {{Nitrogen line spectroscopy of O-stars. I. Nitrogen III emission line formation revisited}},
 volume = {536},
 year = {2011}
}

@article{Rodriguez-Ardilla_AGN_2004,
 adsurl = {https://ui.adsabs.harvard.edu/abs/2004A&A...425..457R},
 archiveprefix = {arXiv},
 author = {{Rodr{\'\i}guez-Ardila}, A. and {Pastoriza}, M.~G. and {Viegas}, S. and {Sigut}, T.~A.~A. and {Pradhan}, A.~K.},
 doi = {10.1051/0004-6361:20034285},
 eprint = {astro-ph/0406402},
 journal = {\aap},
 month = {October},
 pages = {457-474},
 primaryclass = {astro-ph},
 title = {{Molecular hydrogen and [Fe II] in Active Galactic Nuclei}},
 volume = {425},
 year = {2004}
}

@article{Rodriguez_Fe_SNIa_2023,
 adsurl = {https://ui.adsabs.harvard.edu/abs/2023ApJ...955...71R},
 archiveprefix = {arXiv},
 author = {{Rodr{\'\i}guez}, {\'O}smar and {Maoz}, Dan and {Nakar}, Ehud},
 doi = {10.3847/1538-4357/ace2bd},
 eid = {71},
 eprint = {2209.05552},
 journal = {\apj},
 month = {September},
 number = {1},
 pages = {71},
 primaryclass = {astro-ph.HE},
 title = {{The Iron Yield of Core-collapse Supernovae}},
 volume = {955},
 year = {2023}
}

@article{Rodriguez_Rubin_Fe_ICF_2005,
 adsurl = {https://ui.adsabs.harvard.edu/abs/2005ApJ...626..900R},
 archiveprefix = {arXiv},
 author = {{Rodr{\'\i}guez}, M{\'o}nica and {Rubin}, Robert H.},
 doi = {10.1086/429958},
 eprint = {astro-ph/0504131},
 journal = {\apj},
 month = {June},
 number = {2},
 pages = {900-908},
 primaryclass = {astro-ph},
 title = {{The [Fe IV] Discrepancy: Constraining the Iron Abundances in Nebulae}},
 volume = {626},
 year = {2005}
}

@article{Rogers_Cecilia_z3_2024,
 adsurl = {https://ui.adsabs.harvard.edu/abs/2024ApJ...964L..12R},
 archiveprefix = {arXiv},
 author = {{Rogers}, Noah S.~J. and {Strom}, Allison L. and {Rudie}, Gwen C. and {Trainor}, Ryan F. and {Raptis}, Menelaos and {von Raesfeld}, Caroline},
 doi = {10.3847/2041-8213/ad2f37},
 eid = {L12},
 eprint = {2312.08427},
 journal = {\apjl},
 month = {March},
 number = {1},
 pages = {L12},
 primaryclass = {astro-ph.GA},
 title = {{CECILIA: Direct O, N, S, and Ar Abundances in Q2343-D40, a Galaxy at z {\ensuremath{\sim}} 3}},
 volume = {964},
 year = {2024}
}

@article{Roman-Duval_METAL_dust_depletion_2022,
 adsurl = {https://ui.adsabs.harvard.edu/abs/2022ApJ...928...90R},
 archiveprefix = {arXiv},
 author = {{Roman-Duval}, Julia and {Jenkins}, Edward B. and {Tchernyshyov}, Kirill and {Clark}, Christopher J.~R. and {De Cia}, Annalisa and {Gordon}, Karl D. and {Hamanowicz}, Aleksandra and {Lebouteiller}, Vianney and {Rafelski}, Marc and {Sandstrom}, Karin and {Werk}, Jessica and {Yanchulova Merica-Jones}, Petia},
 doi = {10.3847/1538-4357/ac5248},
 eid = {90},
 eprint = {2202.04765},
 journal = {\apj},
 month = {March},
 number = {1},
 pages = {90},
 primaryclass = {astro-ph.GA},
 title = {{METAL: The Metal Evolution, Transport, and Abundance in the Large Magellanic Cloud Hubble Program. III. Interstellar Depletions, Dust-to-Metal, and Dust-to-Gas Ratios versus Metallicity}},
 volume = {928},
 year = {2022}
}

@article{Rudy_NGC7027_1992,
 adsurl = {https://ui.adsabs.harvard.edu/abs/1992ApJ...384..536R},
 author = {{Rudy}, Richard J. and {Erwin}, Peter and {Rossano}, George S. and {Puetter}, R.~C.},
 doi = {10.1086/170896},
 journal = {\apj},
 month = {January},
 pages = {536},
 title = {{0.8--1.6 Micron Spectroscopy of the Planetary Nebula NGC 7027}},
 volume = {384},
 year = {1992}
}

@article{Rudy_OI_AGN_1992,
 adsurl = {https://ui.adsabs.harvard.edu/abs/1989ApJ...342..235R},
 author = {{Rudy}, Richard J. and {Rossano}, George S. and {Puetter}, R.~C.},
 doi = {10.1086/167587},
 journal = {\apj},
 month = {July},
 pages = {235},
 title = {{Detection of the O i 11287 Angstrom Line in the Seyfert 1 Galaxy I ZW 1}},
 volume = {342},
 year = {1989}
}

@article{Rupke_outflows_2005,
 adsurl = {https://ui.adsabs.harvard.edu/abs/2005ApJ...632..751R},
 archiveprefix = {arXiv},
 author = {{Rupke}, David S. and {Veilleux}, Sylvain and {Sanders}, D.~B.},
 doi = {10.1086/444451},
 eprint = {astro-ph/0507037},
 journal = {\apj},
 month = {October},
 number = {2},
 pages = {751-780},
 primaryclass = {astro-ph},
 title = {{Outflows in Active Galactic Nucleus/Starburst-Composite Ultraluminous Infrared Galaxies1,}},
 volume = {632},
 year = {2005}
}

@article{Sabhahit_VMS_2023,
 adsurl = {https://ui.adsabs.harvard.edu/abs/2023MNRAS.524.1529S},
 archiveprefix = {arXiv},
 author = {{Sabhahit}, Gautham N. and {Vink}, Jorick S. and {Sander}, Andreas A.~C. and {Higgins}, Erin R.},
 doi = {10.1093/mnras/stad1888},
 eprint = {2306.11785},
 journal = {\mnras},
 month = {September},
 number = {1},
 pages = {1529-1546},
 primaryclass = {astro-ph.SR},
 title = {{Very massive stars and pair-instability supernovae: mass-loss framework for low metallicity}},
 volume = {524},
 year = {2023}
}

@article{Sander_WR_2012,
 adsurl = {https://ui.adsabs.harvard.edu/abs/2012A&A...540A.144S},
 archiveprefix = {arXiv},
 author = {{Sander}, A. and {Hamann}, W. -R. and {Todt}, H.},
 doi = {10.1051/0004-6361/201117830},
 eid = {A144},
 eprint = {1201.6354},
 journal = {\aap},
 month = {April},
 pages = {A144},
 primaryclass = {astro-ph.SR},
 title = {{The Galactic WC stars. Stellar parameters from spectral analyses indicate a new evolutionary sequence}},
 volume = {540},
 year = {2012}
}

@article{Sanders_aurora_calibrations_2025,
 adsurl = {https://ui.adsabs.harvard.edu/abs/2025arXiv250810099S},
 archiveprefix = {arXiv},
 author = {{Sanders}, Ryan L. and {Shapley}, Alice E. and {Topping}, Michael W. and {Reddy}, Naveen A. and {Berg}, Danielle A. and {Khostovan}, Ali Ahmad and {Bouwens}, Rychard J. and {Brammer}, Gabriel and {Carnall}, Adam C. and {Cullen}, Fergus and {Dav{\'e}}, Romeel and {Dunlop}, James S. and {Ellis}, Richard S. and {F{\"o}rster Schreiber}, N.~M. and {Furlanetto}, Steven R. and {Glazebrook}, Karl and {Illingworth}, Garth D. and {Jones}, Tucker and {Kriek}, Mariska and {McLeod}, Derek J. and {McLure}, Ross J. and {Narayanan}, Desika and {Oesch}, Pascal A. and {Pahl}, Anthony J. and {Pettini}, Max and {Schaerer}, Daniel and {Stark}, Daniel P. and {Steidel}, Charles C. and {Tang}, Mengtao and {Clarke}, Leonardo and {Donnan}, Callum T. and {Kehoe}, Emily},
 doi = {10.48550/arXiv.2508.10099},
 eid = {arXiv:2508.10099},
 eprint = {2508.10099},
 journal = {arXiv e-prints},
 month = {August},
 pages = {arXiv:2508.10099},
 primaryclass = {astro-ph.GA},
 title = {{The AURORA Survey: High-Redshift Empirical Metallicity Calibrations from Electron Temperature Measurements at z=2-10}},
 year = {2025}
}

@article{sanders_calibrations_2023,
 adsurl = {https://ui.adsabs.harvard.edu/abs/2024ApJ...962...24S},
 archiveprefix = {arXiv},
 author = {{Sanders}, Ryan L. and {Shapley}, Alice E. and {Topping}, Michael W. and {Reddy}, Naveen A. and {Brammer}, Gabriel B.},
 doi = {10.3847/1538-4357/ad15fc},
 eid = {24},
 eprint = {2303.08149},
 journal = {\apj},
 month = {February},
 number = {1},
 pages = {24},
 primaryclass = {astro-ph.GA},
 title = {{Direct T $_{e}$-based Metallicities of z = 2{\textendash}9 Galaxies with JWST/NIRSpec: Empirical Metallicity Calibrations Applicable from Reionization to Cosmic Noon}},
 volume = {962},
 year = {2024}
}

@article{sanders_mosdef_2015,
 author = {Sanders, R. L. and Shapley, A. E. and Kriek, M. and Reddy, N. A. and Freeman, W. R. and Coil, A. L. and Siana, B. and Mobasher, B. and Shivaei, I. and Price, S. H. and de Groot, L.},
 doi = {10.1088/0004-637X/799/2/138},
 journal = {\apj},
 month = {February},
 pages = {138},
 title = {The {MOSDEF} {Survey}: {Mass}, {Metallicity}, and {Star}-formation {Rate} at z ∼ 2.3},
 volume = {799},
 year = {2015}
}

@article{sanders_mosdef_2016-1,
 author = {Sanders, R. L. and Shapley, A. E. and Kriek, M. and Reddy, N. A. and Freeman, W. R. and Coil, A. L. and Siana, B. and Mobasher, B. and Shivaei, I. and Price, S. H. and de Groot, L.},
 doi = {10.3847/2041-8205/825/2/L23},
 journal = {\apjl},
 month = {July},
 pages = {L23},
 title = {The {MOSDEF} {Survey}: {Detection} of [{O} {III}]λ4363 and the {Direct}-method {Oxygen} {Abundance} of a {Star}-forming {Galaxy} at z = 3.08},
 volume = {825},
 year = {2016}
}

@article{SawadaSuwa2023,
 author = {Sawada, Ryo and Suwa, Yudai},
 eprint = {arXiv:2301.03610},
 journal = {ApJ (submitted)},
 title = {Updating the $^{56}$Ni Problem in Core-collapse Supernova Explosion},
 year = {2023}
}

@article{Schaye_EAGLE_2015,
 adsurl = {https://ui.adsabs.harvard.edu/abs/2015MNRAS.446..521S},
 archiveprefix = {arXiv},
 author = {{Schaye}, Joop and {Crain}, Robert A. and {Bower}, Richard G. and {Furlong}, Michelle and {Schaller}, Matthieu and {Theuns}, Tom and {Dalla Vecchia}, Claudio and {Frenk}, Carlos S. and {McCarthy}, I.~G. and {Helly}, John C. and {Jenkins}, Adrian and {Rosas-Guevara}, Y.~M. and {White}, Simon D.~M. and {Baes}, Maarten and {Booth}, C.~M. and {Camps}, Peter and {Navarro}, Julio F. and {Qu}, Yan and {Rahmati}, Alireza and {Sawala}, Till and {Thomas}, Peter A. and {Trayford}, James},
 doi = {10.1093/mnras/stu2058},
 eprint = {1407.7040},
 journal = {\mnras},
 month = {January},
 number = {1},
 pages = {521-554},
 primaryclass = {astro-ph.GA},
 title = {{The EAGLE project: simulating the evolution and assembly of galaxies and their environments}},
 volume = {446},
 year = {2015}
}

@article{Scholte_EXCELS_2025,
 adsurl = {https://ui.adsabs.harvard.edu/abs/2025MNRAS.540.1800S},
 archiveprefix = {arXiv},
 author = {{Scholte}, D. and {Cullen}, F. and {Carnall}, A.~C. and {Arellano-C{\'o}rdova}, K.~Z. and {Stanton}, T.~M. and {Barrufet}, L. and {Begley}, R. and {Bondestam}, C. and {Donnan}, C.~T. and {Dunlop}, J.~S. and {Leung}, H.-H. and {McLeod}, D.~J. and {McLure}, R.~J. and {Moustakas}, J.~M. and {Pollock}, C.~L. and {Shapley}, A.~E. and {Stevenson}, S. and {Zou}, H.},
 doi = {10.1093/mnras/staf834},
 eid = {arXiv:2502.10499},
 eprint = {2502.10499},
 journal = {\mnras},
 month = {June},
 number = {2},
 pages = {1800-1826},
 primaryclass = {astro-ph.GA},
 title = {{The JWST EXCELS survey: probing strong-line diagnostics and the chemical evolution of galaxies over cosmic time using T$_{e}$-metallicities}},
 volume = {540},
 year = {2025}
}

@article{Scholtz_JADES_DR4_2025,
 adsurl = {https://ui.adsabs.harvard.edu/abs/2025arXiv251001034S},
 archiveprefix = {arXiv},
 author = {{Scholtz}, J. and {Carniani}, S. and {Parlanti}, E. and {D'Eugenio}, F. and {Curtis-Lake}, E. and {Jakobsen}, P. and {Bunker}, A.~J. and {Cameron}, A.~J. and {Arribas}, S. and {Baker}, W.~M. and {Charlot}, S. and {Chevellard}, J. and {Circosta}, C. and {Curti}, M. and {Duan}, Q. and {Eisenstein}, D.~J. and {Hainline}, K. and {Ji}, Z. and {Johnson}, B.~D. and {Jones}, G.~C. and {Kumari}, N. and {Maiolino}, R. and {Maseda}, M.~V. and {Perna}, M. and {P{\'e}rez-Gonz{\'a}lez}, P.~G. and {Rawle}, T. and {Rieke}, M. and {Rinaldi}, P. and {Robertson}, B. and {Saxena}, A. and {Shivaei}, I. and {Silcock}, M.~S. and {Sun}, Y. and {Rodr{\'\i}guez Del Pino}, B. and {Tacchella}, S. and {{\"U}bler}, H. and {Venturi}, G. and {Williams}, C.~C. and {Willmer}, C.~N.~A. and {Willott}, C. and {Witstok}, J.},
 doi = {10.48550/arXiv.2510.01034},
 eid = {arXiv:2510.01034},
 eprint = {2510.01034},
 journal = {arXiv e-prints},
 month = {October},
 pages = {arXiv:2510.01034},
 primaryclass = {astro-ph.GA},
 title = {{JADES Data Release 4 -- Paper II: Data reduction, analysis and emission-line fluxes of the complete spectroscopic sample}},
 year = {2025}
}

@article{Schwarz_BIC_1978,
 adsurl = {https://ui.adsabs.harvard.edu/abs/1978AnSta...6..461S},
 author = {{Schwarz}, Gideon},
 journal = {Annals of Statistics},
 month = {July},
 number = {2},
 pages = {461-464},
 title = {{Estimating the Dimension of a Model}},
 volume = {6},
 year = {1978}
}

@article{Shapley_AURORA_BPT_2025,
 adsurl = {https://ui.adsabs.harvard.edu/abs/2025ApJ...980..242S},
 archiveprefix = {arXiv},
 author = {{Shapley}, Alice E. and {Sanders}, Ryan L. and {Topping}, Michael W. and {Reddy}, Naveen A. and {Berg}, Danielle A. and {Bouwens}, Rychard J. and {Brammer}, Gabriel and {Carnall}, Adam C. and {Cullen}, Fergus and {Dav{\'e}}, Romeel and {Dunlop}, James S. and {Ellis}, Richard S. and {F{\"o}rster Schreiber}, N.~M. and {Furlanetto}, Steven R. and {Glazebrook}, Karl and {Illingworth}, Garth D. and {Jones}, Tucker and {Kriek}, Mariska and {McLeod}, Derek J. and {McLure}, Ross J. and {Narayanan}, Desika and {Oesch}, Pascal and {Pahl}, Anthony J. and {Pettini}, Max and {Schaerer}, Daniel and {Stark}, Daniel P. and {Steidel}, Charles C. and {Tang}, Mengtao and {Clarke}, Leonardo and {Donnan}, Callum T. and {Kehoe}, Emily},
 doi = {10.3847/1538-4357/adad68},
 eid = {242},
 eprint = {2407.00157},
 journal = {\apj},
 month = {February},
 number = {2},
 pages = {242},
 primaryclass = {astro-ph.GA},
 title = {{The AURORA Survey: A New Era of Emission-line Diagrams with JWST/NIRSpec}},
 volume = {980},
 year = {2025}
}

@article{shapley_mosdef_2015,
 author = {Shapley, A. E. and Reddy, N. A. and Kriek, M. and Freeman, W. R. and Sanders, R. L. and Siana, B. and Coil, A. L. and Mobasher, B. and Shivaei, I. and Price, S. H. and de Groot, L.},
 doi = {10.1088/0004-637X/801/2/88},
 journal = {\apj},
 month = {March},
 pages = {88},
 title = {The {MOSDEF} {Survey}: {Excitation} {Properties} of z \tilde 2.3 {Star}-forming {Galaxies}},
 volume = {801},
 year = {2015}
}

@article{shapley_physical_2011,
 author = {Shapley, A. E.},
 doi = {10.1146/annurev-astro-081710-102542},
 journal = {\araa},
 month = {September},
 pages = {525--580},
 title = {Physical {Properties} of {Galaxies} from z = 2-4},
 volume = {49},
 year = {2011}
}

@article{sommariva_stellar_2012,
 author = {Sommariva, V. and Mannucci, F. and Cresci, G. and Maiolino, R. and Marconi, A. and Nagao, T. and Baroni, A. and Grazian, A.},
 doi = {10.1051/0004-6361/201118134},
 journal = {\aap},
 month = {March},
 pages = {A136},
 title = {Stellar metallicity of star-forming galaxies at z ∼ 3},
 volume = {539},
 year = {2012}
}

@article{Stanton_excels_2025,
 adsurl = {https://ui.adsabs.harvard.edu/abs/2025MNRAS.537.1735S},
 archiveprefix = {arXiv},
 author = {{Stanton}, T.~M. and {Cullen}, F. and {Carnall}, A.~C. and {Scholte}, D. and {Arellano-C{\'o}rdova}, K.~Z. and {McLeod}, D.~J. and {Begley}, R. and {Donnan}, C.~T. and {Dunlop}, J.~S. and {Hamadouche}, M.~L. and {McLure}, R.~J. and {Shapley}, A.~E. and {Bondestam}, C. and {Stevenson}, S.},
 doi = {10.1093/mnras/staf106},
 eprint = {2411.11837},
 journal = {\mnras},
 month = {February},
 number = {2},
 pages = {1735-1748},
 primaryclass = {astro-ph.GA},
 title = {{The JWST EXCELS survey: tracing the chemical enrichment pathways of high-redshift star-forming galaxies with O, Ar, and Ne abundances}},
 volume = {537},
 year = {2025}
}

@article{Stanway_Eldrige_BPASS_2018,
 adsurl = {https://ui.adsabs.harvard.edu/abs/2018MNRAS.479...75S},
 archiveprefix = {arXiv},
 author = {{Stanway}, E.~R. and {Eldridge}, J.~J.},
 doi = {10.1093/mnras/sty1353},
 eprint = {1805.08784},
 journal = {\mnras},
 month = {September},
 number = {1},
 pages = {75-93},
 primaryclass = {astro-ph.GA},
 title = {{Re-evaluating old stellar populations}},
 volume = {479},
 year = {2018}
}

@article{steidel_lyman_2003,
 author = {Steidel, C. C. and Adelberger, K. L. and Shapley, A. E. and Pettini, M. and Dickinson, M. and Giavalisco, M.},
 doi = {10.1086/375772},
 journal = {\apj},
 month = {August},
 pages = {728--754},
 title = {Lyman {Break} {Galaxies} at {Redshift} z ∼ 3: {Survey} {Description} and {Full} {Data} {Set}},
 volume = {592},
 year = {2003}
}

@article{steidel_reconciling_2016,
 author = {Steidel, C. C. and Strom, A. L. and Pettini, M. and Rudie, G. C. and Reddy, N. A. and Trainor, R. F.},
 doi = {10.3847/0004-637X/826/2/159},
 journal = {\apj},
 month = {August},
 pages = {159},
 title = {Reconciling the {Stellar} and {Nebular} {Spectra} of {High}-redshift {Galaxies}},
 volume = {826},
 year = {2016}
}

@article{steidel_strong_2014,
 author = {Steidel, C. C. and Rudie, G. C. and Strom, A. L. and Pettini, M. and Reddy, N. A. and Shapley, A. E. and Trainor, R. F. and Erb, D. K. and Turner, M. L. and Konidaris, N. P. and Kulas, K. R. and Mace, G. and Matthews, K. and McLean, I. S.},
 doi = {10.1088/0004-637X/795/2/165},
 journal = {\apj},
 month = {November},
 pages = {165},
 title = {Strong {Nebular} {Line} {Ratios} in the {Spectra} of z ∼ 2-3 {Star} {Forming} {Galaxies}: {First} {Results} from {KBSS}-{MOSFIRE}},
 volume = {795},
 year = {2014}
}

@article{Stern_AGN_I_2012,
 adsurl = {https://ui.adsabs.harvard.edu/abs/2012MNRAS.423..600S},
 archiveprefix = {arXiv},
 author = {{Stern}, Jonathan and {Laor}, Ari},
 doi = {10.1111/j.1365-2966.2012.20901.x},
 eprint = {1203.3158},
 journal = {\mnras},
 month = {June},
 number = {1},
 pages = {600-631},
 primaryclass = {astro-ph.CO},
 title = {{Type 1 AGN at low z- I. Emission properties}},
 volume = {423},
 year = {2012}
}

@article{StricklandHeckman2009,
 author = {Strickland, David K. and Heckman, Timothy M.},
 doi = {10.1088/0004-637X/697/2/2030},
 eprint = {arXiv:0903.4175},
 journal = {ApJ},
 number = {2},
 pages = {2030--2056},
 title = {Supernova Feedback Efficiency and Mass Loading in the Starburst and Galactic Superwind Exemplar M82},
 volume = {697},
 year = {2009}
}

@article{Strom_Cecilia_2023,
 adsurl = {https://ui.adsabs.harvard.edu/abs/2023ApJ...958L..11S},
 archiveprefix = {arXiv},
 author = {{Strom}, Allison L. and {Rudie}, Gwen C. and {Trainor}, Ryan F. and {Brammer}, Gabriel B. and {Maseda}, Michael V. and {Raptis}, Menelaos and {Rogers}, Noah S.~J. and {Steidel}, Charles C. and {Chen}, Yuguang and {Law}, David R.},
 doi = {10.3847/2041-8213/ad07dc},
 eid = {L11},
 eprint = {2308.13508},
 journal = {\apjl},
 month = {November},
 number = {1},
 pages = {L11},
 primaryclass = {astro-ph.GA},
 title = {{CECILIA: The Faint Emission Line Spectrum of z   2-3 Star-forming Galaxies}},
 volume = {958},
 year = {2023}
}

@article{strom_measuring_2018,
 adsurl = {https://ui.adsabs.harvard.edu/abs/2018ApJ...868..117S},
 archiveprefix = {arXiv},
 author = {{Strom}, Allison L. and {Steidel}, Charles C. and {Rudie}, Gwen C. and {Trainor}, Ryan F. and {Pettini}, Max},
 doi = {10.3847/1538-4357/aae1a5},
 eid = {117},
 eprint = {1711.08820},
 journal = {\apj},
 month = {December},
 number = {2},
 pages = {117},
 primaryclass = {astro-ph.GA},
 title = {{Measuring the Physical Conditions in High-redshift Star-forming Galaxies: Insights from KBSS-MOSFIRE}},
 volume = {868},
 year = {2018}
}

@article{tominaga_2007,
 adsurl = {https://ui.adsabs.harvard.edu/abs/2007ApJ...660..516T},
 archiveprefix = {arXiv},
 author = {{Tominaga}, Nozomu and {Umeda}, Hideyuki and {Nomoto}, Ken'ichi},
 doi = {10.1086/513063},
 eprint = {astro-ph/0701381},
 journal = {\apj},
 month = {May},
 number = {1},
 pages = {516-540},
 primaryclass = {astro-ph},
 title = {{Supernova Nucleosynthesis in Population III 13-50 M$_{solar}$ Stars and Abundance Patterns of Extremely Metal-poor Stars}},
 volume = {660},
 year = {2007}
}

@article{topping_mosdef-lris_2020_i,
 adsurl = {https://ui.adsabs.harvard.edu/abs/2020MNRAS.495.4430T},
 archiveprefix = {arXiv},
 author = {{Topping}, Michael W. and {Shapley}, Alice E. and {Reddy}, Naveen A. and {Sanders}, Ryan L. and {Coil}, Alison L. and {Kriek}, Mariska and {Mobasher}, Bahram and {Siana}, Brian},
 doi = {10.1093/mnras/staa1410},
 eprint = {1912.10243},
 journal = {\mnras},
 month = {July},
 number = {4},
 pages = {4430-4444},
 primaryclass = {astro-ph.GA},
 title = {{The MOSDEF-LRIS Survey: the interplay between massive stars and ionized gas in high-redshift star-forming galaxies}},
 volume = {495},
 year = {2020}
}

@article{topping_mosdef-lris_2020_ii,
 author = {Topping, Michael W. and Shapley, Alice E. and Reddy, Naveen A. and Sanders, Ryan L. and Coil, Alison L. and Kriek, Mariska and Mobasher, Bahram and Siana, Brian},
 journal = {arXiv:2008.02282 [astro-ph]},
 language = {en},
 month = {August},
 note = {arXiv: 2008.02282},
 shorttitle = {The {MOSDEF}-{LRIS} {Survey}},
 title = {The {MOSDEF}-{LRIS} {Survey}: {The} {Connection} {Between} {Massive} {Stars} and {Ionized} {Gas} in {Individual} {Galaxies} at \$z\sim2\$},
 url = {http://arxiv.org/abs/2008.02282},
 urldate = {2020-08-17},
 year = {2020}
}

@article{topping_z6_lens_2024,
 adsurl = {https://ui.adsabs.harvard.edu/abs/2024MNRAS.529.3301T},
 archiveprefix = {arXiv},
 author = {{Topping}, Michael W. and {Stark}, Daniel P. and {Senchyna}, Peter and {Plat}, Adele and {Zitrin}, Adi and {Endsley}, Ryan and {Charlot}, St{\'e}phane and {Furtak}, Lukas J. and {Maseda}, Michael V. and {Smit}, Renske and {Mainali}, Ramesh and {Chevallard}, Jacopo and {Molyneux}, Stephen and {Rigby}, Jane R.},
 doi = {10.1093/mnras/stae682},
 eprint = {2401.08764},
 journal = {\mnras},
 month = {April},
 number = {4},
 pages = {3301-3322},
 primaryclass = {astro-ph.GA},
 title = {{Metal-poor star formation at z > 6 with JWST: new insight into hard radiation fields and nitrogen enrichment on 20 pc scales}},
 volume = {529},
 year = {2024}
}

@article{Tramper_WR-WO_2015,
 adsurl = {https://ui.adsabs.harvard.edu/abs/2015A&A...581A.110T},
 archiveprefix = {arXiv},
 author = {{Tramper}, F. and {Straal}, S.~M. and {Sanyal}, D. and {Sana}, H. and {de Koter}, A. and {Gr{\"a}fener}, G. and {Langer}, N. and {Vink}, J.~S. and {de Mink}, S.~E. and {Kaper}, L.},
 doi = {10.1051/0004-6361/201425390},
 eid = {A110},
 eprint = {1507.00839},
 journal = {\aap},
 month = {September},
 pages = {A110},
 primaryclass = {astro-ph.SR},
 title = {{Massive stars on the verge of exploding: the properties of oxygen sequence Wolf-Rayet stars}},
 volume = {581},
 year = {2015}
}

@article{TumlinsonPeeplesWerk2017,
 author = {Tumlinson, Jason and Peeples, Molly S. and Werk, Jessica K.},
 doi = {10.1146/annurev-astro-091916-055240},
 eprint = {arXiv:1709.09180},
 journal = {Annual Review of Astronomy and Astrophysics},
 pages = {389--432},
 title = {The Circumgalactic Medium},
 volume = {55},
 year = {2017}
}

@article{vale_asari_bond_2016,
 author = {Vale Asari, N. and Stasińska, G. and Morisset, C. and Cid Fernandes, R.},
 doi = {10.1093/mnras/stw971},
 journal = {\mnras},
 month = {August},
 pages = {1739--1757},
 title = {{BOND}: {Bayesian} {Oxygen} and {Nitrogen} abundance {Determinations} in giant {H} {II} regions using strong and semistrong lines},
 volume = {460},
 year = {2016}
}

@inproceedings{vink_vms_2024,
 adsurl = {https://ui.adsabs.harvard.edu/abs/2025IAUS..391..106V},
 archiveprefix = {arXiv},
 author = {{Vink}, Jorick S.},
 booktitle = {IAU Symposium},
 doi = {10.1017/S1743921324002539},
 editor = {{Nanayakkara}, T. and {Maseda}, M.},
 eid = {arXiv:2410.18980},
 eprint = {2410.18980},
 journal = {arXiv e-prints},
 month = {December},
 pages = {106-111},
 primaryclass = {astro-ph.GA},
 series = {IAU Symposium},
 title = {{Very massive stars and Nitrogen-emitting galaxies}},
 volume = {19},
 year = {2025}
}

@article{Vink_WR_massloss_2011,
 adsurl = {https://ui.adsabs.harvard.edu/abs/2011A&A...531A.132V},
 archiveprefix = {arXiv},
 author = {{Vink}, Jorick S. and {Muijres}, L.~E. and {Anthonisse}, B. and {de Koter}, A. and {Gr{\"a}fener}, G. and {Langer}, N.},
 doi = {10.1051/0004-6361/201116614},
 eid = {A132},
 eprint = {1105.0556},
 journal = {\aap},
 month = {July},
 pages = {A132},
 primaryclass = {astro-ph.SR},
 title = {{Wind modelling of very massive stars up to 300 solar masses}},
 volume = {531},
 year = {2011}
}

@article{Vink_WR_massloss_2012,
 adsurl = {https://ui.adsabs.harvard.edu/abs/2012ApJ...751L..34V},
 archiveprefix = {arXiv},
 author = {{Vink}, Jorick S. and {Gr{\"a}fener}, G{\"o}tz},
 doi = {10.1088/2041-8205/751/2/L34},
 eid = {L34},
 eprint = {1205.0394},
 journal = {\apjl},
 month = {June},
 number = {2},
 pages = {L34},
 primaryclass = {astro-ph.SR},
 title = {{The Transition Mass-loss Rate: Calibrating the Role of Line-driven Winds in Massive Star Evolution}},
 volume = {751},
 year = {2012}
}

@article{Waxman_dust_sublimation_2000,
 adsurl = {https://ui.adsabs.harvard.edu/abs/2000ApJ...537..796W},
 archiveprefix = {arXiv},
 author = {{Waxman}, E. and {Draine}, B.~T.},
 doi = {10.1086/309053},
 eprint = {astro-ph/9909020},
 journal = {\apj},
 month = {July},
 number = {2},
 pages = {796-802},
 primaryclass = {astro-ph},
 title = {{Dust Sublimation by Gamma-ray Bursts and Its Implications}},
 volume = {537},
 year = {2000}
}

@article{Webbink1984,
 adsurl = {https://ui.adsabs.harvard.edu/abs/1984ApJ...277..355W},
 author = {{Webbink}, R.~F.},
 doi = {10.1086/161701},
 journal = {\apj},
 month = {February},
 pages = {355-360},
 title = {{Double white dwarfs as progenitors of R Coronae Borealis stars and type I supernovae.}},
 volume = {277},
 year = {1984}
}

@article{Welch_Sunburst_arc_2025,
 adsurl = {https://ui.adsabs.harvard.edu/abs/2025ApJ...980...33W},
 archiveprefix = {arXiv},
 author = {{Welch}, Brian and {Rivera-Thorsen}, T. Emil and {Rigby}, Jane R. and {Hutchison}, Taylor A. and {Olivier}, Grace M. and {Berg}, Danielle A. and {Sharon}, Keren and {Dahle}, H{\r{a}}kon and {Owens}, M. Riley and {Bayliss}, Matthew B. and {Khullar}, Gourav and {Chisholm}, John and {Hayes}, Matthew and {Kim}, Keunho J.},
 doi = {10.3847/1538-4357/ada76c},
 eid = {33},
 eprint = {2405.06631},
 journal = {\apj},
 month = {February},
 number = {1},
 pages = {33},
 primaryclass = {astro-ph.GA},
 title = {{The Sunburst Arc with JWST. III. An Abundance of Direct Chemical Abundances}},
 volume = {980},
 year = {2025}
}

@article{WoosleyHeger2007,
 author = {Woosley, S. E. and Heger, A.},
 doi = {10.1016/j.physrep.2007.02.009},
 journal = {Phys. Rep.},
 pages = {269-283},
 title = {Nucleosynthesis and Remnants in Massive Stars of Solar Metallicity},
 volume = {442},
 year = {2007}
}
\bibliographystyle{aa}

\begin{appendix}

\section{Comparing synthetic to in-slit photometry}
\label{sec:appendix_A}

\begin{figure*}[!h]
    \centering
    \includegraphics[width=0.99\textwidth]{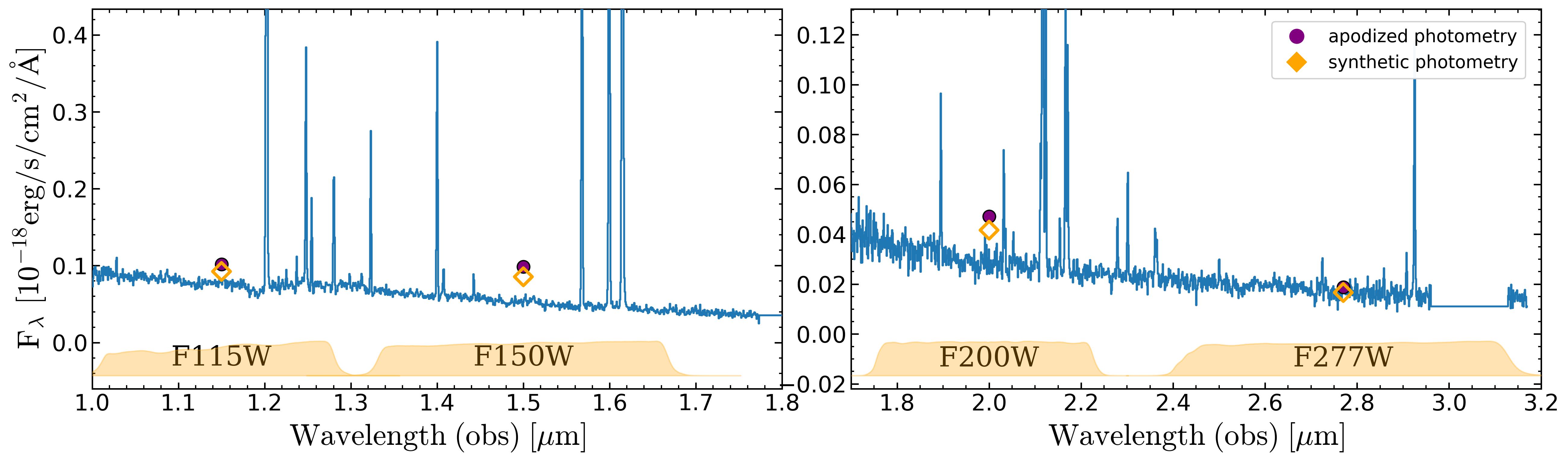}
    \caption{Comparison between synthetic and apodized (in-slit) photometry.
    The synthetic photometry is obtained from the NIRSpec spectrum of \sourceshort as delivered by the data reduction pipeline (with point-source-like path-losses correction) by integrating over the NIRCam F115W, F150W, F200W, and F277W bandpasses. The apodized photometry is instead measured from the area of the NIRCam images in the same filters that overlaps with the central shutter of the NIRSpec slitlet (i.e., the region from where the 1D spectrum is extracted).}
    \label{fig:compare_phot}
\end{figure*}

As discussed in Sect.~\ref{sec:data}, and more extensively in \cite{Cataldi_MARTA_2025}, at the data processing stage MARTA galaxies are corrected for path-losses assuming a point-like source shape.
Although in many cases such an assumption works well for (clumpy) high-z systems than assuming a uniformly illuminated slit, galaxies at Cosmic Noon often show extended and resolved morphologies, and deviations from the point-like correction curve can impact the measured emission line ratios, especially for lines widely spaced in wavelength. 
At the same time, to correct the extracted spectrum to match the full galaxy photometry, including regions not probed by the NIRSpec slit and possibly characterized by different color gradients, might introduce unknown systematics in the analysis.
Therefore, we have tested the consistency between the flux calibrated NIRSpec spectrum of \sourceshort in both G140M/F100LP and G235M/F170LP gratings/filters and the photometry extracted from the region of the galaxy that is `sampled' by the same area of the slit from which the spectrum is extracted, i.e., in our case the central shutter.

We computed the synthetic photometry from the spectrum by applying the bandpass of NIRCAM F115W, F150W, F200W, and F277W filters, i.e., the available broadband filters from the PRIMER programme \citep{Dunlop_Primer_proposal_2021} that overlaps with our NIRSpec  grating/filter configurations. 
To extract the `in-slit' photometry instead, we first generated a segmentation map from each NIRCAM image, and identified the overlapping area  between the segmentation maps and the overlaid central shutter of the NIRSpec slit, at sub-pixel precision.
We then extracted the photometry for each filter weighting each image pixel by the fraction of its area that is intercepted by the central NIRSpec shutter.

The comparison between the reduced NIRSpec spectrum of \sourceshort (as delivered by the pipeline with point-source path-losses corrections), the derived synthetic photometry, and the in-slit photometry extracted from F115W, F150W, F200W, and F277W images is shown in Figure~\ref{fig:compare_phot}. The two measurements appear in good agreement, and only a minor correction to the extracted 1D spectrum is required to match the synthetic to the in-slit photometric points, with a small wavelength dependence.
For this purpose, we derived the correction as a function of wavelength by interpolating the ratio between the two photometric measurements at the four available pivot wavelengths across the full wavelength array, which we then apply to the 1D NIRSpec spectrum.

\section{More details on the derivation of physical parameters}
\label{sec:appendix_B}

\subsection{Temperature, density, and \Av}
\label{ssec:appendix_B_te_ne_av}

\begin{figure}[!h]
    \centering
    \includegraphics[width=0.99\columnwidth]{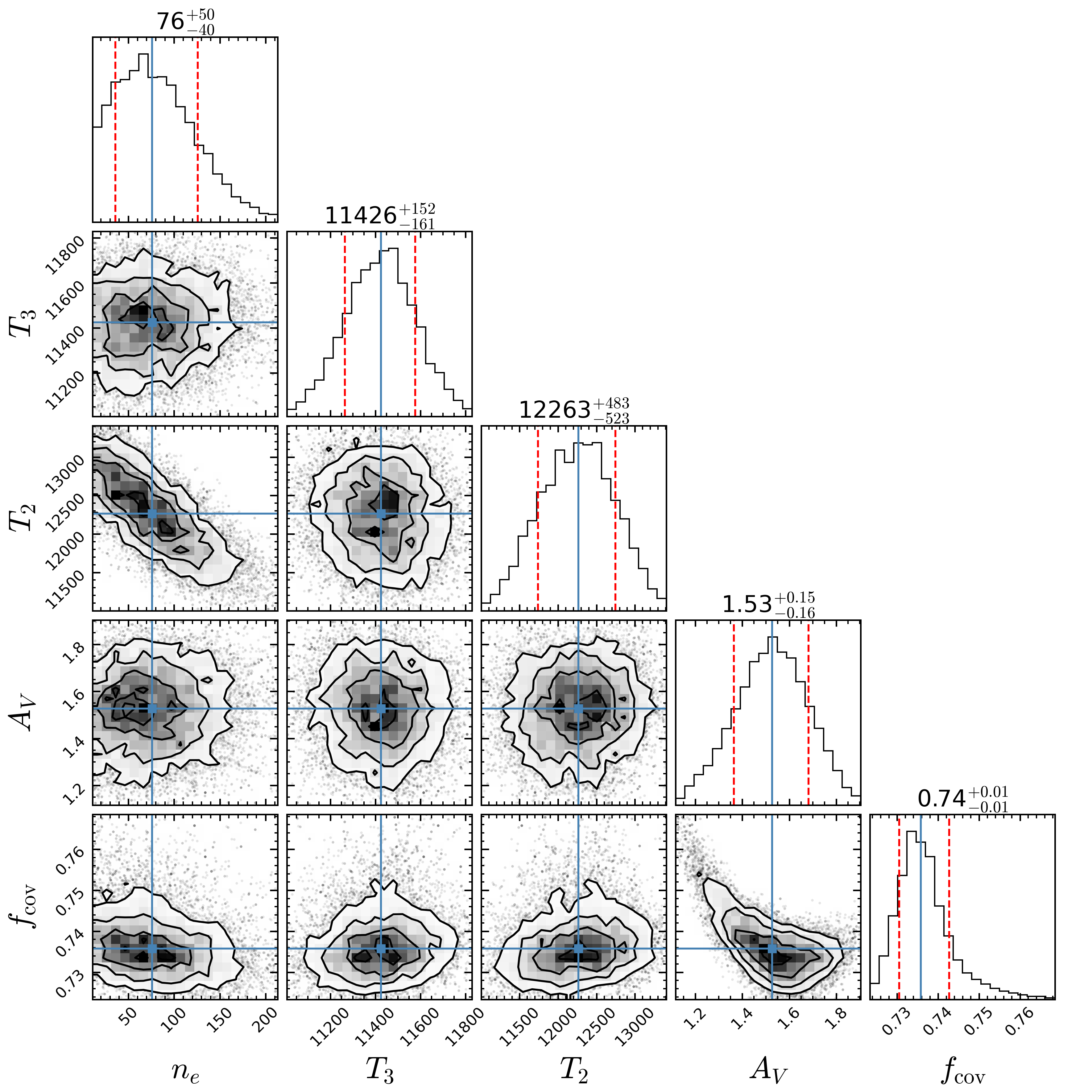} \\
    \includegraphics[width=0.99\columnwidth]{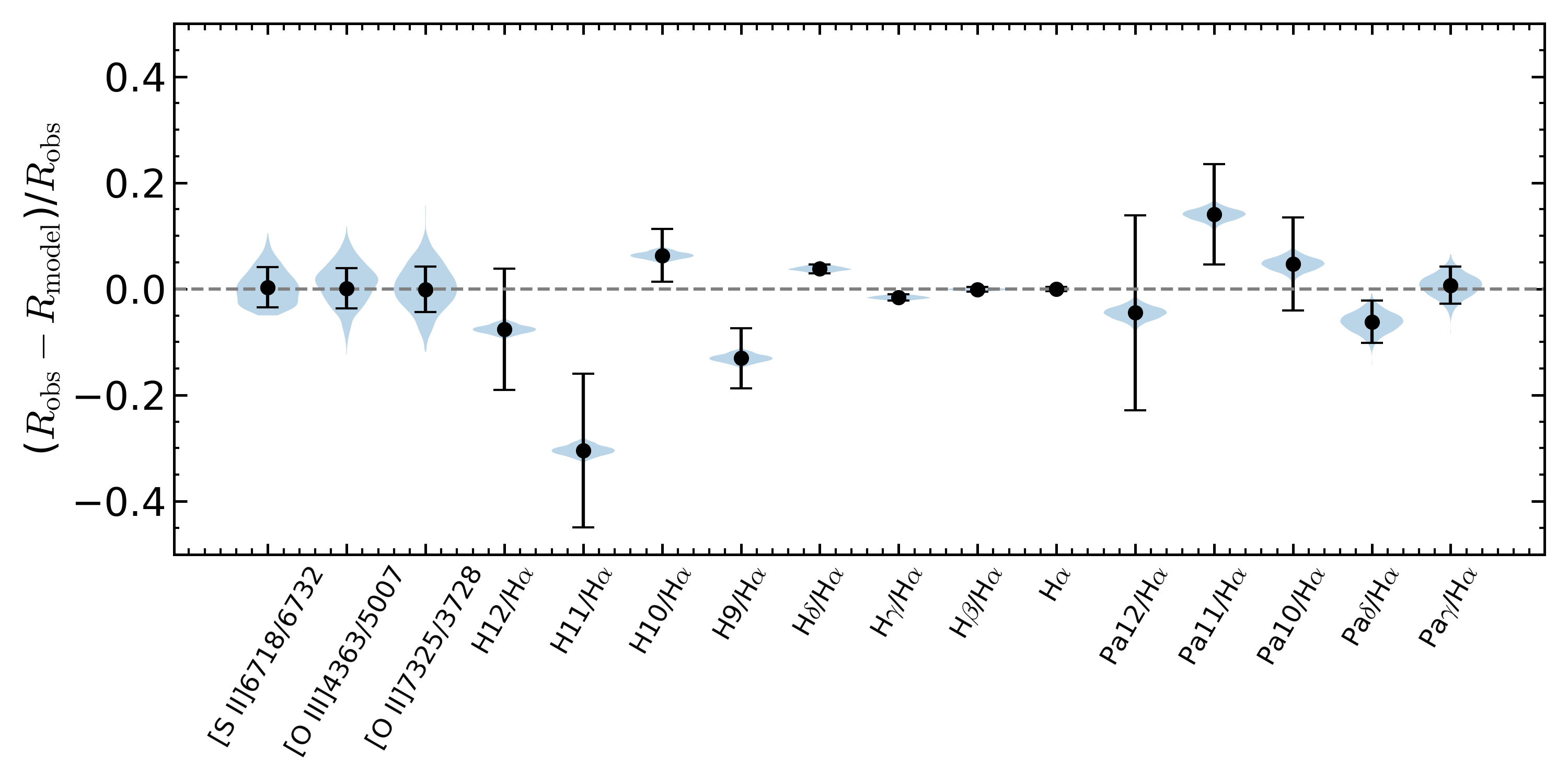}
    \caption{Results of the MCMC sampling analysis for the derivation of \Te, \Ne, and \Av. The top panel reports the corner plots with 2D and 1D marginalized posterior distributions for the free parameters in our model, namely \Ne, $T_2$, $T_3$, \Av, and $f_{cov}$. 
    The PDF medians, $16$th, and $84$th percentiles are marked.
    In the bottom panel, we compare the observed to modeled line ratios. Black points with errorbars represent observed values, whereas blue violin plots show the full posterior distribution for each line ratio as sampled from the MCMC chain.
    }
    \label{fig:MCMC_Te_ne_AV_fiducial}
\end{figure}

Here, we show the results of the MCMC analysis described in Sect.~\ref{ssec:te_ne_av}. 
Fig.~\ref{fig:MCMC_Te_ne_AV_fiducial} reports the marginalized posterior probability distribution function (PDF) for our parameters, and we mark the median values (taken as fiducial estimates) together with the $16$th-84$th$ percentiles (taken as formal uncertainties) of each distribution.

In the bottom panel of the same Figure, we compare the observed and predicted line ratios, the latter modeled based on the fiducial values inferred for the set of parameters explored by the MCMC analysis. Black points and errorbars mark observed values and their uncertainties, whereas the blue shaded areas span the range of values explored for each line ratio by the MCMC sampler.
Most line ratios are reproduced with excellent accuracy and within their $1\sigma$ uncertainty, with more significant deviations only for \Hdelta, H9, H11, and Pa11.

\subsection{Metal abundances}
\label{ssec:appendix_B_abundances}

\begin{figure}[!h]
    \centering
    \includegraphics[width=0.99\columnwidth]{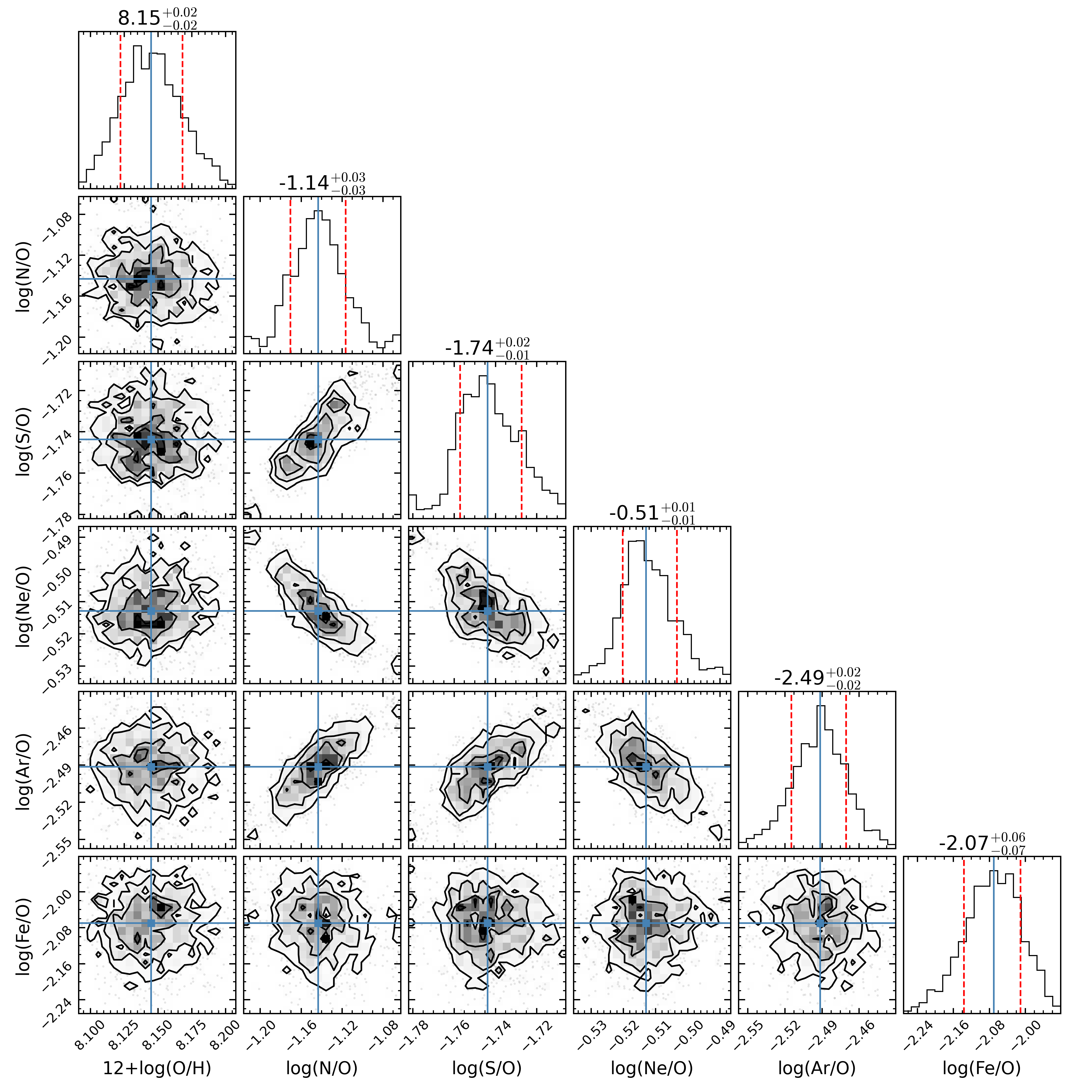}
    \caption{Corner plots and marginalized PDF for different relative gas-phase elemental abundances of metal ions, as obtained by post-processing the full MCMC chain generated in the derivation of \Te, \Ne, and \Av as discussed in Sect.~\ref{ssec:abundances}.
    The medians, $16$th, and $84$th percentiles of the marginalized PDFs  are marked.}
    \label{fig:MCMC_abundances_fiducial}
\end{figure}

Here we report the final corner plots and marginalized posterior distributions for elemental abundances derived as described in Sect.~\ref{ssec:abundances}: these are shown in Fig.~\ref{fig:MCMC_abundances_fiducial}.

\subsection{Helium abundance}
\label{ssec:appendix_B_helium}

In Sect.~\ref{ssec:helium_abund} we have described the approach followed to measure the helium abundance in \sourceshort.
Here, we provide more details on the implemented emission line modeling of Eq.~(3).
To account for the contribution of collisional excitation to the emissivity of the hydrogen and helium lines, we followed the methodology outlined in \cite{Aver_He_2010} and implemented as in \cite{Hsyu_PLEK_2020, Aver_Helium_LeoP_2021} (see, e.g., equation B.5 of the latter), where the collisional-to-recombination correction factor $\frac{C}{R} (\lambda)$ is derived as a function of the collisional transition rate from the ground state to the level $i$ above the level of interest $j$ (with transition from $j \rightarrow 2$ emitted at wavelength $\lambda$), the various possible branching paths for transitioning to \textit{j} and their relative fraction $BR_{i \rightarrow j}$, and the ratio between the (residual) neutral and ionized hydrogen densities $\xi = n(\text{H} \rm I) / n (\text{H} \rm II)$.
Ultimately, $\frac{C}{R} (\lambda)$ is expressed as a function of $\xi$ and electron temperature as
\begin{equation}
\frac{C}{R} (\lambda) = \xi \cdot 10^4 \times \sum_{i} a_{i}~\text{exp}(-\frac{b_{i}}{T_{4}})T_4^{c_{i}} \, 
\end{equation}
where $T{_4}$ is the temperature in units of $10^4$~K and the coefficients $a_i$, $b_i$, $c_i$ for the different hydrogen lines are taken from Table 8 of \cite{Aver_Helium_LeoP_2021}, noting that the contribution of collisional excitation to helium lines emissivity is already included in our assumed set of atomic parameters from \cite{Porter_He_emiss_2013}.
We also note that, for small residual fractions of neutral hydrogen as typical in H II regions ($\xi \approx 10^{-4}$), the relative contribution to hydrogen emissivity is predicted to be low, being e.g. larger than $1\%$ for \Halpha only at temperatures T~$\gtrsim 18,000$~K (see e.g. Figure~2 in \citealt{Hsyu_PLEK_2020}). Therefore, we do not expect such a contribution to be highly relevant in our case study\footnote{We also note that temperature has a competing effect here in the modeling of Balmer line ratios as, on the one hand, lower temperatures implies a higher intrinsic \Halpha/\Hbeta (for the Case B) at fixed \AV (and vice versa), whereas on the other hand to boost the predicted \Halpha/\Hbeta ratio via collisional excitation one needs to substantially increase \Te at fixed \Av and $\xi$.}.
The optical depth correction term $f$(\Ne, T, $\tau$), normalised to the value at He I$\lambda 3889$, is instead modeled as 
\begin{equation}
\label{eq:f_tau}
    f_{\lambda}(n_{e}, T, \tau) = 1 + \frac{\tau}{2} \ [a + (b_0 + b_1 n_{e} + b_2 n_{e}^{2})\ (T/10^4~K)] \,
\end{equation}
with coefficients taken from \cite{Olive_Skillman_prim_He_2004}.

\begin{figure}[!h]
    \centering
    \includegraphics[width=0.95\columnwidth]{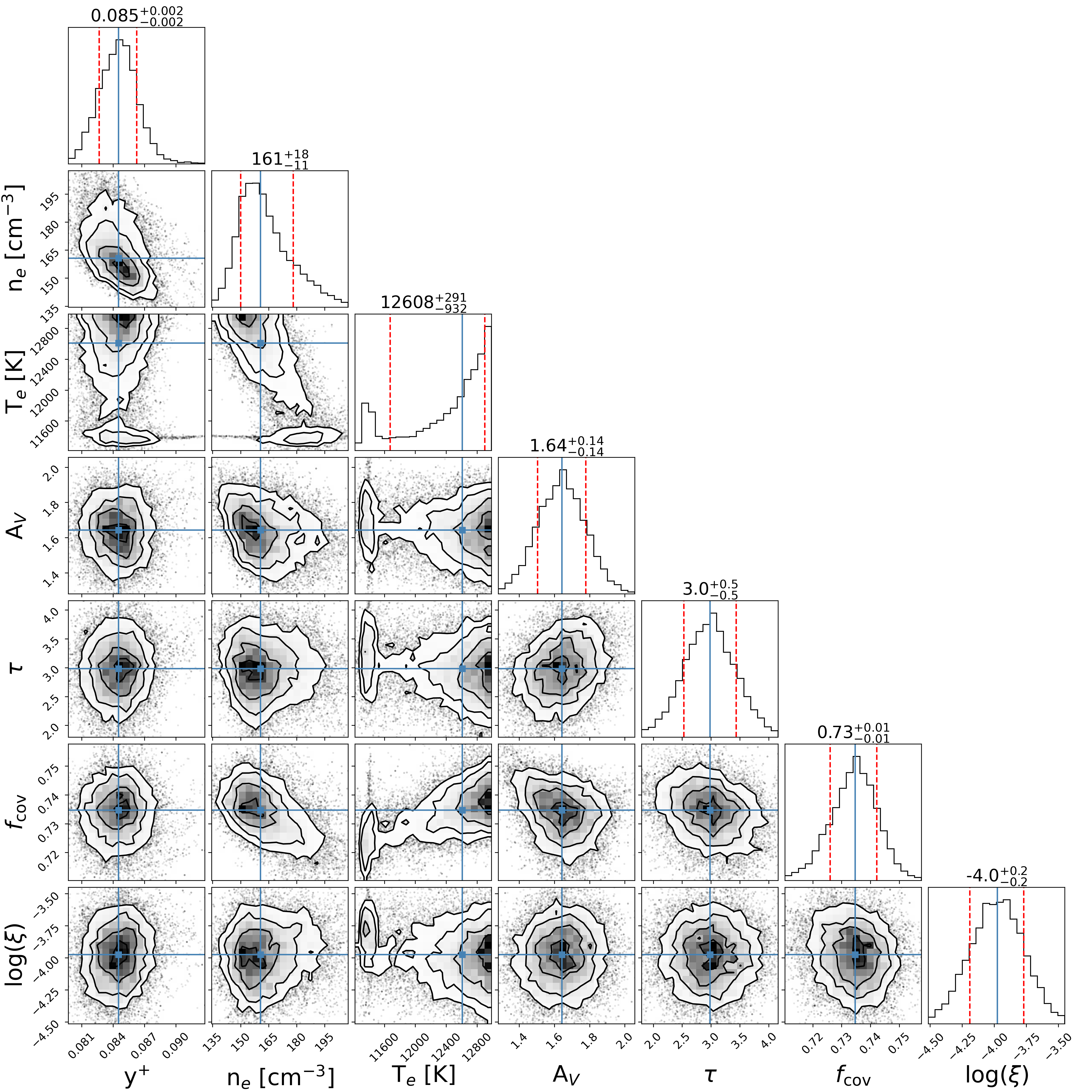}
    \includegraphics[width=0.95\columnwidth]{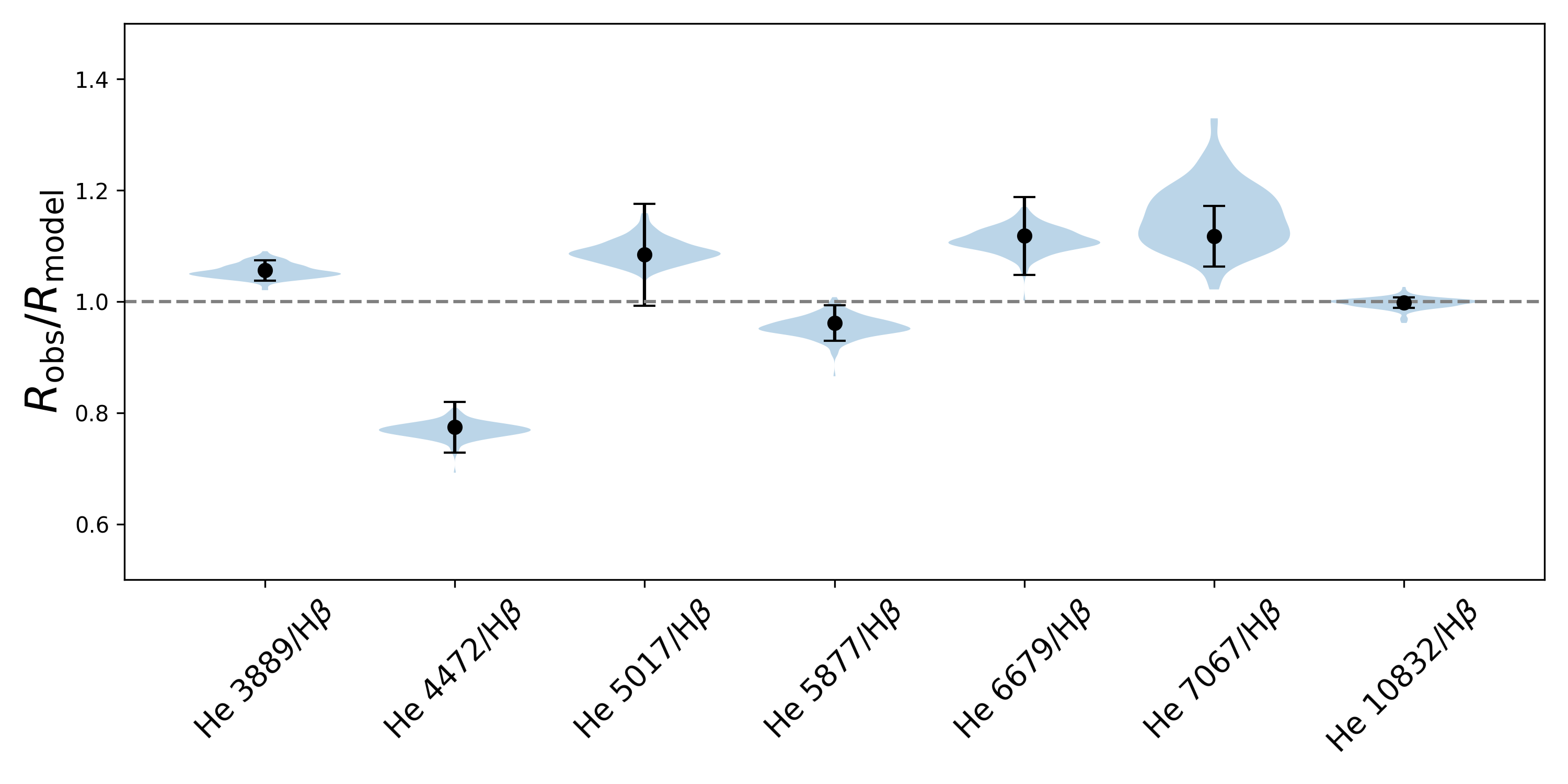}
    \caption{Results of the MCMC run for He abundance derivation (y$^{+}$). \textit{Top panel:} Corner plots and marginalized PDF for the parameters in our model. Compared to the framework adopted in Sect.~\ref{ssec:te_ne_av}, here we included corrections for the optical depth $\tau$ and for the collisional excitation of hydrogen and helium lines (expressed as a function of temperature and the residual fraction of neutral hydrogen $\xi$) as additional terms, as discussed in the text of Appendix~\ref{sec:appendix_B}.
    \textit{Bottom panel:} Comparison between modeled and observed He-to-\Hbeta line ratios included in our fit.
    }
    \label{fig:He_MCMC}
\end{figure}

Beyond \Te, \Ne, \AV, and $f_{cov}$, the free parameters in our model are therefore y$^{+}$ and $\tau$, for which we impose uniform priors within 0.05 < y$^{+}$ < 0.15 and 0 < $\tau$ < 10, respectively, and $\xi$, for which we assume a Gaussian prior centered on log($\xi$)=$-4$ and with $\sigma=0.5$ to prevent the exploration of regions of the parameter space non compatible with the typical ionization conditions of HII regions (e.g. with residual neutral fractions much higher than 1\%).
To model the temperature of the helium emitting zone, we impose a weak Gaussian prior centered on the independent measurement available from the \OIIIoptL[4363]/5007 ratio, and assuming a conservative 20\% uncertainty \citep[e.g.][]{Aver_He_mcmc_2011}.
For the electron density instead, we leave \Ne free to vary to exploit the sensitivity of the HeI$\lambda10832$ to such parameter, with no imposed priors based on the sulfur doublet line ratio \citep[see e.g.] [for a discussion on the effects of imposing such a prior in the He abundance derivation]{berg_aurora_helium_2025}.
The corner plots resulting from our MCMC run are displayed in Figure~\ref{fig:He_MCMC}, whereas the results are discussed in Sect.~\ref{ssec:helium_abund} of the main text.

Aside helium abundance derivation, we note that in the \sourceshort spectrum we have access to helium line ratios sensitive to electron temperature, i.e., He I $\lambda$7281/6678 and $\lambda$7281/5876, providing complementary information to auroral lines of metals.
Following \cite{Mendez-Delgado_Helium_temp_2024}, we interpolated over a fine (10~K) grid of temperature versus emissivity computed with the \textsc{pyneb.getEmissivity} routine, and compared predictions with observed line ratios to infer the temperature values.
We measure \Te(He I $\lambda$7281/6678) = 9828 $\pm$ 4118~K, and \Te(He I $\lambda$7281/5876) = 10002 $\pm$ 4619~K, and we take the average \Te(He I) = 9914 $\pm$ 3073~K as our fiducial estimate.
This value is consistent (within the uncertainties) with that derived from the auroral lines of oxygen, suggesting no significant (or only minor) temperature fluctuations nor strong deviations from Case B recombination, at least in the region probed by (highly) ionized gas \citep[but see the discussion in][where discrepancies between the two temperatures are observed in large samples of nearby star-forming galaxies and planetary nebulae]{Mendez-Delgado_Helium_temp_2024}.

\section{Alternative fits to the blue and red bumps}
\label{sec:appendix_C}

\begin{figure*}
    \centering
    \includegraphics[width=0.98\textwidth]{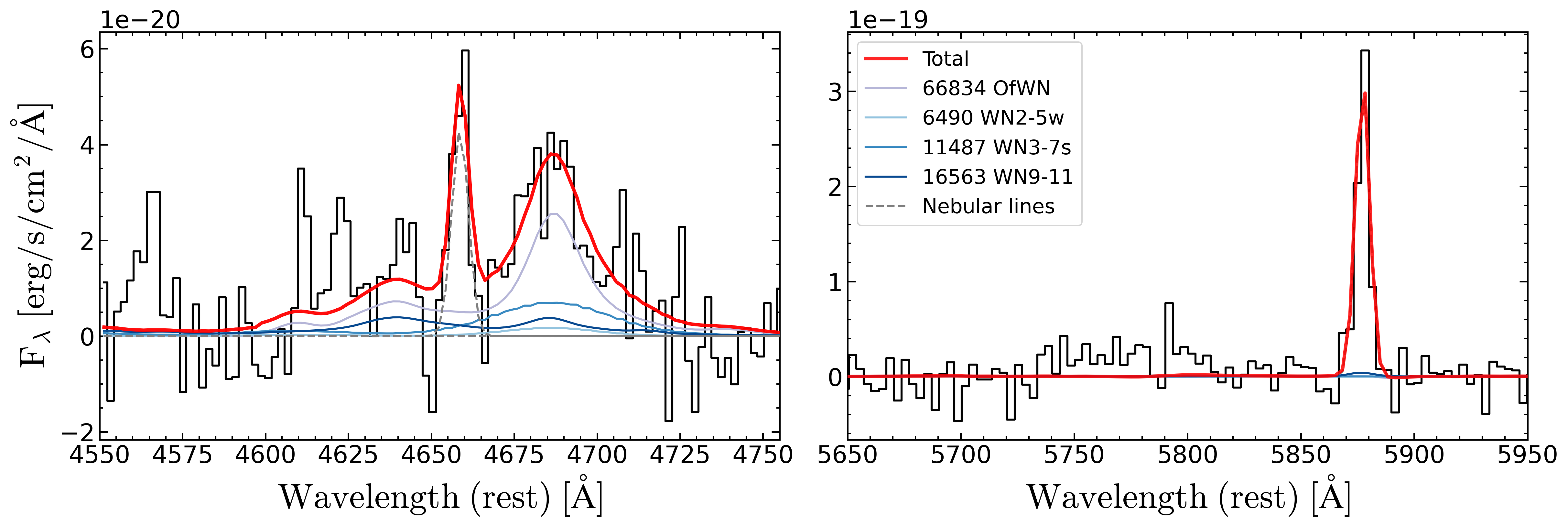}\\
    \includegraphics[width=0.98\textwidth]{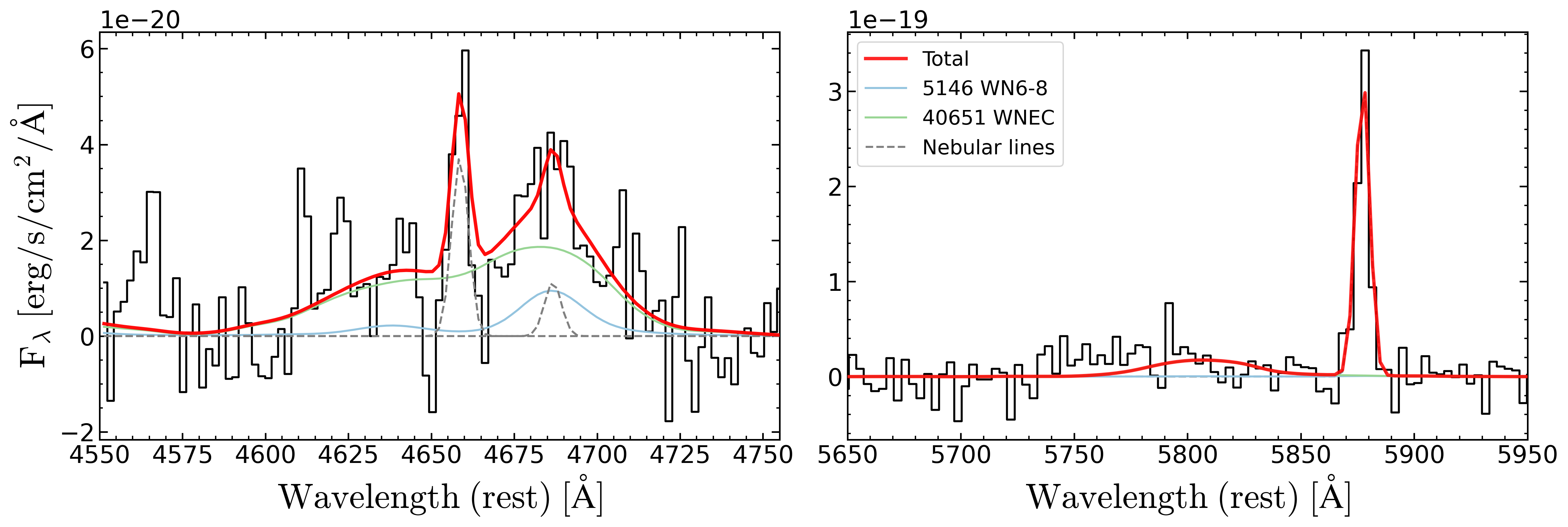}\\   
    \includegraphics[width=0.98\textwidth]{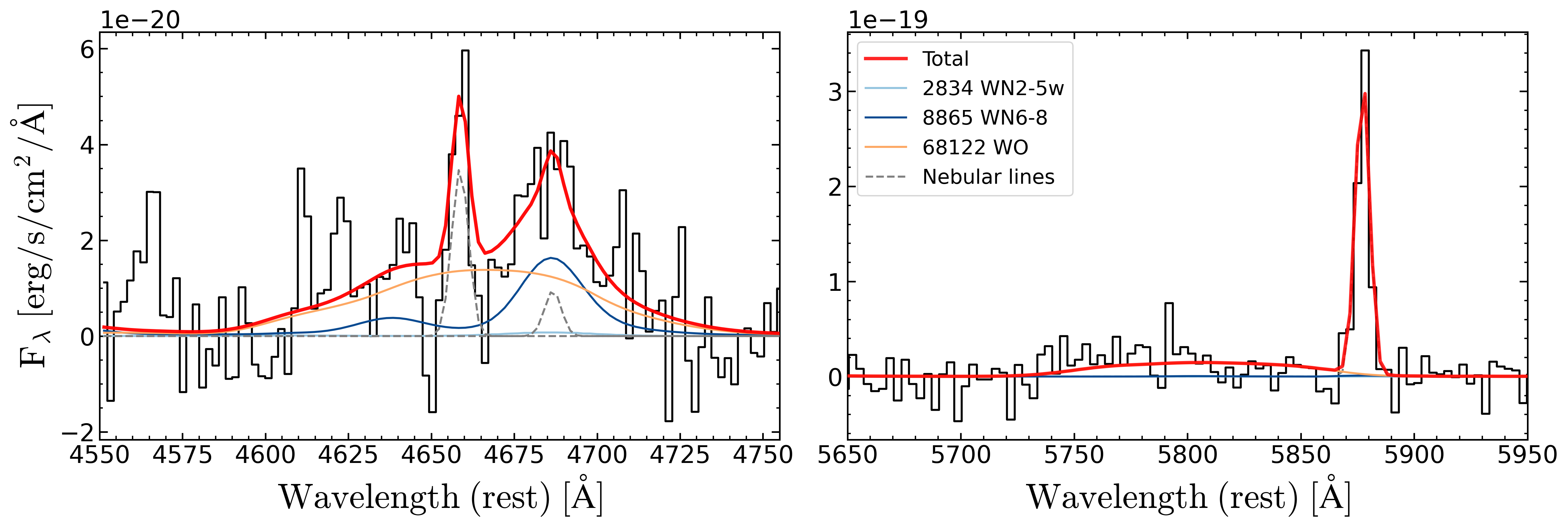}\\   
    \caption{Alternative fits to the blue and red bump in \sourceshort.
    The best-fit models to the combined spectral regions hosting the blue and red bumps are built from a combination of LMC empirical templates from \citet{Crowther_WR_LMC_2023}. While all models contain WN stars, each row differs for the set of complementary templates adopted, namely Of/WN, WNE/C and WNE/L, and WO for the upper, middle, and bottom panels, respectively.
     }
    \label{fig:alternative_WR_fits}
\end{figure*}

As discussed in Sect.~\ref{sssec:WR_bumps}, we attempted to fit the spectral regions hosting the blue and red bumps by complementing WN templates with a combination of templates of transition WR stars, namely OfWN, WNE/C, and WNE/L, as well as stars of the oxygen sequence (WO). 
In Fig.~\ref{fig:alternative_WR_fits} we plot the results of this fitting exercise.

We find that a combination of Of/WN (hydrogen-rich) and early-type WN templates (WN2-5 and WN3-7) produces a match to the blue bump as good as that delivered by our fiducial combination of early- and late-type WN stars , whereas WNE/C and WNE/L types provide results similar to including WC templates only. 
Despite the very large number ($\sim85,000$) of Of/WN templates required, we argue that such a scenario might not be completely un-physical when analyzing integrated spectra of z$\sim2$ galaxies experiencing young bursts of star-formation at sub-solar metallicity\footnote{We note that because of degeneracies among templates, Of/WN could possibly be absorbing some early-WN contribution, hence we warn against any straightforward physical interpretation of the inferred number of templates for each stellar type}. 

Finally, we tried to include WR stars of the oxygen sequence (WO) in the fit. These stars are thought to represent the very final stages of the evolution of massive stars ($\approx 45-60$~\MSun) after the WC phase, but differ from the latter by means of a prominent, broad O VI~$\lambda$ emission at 3811-34~$\AA$ which reflects the high oxygen abundance expected near the end of core-helium burning and high stellar temperatures \citep{Sander_WR_2012, Tramper_WR-WO_2015}; alternative explanations invoke higher excitation conditions without necessarily requiring an enhanced oxygen abundance \citep{HillierMiller1999, Aadland2022}. 
Interestingly, WO templates are the only ones in the set producing a contribution to the $\approx5760 \AA$ red bump feature without significantly over-predicting the C IV features at $\approx5810 \AA$; at the same time, their contribution to the blue bump is not dissimilar to that of WC4-5 stars.
However, given their very short-lived phase, which makes them extremely rare (in the LMC there are only 3 WO stars currently known), and the absence of any evident O\textsc{iv}$\lambda 3811,34\AA$ emission in the \sourceshort spectrum, we consider the number of 64317 WO stars as inferred in this scenario largely implausible, even when accounting for resolution effects and possible strong degeneracies among the different templates that can cause significant overfitting.

\section{An intermediate-mass black hole?}
\label{sec:appendix_D}

Here, we discuss an alternative scenario to explain the observed broad \Halpha component, pointing toward the possible presence of a small, vigorously accreting black hole.
Under such a scenario, we assume that the broad H$\alpha$ emission arises from the broad--line region (BLR), and we applied the single--epoch virial method of \cite{Reines_BH_mass_2013} to derive the black-hole mass:
\begin{equation}
\log\left(\frac{M_{\mathrm{BH}}}{M_\odot}\right) = 6.57
+ 0.47\,\log\left(\frac{L_{\mathrm{H\alpha}}}{10^{42}\ \mathrm{erg\ s^{-1}}}\right)
+ 2.06\,\log\left(\frac{\mathrm{FWHM}_{\mathrm{H\alpha}}}{10^{3}\ \mathrm{km\ s^{-1}}}\right),
\end{equation}
where $L_{\mathrm{H\alpha}}$ is the dust--corrected broad H$\alpha$ luminosity and $\mathrm{FWHM}_{\mathrm{H\alpha}}$ is the intrinsic FWHM of the broad \Halpha component, corrected for instrumental resolution. With $L_{\mathrm{H\alpha,br}}\approx 2.31\times 10^{41}\ \mathrm{erg\ s^{-1}}$ and $\mathrm{FWHM}_{\mathrm{H\alpha,int}}=425\ \mathrm{km\ s^{-1}}$, we obtain $M_{\mathrm{BH}} \approx 3\times 10^{5}\ M_\odot$.
We then estimated the bolometric luminosity using the empirical scaling of \cite{Greene_Ho_BHmass_Ha_2005, Greene_AGN_mass_function_2007, Stern_AGN_I_2012}, $L_{\mathrm{bol}} \approx 130 \times L_{\mathrm{H\alpha,br}}$,
which yields $L_{\mathrm{bol}} \approx 3.0\times 10^{43}$ erg s$^{-1}$. The corresponding Eddington luminosity is
\begin{equation}
L_{\mathrm{Edd}} = 1.26\times 10^{38}\ \left(\frac{M_{\mathrm{BH}}}{M_\odot}\right)\ \mathrm{erg\ s^{-1}} \approx 4.03\times 10^{43}\ \mathrm{erg\ s^{-1}},
\end{equation}
and thus the inferred Eddington ratio is $\lambda_{\mathrm{Edd}} \equiv \frac{L_{\mathrm{bol}}}{L_{\mathrm{Edd}}} \approx 0.74$.

Such a vigorously accreting intermediate--mass black hole could, in principle, produce the observed broad component without dominating the narrow--line ISM diagnostics if the narrow--line region is weak, obscured, or strongly diluted by star-formation. Furthermore, it could represent a significant (if not the main) powering source of the OI$\lambda8446$ emission discussed in Sect.~\ref{ssec:discussion_OI8446}.

However, the modest $\mathrm{FWHM}\ (\sim 425$ km s$^{-1})$ and the lack of other strong AGN indicators still favor a stellar--feedback origin, and deeper multi-wavelength observations would be required to bring further evidence in favor of the faint-AGN scenario.
Finally, we note that the BH mass hereby estimated for \sourceshort leverages calibration relations obtained from galaxy samples with typical $\mathrm{FWHM}_{\Halpha} > 1000~ \mathrm{km~s}^{-1}$ and brighter luminosities ($L_{\mathrm{H\alpha}} > 10^{42}~\mathrm{erg~s^{-1}}$): therefore, the applicability of such BH mass calibrations to the case of \source is not well validated yet, due to the poor available statistics of intermediate-mass black holes.

\end{appendix}

\end{document}